\documentclass[11pt]{article}
\usepackage{hyperref,url}
\usepackage[utf8x]{inputenc} 
\usepackage{latexsym,graphicx,mathrsfs,amsfonts}
\usepackage{slashed, color}
\usepackage{multirow}
\usepackage{tikz}
\usepackage{empheq}
\usepackage{amsmath}
\allowdisplaybreaks[2]
\usepackage{color}
\usepackage[multiple]{footmisc}%for multiple footnotes
\usepackage{textcomp}
\usepackage{amssymb}%for semidirect product symbol
\usetikzlibrary{positioning}
\usepackage{amscd}
\usepackage{amsthm}
\usepackage{array} 

\renewcommand{\thefootnote}{\fnsymbol{footnote}}
\numberwithin{equation}{section}
\usepackage{dsfont}
\usepackage[toc,page]{appendix}

\newcommand*\widefbox[1]{\fbox{\hspace{2em}#1\hspace{2em}}}%Used together with the empheq package

\DeclareFontFamily{U}{MnSymbolC}{}
\DeclareSymbolFont{MnSyC}{U}{MnSymbolC}{m}{n}
\DeclareFontShape{U}{MnSymbolC}{m}{n}{
	<-6>  MnSymbolC5
	<6-7>  MnSymbolC6
	<7-8>  MnSymbolC7
	<8-9>  MnSymbolC8
	<9-10> MnSymbolC9
	<10-12> MnSymbolC10
	<12->   MnSymbolC12}{}
\DeclareMathSymbol{\intprod}{\mathbin}{MnSyC}{'270}
\newcommand{\ov}{\overline}

\newcommand{\CP}{\C P}

\newcommand{\gf}{\mathfrak{g}}

\newcommand{\ket}[1]{|#1\rangle}

\newcommand{\vev}[1]{\langle #1 \rangle}

\newcommand{\til}{\widetilde}

\let\nc\newcommand
\let\renc\renewcommand
\nc{\wbar}{\overline}
\let\td\tilde
\let\wtd\widetilde
\let\wht\widehat
\let\mcl\mathcal

\nc{\ab}{{\bar{a}}} \nc{\at}{\tilde{a}} \nc{\ah}{\hat{a}}
\nc{\bb}{{\bar{b}}} 
\nc{\bh}{\hat{b}}
\nc{\cb}{{\bar{c}}} \nc{\ct}{\tilde{c}} %\nc{\ch}{\hat{c}}
\nc{\db}{{\bar{d}}} \nc{\dt}{\tilde{d}} \renc{\dh}{\hat{d}}
\nc{\eb}{{\bar{e}}} \nc{\et}{\tilde{e}} \nc{\eh}{\hat{e}}
\nc{\fb}{{\bar{f}}} \nc{\ft}{\tilde{f}} \nc{\fh}{\hat{f}}
\nc{\gb}{{\bar{g}}} \nc{\gt}{\tilde{g}} \nc{\gh}{\hat{g}}
\nc{\ib}{{\bar{\imath}}} \nc{\ih}{\hat{\imath}} %\nc{\it}{\tilde{\imath}}
\nc{\jb}{{\bar{\jmath}}} \nc{\jt}{\tilde{\jmath}} \nc{\jh}{\hat{\jmath}}
\nc{\kb}{{\bar{k}}} \nc{\kt}{\tilde{k}} \nc{\kh}{\hat{k}}
\nc{\lb}{{\bar{l}}} \nc{\lt}{\tilde{l}} \nc{\lh}{\hat{l}}
\nc{\mb}{{\bar{m}}} \nc{\mt}{\tilde{m}} \nc{\mh}{\hat{m}}
\nc{\nb}{{\bar{n}}} \nc{\nt}{\tilde{n}} \nc{\nh}{\hat{n}}
\nc{\ob}{{\bar{o}}} \nc{\ot}{\tilde{o}} \nc{\oh}{\hat{o}}
\nc{\pb}{{\bar{p}}} \nc{\pt}{\tilde{p}} \nc{\ph}{\hat{p}}
\nc{\qb}{{\bar{q}}} \nc{\qt}{\tilde{q}} \nc{\qh}{\hat{q}}
\nc{\rb}{{\bar{r}}} \nc{\rt}{\tilde{r}} \nc{\rh}{\hat{r}}
\renc{\sb}{{\bar{s}}} \nc{\st}{\tilde{s}} \nc{\sh}{\hat{s}}
\nc{\tb}{{\bar{t}}} \renc{\th}{\hat{t}} %\nc{\tt}{\tilde{t}}
\nc{\ub}{{\bar{u}}} \nc{\ut}{\tilde{u}} \nc{\uh}{\hat{u}}
\nc{\vb}{{\bar{v}}} \nc{\vt}{\tilde{v}} \nc{\vh}{\hat{v}}
\nc{\wt}{\tilde{w}} \nc{\wh}{\hat{w}}
\nc{\xb}{{\bar{x}}} \nc{\xt}{\tilde{x}} \nc{\xh}{\hat{x}}
\nc{\yb}{{\bar{y}}} \nc{\yt}{\tilde{y}} \nc{\yh}{\hat{y}}
\nc{\zb}{{\bar{z}}} \nc{\zt}{\tilde{z}} 
\nc{\Ab}{\wbar{A}} \nc{\At}{\wtd{A}} \nc{\Ah}{\wht{A}}
\nc{\Bb}{\wbar{B}} \nc{\Bt}{\wtd{B}} \nc{\Bh}{\wht{B}}
\nc{\Cb}{\wbar{C}} \nc{\Ct}{\wtd{C}} \nc{\Ch}{\wht{C}}
\nc{\Db}{\wbar{D}} \nc{\Dt}{\wtd{D}} \nc{\Dh}{\wht{D}}
\nc{\Eb}{\wbar{E}} \nc{\Et}{\wtd{E}} \nc{\Eh}{\wht{E}}
\nc{\Fb}{\wbar{F}} \nc{\Ft}{\wtd{F}} \nc{\Fh}{\wht{F}}
\nc{\Gb}{\wbar{G}} \nc{\Gt}{\wtd{G}} \nc{\Gh}{\wht{G}}
\nc{\Hb}{\wbar{H}} \nc{\Ht}{\wtd{H}} \nc{\Hh}{\wht{H}}
\nc{\Ib}{\wbar{I}} \nc{\It}{\wtd{I}} \nc{\Ih}{\wht{I}}
\nc{\Jb}{\wbar{J}} \nc{\Jt}{\wtd{J}} \nc{\Jh}{\wht{J}}
\nc{\Kb}{\wbar{K}} \nc{\Kt}{\wtd{K}} \nc{\Kh}{\wht{K}}
\nc{\Lb}{\wbar{L}} \nc{\Lt}{\wtd{L}} \nc{\Lh}{\wht{L}}
\nc{\Mb}{\wbar{M}} \nc{\Mt}{\wtd{M}} \nc{\Mh}{\wht{M}}
\nc{\Nb}{\wbar{N}} \nc{\Nt}{\wtd{N}} \nc{\Nh}{\wht{N}}
\nc{\Ob}{\wbar{O}} \nc{\Ot}{\wtd{O}} \nc{\Oh}{\wht{O}}
\nc{\Pb}{\wbar{P}} \nc{\Pt}{\wtd{P}} \nc{\Ph}{\wht{P}}
\nc{\Qb}{\wbar{Q}} \nc{\Qt}{\wtd{Q}} \nc{\Qh}{\wht{Q}}
\nc{\Rb}{\wbar{R}} \nc{\Rt}{\wtd{R}} \nc{\Rh}{\wht{R}}
\nc{\Sb}{\wbar{S}} \nc{\St}{\wtd{S}} \nc{\Sh}{\wht{S}}
\nc{\Tb}{\wbar{T}} \nc{\Tt}{\wtd{T}} \nc{\Th}{\wht{T}}
\nc{\Ub}{\wbar{U}} \nc{\Ut}{\wtd{U}} \nc{\Uh}{\wht{U}}
\nc{\Vb}{\wbar{V}} \nc{\Vt}{\wtd{V}} \nc{\Vh}{\wht{V}}
\nc{\Wb}{\wbar{W}} \nc{\Wt}{\wtd{W}} \nc{\Wh}{\wht{W}}
\nc{\Xb}{\wbar{X}} \nc{\Xt}{\wtd{X}} \nc{\Xh}{\wht{X}}
\nc{\Yb}{\wbar{Y}} \nc{\Yt}{\wtd{Y}} \nc{\Yh}{\wht{Y}}
\nc{\Zb}{\wbar{Z}} \nc{\Zt}{\wtd{Z}} \nc{\Zh}{\wht{Z}}

\nc{\CA}{\mcl{A}} \nc{\CAb}{\wbar{\CA}} \nc{\CAt}{\wtd{\CA}} \nc{\CAh}{\wht{\CA}}
\nc{\CB}{\mcl{B}} \nc{\CBb}{\wbar{\CB}} \nc{\CBt}{\wtd{\CB}} \nc{\CBh}{\wht{\CB}}
\nc{\CC}{\mcl{C}} \nc{\CCb}{\wbar{\CC}} \nc{\CCt}{\wtd{\CC}} \nc{\CCh}{\wht{\CC}}
\nc{\cDt}{\wtd{\cC}} \nc{\cDh}{\wht{\cD}}
\nc{\CE}{\mcl{E}} \nc{\CEb}{\wbar{\CE}} \nc{\CEt}{\wtd{\CE}} \nc{\CEh}{\wht{\CE}}
\nc{\CF}{\mcl{F}} \nc{\CFb}{\wbar{\CF}} \nc{\CFt}{\wtd{\CF}} \nc{\CFh}{\wht{\CF}}
\nc{\CG}{\mcl{G}} \nc{\CGb}{\wbar{\CG}} \nc{\CGt}{\wtd{\CG}} \nc{\CGh}{\wht{\CG}}
\nc{\CH}{\mcl{H}} \nc{\CHb}{\wbar{\CH}} \nc{\CHt}{\wtd{\CH}} \nc{\CHh}{\wht{\CH}}
\nc{\CI}{\mcl{I}} \nc{\CIb}{\wbar{\CI}} \nc{\CIt}{\wtd{\CI}} \nc{\CIh}{\wht{\CI}}
\nc{\CJ}{\mcl{J}} \nc{\CJb}{\wbar{\CJ}} \nc{\CJt}{\wtd{\CJ}} \nc{\CJh}{\wht{\CJ}}
\nc{\CK}{\mcl{K}} \nc{\CKb}{\wbar{\CK}} \nc{\CKt}{\wtd{\CK}} \nc{\CKh}{\wht{\CK}}
\nc{\CL}{\mcl{L}} \nc{\CLb}{\wbar{\CL}} \nc{\CLt}{\wtd{\CL}} \nc{\CLh}{\wht{\CL}}
\nc{\CM}{\mcl{M}} \nc{\CMb}{\wbar{\CM}} \nc{\CMt}{\wtd{\CM}} \nc{\CMh}{\wht{\CM}}
\nc{\CN}{\mcl{N}} \nc{\CNb}{\wbar{\CN}} \nc{\CNt}{\wtd{\CN}} \nc{\CNh}{\wht{\CN}}
\nc{\CO}{\mcl{O}} \nc{\COb}{\wbar{\CO}} \nc{\COt}{\wtd{\CO}} \nc{\COh}{\wht{\CO}}
\nc{\CQ}{\mcl{Q}} \nc{\CQb}{\wbar{\CQ}} \nc{\CQt}{\wtd{\CQ}} \nc{\CQh}{\wht{\CQ}}
\nc{\CR}{\mcl{R}} \nc{\CRb}{\wbar{\CR}} \nc{\CRt}{\wtd{\CR}} \nc{\CRh}{\wht{\CR}}
\nc{\CS}{\mcl{S}} \nc{\CSb}{\wbar{\CS}} \nc{\CSt}{\wtd{\CS}} \nc{\CSh}{\wht{\CS}}
\nc{\CT}{\mcl{T}} \nc{\CTb}{\wbar{\CT}} \nc{\CTt}{\wtd{\CT}} \nc{\CTh}{\wht{\CT}}
\nc{\CU}{\mcl{U}} \nc{\CUb}{\wbar{\CU}} \nc{\CUt}{\wtd{\CU}} \nc{\CUh}{\wht{\CU}}
\nc{\CV}{\mcl{V}} \nc{\CVb}{\wbar{\CV}} \nc{\CVt}{\wtd{\CV}} \nc{\CVh}{\wht{\CV}}
\nc{\CW}{\mcl{W}} \nc{\CWb}{\wbar{\CW}} \nc{\CWt}{\wtd{\CW}} \nc{\CWh}{\wht{\CW}}
\nc{\CX}{\mcl{X}} \nc{\CXb}{\wbar{\CX}} \nc{\CXt}{\wtd{\CX}} \nc{\CXh}{\wht{\CX}}
\nc{\CY}{\mcl{Y}} \nc{\CYb}{\wbar{\CY}} \nc{\CYt}{\wtd{\CY}} \nc{\CYh}{\wht{\CY}}
\nc{\CZ}{\mcl{Z}} \nc{\CZb}{\wbar{\CZ}} \nc{\CZt}{\wtd{\CZ}} \nc{\CZh}{\wht{\CZ}}

\let\eps\epsilon
\let\ups\upsilon
\let\veps\varepsilon
\let\vtht\vartheta
\let\vsgm\varsigma
\let\vphi\varphi
\let\vrho\varrho

\nc{\alphab}{\bar{\alpha}} \nc{\alphat}{\td{\alpha}} \nc{\alphah}{\hat{\alpha}}
\nc{\betab}{\bar{\beta}}   \nc{\betat}{\td{\beta}}   \nc{\betah}{\hat{\beta}} 
\nc{\gammab}{\bar{\gamma}} \nc{\gammat}{\td{\gamma}} \nc{\gammah}{\hat{\gamma}} 
\nc{\deltab}{\bar{\delta}} \nc{\deltat}{\td{\delta}} \nc{\deltah}{\hat{\delta}} 
\nc{\epsilonb}{\bar{\eps}} \nc{\epsilont}{\td{\eps}} \nc{\epsilonh}{\hat{\eps}} 
\nc{\vepsb}{\bar{\veps}}   \nc{\vepst}{\td{\veps}}   \nc{\vepsh}{\hat{\veps}} 
\nc{\zetab}{\bar{\zeta}}   \nc{\zetat}{\td{\zeta}}   \nc{\zetah}{\hat{\zeta}} 
\nc{\etab}{\bar{\eta}}     
\nc{\etah}{\hat{\eta}} 
\nc{\thetab}{\bar{\theta}} \nc{\thetat}{\td{\theta}} \nc{\thetah}{\hat{\theta}} 
\nc{\vthetab}{\bar{\vtht}} \nc{\vthetat}{\td{\vtht}} \nc{\vthetah}{\hat{\vtht}} 
\nc{\lambdat}{\td{\lambda}} \nc{\lambdah}{\hat{\lambda}} 
\nc{\iotab}{\bar{\iota}}   \nc{\iotat}{\td{\iota}}   \nc{\iotah}{\hat{\iota}} 
\nc{\kappab}{\bar{\kappa}} \nc{\kappat}{\td{\kappa}} \nc{\kappah}{\hat{\kappa}} 
\nc{\lmdb}{\bar{\lmd}}     \nc{\lmdt}{\td{\lmd}}     \nc{\lmdh}{\hat{\lmd}} 
\nc{\mub}{\bar{\mu}}       \nc{\mut}{\td{\mu}}       \nc{\muh}{\hat{\mu}} 
\nc{\nub}{\bar{\nu}}       \nc{\nut}{\td{\nu}}       \nc{\nuh}{\hat{\nu}} 
\nc{\xib}{\bar{\xi}}       \nc{\xit}{\td{\xi}}       \nc{\xih}{\hat{\xi}} 
\nc{\pib}{\bar{\pi}}       \nc{\pit}{\td{\pi}}       \nc{\pih}{\hat{\pi}} 
\nc{\vpib}{\bar{\vpi}}     \nc{\vpit}{\td{\vpi}}     \nc{\vpih}{\hat{\vpi}} 
\nc{\rhob}{\bar{\rho}}     \nc{\rhot}{\td{\rho}}     \nc{\rhoh}{\hat{\rho}} 
\nc{\vrhob}{\bar{\vrho}}   \nc{\vrhot}{\td{\vrho}}   \nc{\vrhoh}{\hat{\vrho}} 
\nc{\sigmab}{\bar{\sigma}} \nc{\sigmat}{\td{\sigma}} \nc{\sigmah}{\hat{\sigma}} 
\nc{\vsigmab}{\bar{\vsgm}} \nc{\vsigmat}{\td{\vsgm}} \nc{\vsigmah}{\hat{\vsgm}} 
\nc{\taub}{\bar{\tau}}     \nc{\taut}{\td{\tau}}     \nc{\tauh}{\hat{\tau}} 
\nc{\upsb}{\bar{\ups}} \nc{\upst}{\td{\ups}} \nc{\upsh}{\hat{\ups}} 
\nc{\phib}{\bar{\phi}}     \nc{\phit}{\td{\phi}}     \nc{\phih}{\hat{\phi}} 
\nc{\varphib}{\bar{\vphi}}   \nc{\varphit}{\td{\vphi}}   \nc{\varphih}{\hat{\vphi}} 
\nc{\chib}{\bar{\chi}}     
\nc{\chih}{\hat{\chi}} 
\nc{\psib}{\bar{\psi}}     
\nc{\psih}{\hat{\psi}} 
\nc{\omegab}{\bar{\omega}} \nc{\omegat}{\td{\omega}} \nc{\omegah}{\hat{\omega}} 

\nc{\Gammab}{\wbar{\Gamma}}     \nc{\Gammat}{\wtd{\Gamma}}     \nc{\Gammah}{\wht{\Gamma}}
\nc{\Deltab}{\wbar{\Delta}}     \nc{\Deltat}{\wtd{\Delta}}     \nc{\Deltah}{\wht{\Delta}}
\nc{\Thetab}{\wbar{\Theta}}     \nc{\Thetat}{\wtd{\Theta}}     \nc{\Thetah}{\wht{\Theta}}
\nc{\Lambdab}{\wbar{\Lambda}}   \nc{\Lambdat}{\wtd{\Lambda}}   \nc{\Lambdah}{\wht{\Lambda}}
\nc{\Xib}{\wbar{\Xi}}           \nc{\Xit}{\wtd{\Xi}}           \nc{\Xih}{\wht{\Xi}}
\nc{\Pib}{\wbar{\Pi}}           \nc{\Pit}{\wtd{\Pi}}           \nc{\Pih}{\wht{\Pi}}
\nc{\Sigmab}{\wbar{\Sigma}}     \nc{\Sigmat}{\wtd{\Sigma}}     \nc{\Sigmah}{\wht{\Sigma}}
\nc{\Upsilonb}{\wbar{\Upsilon}} \nc{\Upsilont}{\wtd{\Upsilon}} \nc{\Upsilonh}{\wht{\Upsilon}}
\nc{\Phib}{\wbar{\Phi}}         \nc{\Phit}{\wtd{\Phi}}         \nc{\Phih}{\wht{\Phi}}
\nc{\Psib}{\wbar{\Psi}}         \nc{\Psit}{\wtd{\Psi}}         \nc{\Psih}{\wht{\Psi}}
\nc{\Omegab}{\wbar{\Omega}}     \nc{\Omegat}{\wtd{\Omega}}     \nc{\Omegah}{\wht{\Omega}}
\nc{\Varepsilon}{\mathcal{E}}

\newcommand{\wta}{{\widetilde{a}}}
\newcommand{\wtb}{{\widetilde{b}}}
\newcommand{\wtc}{{\widetilde{c}}}

\newcommand{\quar}{\frac{1}{4}}

\newcommand{\cD}{{\cal D}}

\nc{\balpha}{\bar{\alpha}}
\nc{\bbeta}{\bar{\beta}}
\nc{\bgamma}{\bar{\gamma}}
\nc{\bm}{\bar{m}}
\nc{\bn}{\bar{n}}
\nc{\bp}{\bar{p}}
\nc{\al}{\alpha}
\nc{\bt}{\beta}
\nc{\gm}{\gamma}
\nc{\zh}{\wht{z}}
\nc{\zhb}{\ov{\wht{z}}}
\nc{\mbh}{\wht{\ov{m}}}
\nc{\bc}{|_{x^3=0}}

\nc{\tal}{\til{\al}}
\nc{\tbt}{\til{\bt}}
\nc{\tgm}{\til{\gm}}

\nc{\wb}{\ov{w}}
\nc{\teta}{\til{\eta}}
\nc{\tpsi}{\til{\psi}}

\def\IL{\relax{\rm I\kern-.18em L}}
\def\IH{\relax{\rm I\kern-.18em H}}
\def\IB{\relax{\rm I\kern-.18em B}}
\def\ID{\relax{\rm I\kern-.18em D}}
\def\IE{\relax{\rm I\kern-.18em E}}
\def\IF{\relax{\rm I\kern-.18em F}}

\def\IG{\relax\hbox{$\inbar\kern-.3em{\rm G}$}}
\def\IGa{\relax\hbox{${\rm I}\kern-.18em\Gamma$}}
\def\IH{\relax{\rm I\kern-.18em H}}
\def\II{\relax{\rm I\kern-.18em I}}
\def\IK{\relax{\rm I\kern-.18em K}}
\def\IP{\relax{\rm I\kern-.18em P}}
\def\IQ{\relax\hbox{$\inbar\kern-.3em{\rm Q}$}}

\def\hat{\widehat}
\def\CM {{\cal M}}
\def\CN {{\cal N}}
\def\CR {{\cal R}}
\def\CD {{\cal D}}
\def\CF {{\cal F}}
\def\CJ {{\cal J}}
\def\CP {{\cal P }}
\def\CL {{\cal L}}
\def\CV {{\cal V}}
\def\CO {{\cal O}}
\def\CZ {{\cal Z}}
\def\CE {{\cal E}}
\def\CG {{\cal G}}
\def\CH {{\cal H}}
\def\CC {{\cal C}}
\def\CB {{\cal B}}
\def\CS {{\cal S}}
\def\CA{{\cal A}}
\def\CK{{\cal K}}
\def\CQ{{\cal Q}}

\def\p{\partial}
\def\pb{{\bar \p}}

\def\vt#1#2#3{ {\vartheta[{#1 \atop  #2}](#3\vert \tau)} }

\def\jb{{\bar j}}

\def\inbar{\,\vrule height1.5ex width.4pt depth0pt}

\def\half{{1 \over 2}}

\newcommand{\abrac}[1]{\left\langle #1\right\rangle}
\newcommand{\cbrac}[1]{\left(#1\right)}
\newcommand{\sbrac}[1]{\left[#1\right]}
\newcommand{\ccbrac}[1]{\left\{#1\right\}}
\newcommand{\vbrac}[1]{\left|#1\right|}
\newcommand{\mbrac}[1]{\left.\left[#1\right.\right\}}
\newcommand{\tr}{\text{Tr}}

\newcommand{\susy}[1]{$\mathcal{N}=#1$}
\newcommand{\flatmoduli}[1]{\mathcal{M}_{\text{flat}}\left(#1\right)}
\newcommand{\instantonmoduli}[2]{\mathcal{M}^#1_{\text{inst}}\left(#2\right)}
\newcommand{\monopolemoduli}[2]{\mathcal{M}^#1_{\text{mono}}\left(#2\right)}
\newcommand{\mapsmoduli}[2]{\mathcal{M}^#1_{\text{maps}}\left(#2\right)}
\newcommand{\vortexmoduli}[2]{\mathcal{M}^#1_{\text{vort}}\left(#2\right)}
\newcommand{\qcharge}{\mathcal{Q}}
\newcommand{\HFinst}[1]{\text{HF}_*^{\text{inst}}\cbrac{#1}}

\newcommand{\HFlagr}[1]{\text{HF}_*^{\text{Lagr}}\cbrac{#1}}
\newcommand{\HFmono}[1]{\text{HF}_*^{\text{mono}}\cbrac{#1}}
\newcommand{\HFheeg}[1]{\text{HF}_*^{\text{Heeg}}\cbrac{#1}}
\newcommand{\cHFinst}[1]{\text{HF}^*_{\text{inst}}\cbrac{#1}}
\newcommand{\cHFsymp}[1]{\text{HF}^*_{\text{symp}}\cbrac{#1}}

\newcommand{\cHFmono}[1]{\text{HF}^*_{\text{mono}}\cbrac{#1}}

\newcommand{\Qch}[1]{QH^*(#1)}

\newcommand{\id}{\mathds{1}}
\newcommand{\e}[1]{\text{e}^{#1}}

\newcommand{\SU}[1]{SU\cbrac{#1}}
\newcommand{\U}[1]{U\cbrac{#1}}

\newcommand{\real}{\mathbb{R}}

\newcommand{\integer}{\mathbb{Z}}

\newcommand{\mo}{\mathcal{O}}
\newcommand{\maxtorus}{\mathbb{T}}

\let\OLDthebibliography\thebibliography
\renewcommand\thebibliography[1]{
	\OLDthebibliography{#1}
	\setlength{\parskip}{5pt}
	\setlength{\itemsep}{0pt plus 0.3ex}
}
\usepackage{titlesec}
\titleformat*{\section}{\bfseries\large}
\begin{document}
\addtolength{\baselineskip}{1.5mm}

\thispagestyle{empty}
%\begin{flushright}
%hep-th/   \\
%\end{flushright}
\vbox{}
\vspace{3.0cm}

\begin{center}
	\centerline{\LARGE{Boundary \susy{2} Theory, Floer Homologies, Affine Algebras,}}
	\bigskip
	\centerline{\LARGE{and the Verlinde Formula}} 
	%\centerline{\LARGE{Higher-dimensional Gauge Theory and}}
	%\medskip
	%\centerline{\LARGE{Categorifying Integrability}} 

	\vspace{3.0cm}
		
	{Meer~Ashwinkumar\footnote{E-mail: meerashwinkumar@u.nus.edu}, Kee-Seng~Png\footnote{E-mail: pngkeeseng@u.nus.edu}, and Meng-Chwan~Tan\footnote{E-mail: mctan@nus.edu.sg}}
	\\[2mm]
	{\it Department of Physics\\
		National University of Singapore \\%\\[1mm]
		2 Science Drive 3, Singapore 117551} \\[1mm] 
\end{center}

\vspace{2.0cm}

\centerline{\bf Abstract}\smallskip \noindent

Generalizing our ideas in \href{https://arxiv.org/abs/1006.3313}{[arXiv:1006.3313]}, we explain how topologically-twisted \susy{2} gauge theory on a four-manifold with boundary, will allow us to furnish purely physical proofs of (i) the Atiyah-Floer conjecture, (ii) Mu\~noz's theorem relating quantum and instanton Floer cohomology, (iii) their monopole counterparts, and (iv) their higher rank generalizations. In the case where the boundary is a Seifert manifold, one can also relate its instanton Floer homology to modules of an affine algebra via a 2d A-model with target the based loop group. As an offshoot, we will be able to demonstrate an action of the affine algebra on the quantum cohomology of the moduli space of flat connections on a Riemann surface, as well as derive the Verlinde formula.

\newpage

\renewcommand{\thefootnote}{\arabic{footnote}}
\setcounter{footnote}{0}

\tableofcontents
\section{Introduction, Summary and Conventions}

%\bigskip\noindent\textit{Introduction}

Four-dimensional (cohomological) topological quantum field theories (TQFTs), such as Donaldson-Witten (DW) theory and Seiberg-Witten (SW) theory, have been studied extensively over the years since they were first conceived in \cite{witten1988topological}. Such a TQFT possesses an underlying topological supersymmetry, which is generated by a scalar supercharge $\CQ$. Observables of the theory are identified with the $\CQ$-cohomology, and they are invariant under topological deformations. In the aforementioned theories, they generate the celebrated Donaldson and Seiberg-Witten invariants, respectively.

When DW theory is defined on a four-manifold with boundary $Y_3$, the corresponding Donaldson invariants are valued in the instanton Floer homology of $Y_3$ \cite{witten1988topological}. Likewise, for SW theory, the corresponding Seiberg-Witten invariants are valued in the monopole Floer homology of $Y_3$ \cite{marcolli1999seiberg}. 

One may also study DW theory defined on a four-manifold of the form $M_4=\Sigma\times C$, where $\Sigma$ and $C$ are compact Riemann surfaces. Upon shrinking $C$ (or $\Sigma$), it is known that the ensuing 2d theory is an A-twisted sigma model on $\Sigma$ (or $C$) with target being the moduli space of flat connections on $C$ (or $\Sigma$) \cite{bershadsky1995topological}. The relevant observables of the A-model are identified with its $\CQ$-cohomology, which in turn describe the quantum cohomology of the target space \cite{bershadsky1995topological,witten1988topologicalsig}. Thus, the $\CQ$-cohomology of DW theory must also be identified with quantum cohomology, i.e. Donaldson invariants on $M_4$ are related to the quantum cohomology of the space of flat connections on either Riemann surface \cite{gukov2009surface}. 

In \cite{gukov2009surface}, it was suggested that upon shrinking $C$ (or $\Sigma$), SW theory should also descend to an A-model on $\Sigma$ (or $C$), albeit with target being the moduli space of vortices on $C$ (or $\Sigma$). Likewise, the $\CQ$-cohomology of SW theory must also be identified with quantum cohomology, i.e. Seiberg-Witten invariants on $M_4$ are related to the quantum  cohomology of the space of vortices on either Riemann surface \cite{gukov2009surface}. 
 
%It is also known that the moduli space of flat $G$-connections on a disk $D$, is isomorphic to the based loop group $\Omega G$ \cite{popov2015loop,pressley1986loop}. Hence, upon shrinking $D$, DW theory on $M_4=\Sigma\times D$ becomes an A-model on $\Sigma$ with target $\Omega G$, where $G=SU(2)$. Furthermore, it was shown in \cite{ashwinkumar2018little} that such an A-model possess affine symmetry, and that the $\CQ$-cohomology classes of the A-model  correspond to modules of an affine algebra $\mathfrak{g}_{\text{aff}}$. Then, the $\CQ$-cohomology of DW theory can be identified with the space of $\mathfrak{g}_{\text{aff}}$-modules on $\Sigma$, i.e. Donaldson invariants valued in instanton Floer homology ought to be related to $\mathfrak{g}_{\text{aff}}$-modules on $\Sigma$.

The main aim of this paper is to exploit properties of 4d ${\cal N} =2$ TQFT in the manner described above, so that physical proofs of known mathematical conjectures and theorems as well as derivations of mathematically novel identities between 3d and 2d invariants, and more, can be obtained. %Specifically, physical states of a 4d TQFT, must be identified with those on its spatial boundary, and also with those obtained upon shrinking $C$ away. Hence, studying their mathematical analogs -- i.e. the 2d and 3d invariants -- relevant identifications can be established. In particular, we will study DW and SW theory on a suitable four-manifold and carry out topological deformations to obtain identities for the corresponding mathematical invariants.

%As a secondary aim, we will also verify the claim in \cite{gukov2009surface} that SW theory leads to an A-model with a target being the moduli space of vortices, upon shrinking $C$. 

%We further take $M_4$ to be a spin manifold -- i.e. the second Stiefel-Whitney class $w_2(M_4)=0$.

Let us now give a brief plan and summary of the paper.

\bigskip\noindent\textit{A Brief Plan and Summary of the Paper}

In \S\ref{sec:2}, we provide a relevant review of DW theory, and its relation to instanton Floer homology when the underlying four-manifold $M_4$ has boundary $Y_3$. In particular, if the moduli space of instantons is zero-dimensional, the partition function on $M_4$ is expressed as a sum of instanton Floer homology classes $\Psi_{\text{inst}}$ %(in (\ref{ob:DWfloer}))
\begin{equation}
	\boxed{Z_{M_4}=\vev{1}_{\Psi(\Phi_{Y_3})} = \sum_i  \Psi_{\textrm{inst}}(\Phi^i_{Y_3})}
\end{equation}

We will also take $M_4=\Sigma\times C$, and shrink $C$, to obtain a sigma model on $\Sigma$ with target $\CM(C)$. This space is characterized by %(\ref{con:flatnessgen})
\begin{equation}
\boxed{F_{ab}=0}
\end{equation}
which corresponds to the moduli space of flat connections on $C$, which we will denote by $\flatmoduli{C}$. Moreover, upon shrinking $C$, since the topological term of the form $F\wedge F$ leads to the pullback, i.e % (\ref{charge:degreepullback}) -- i.e. 
\begin{equation}
\boxed{{1\over 8\pi^2}\int_{M_4}\ \tr \cbrac{F\wedge F}=\int_{\Sigma}\ X^*\omega_{\text{flat}}}
\end{equation}
the ensuing 2d model on the remaining Riemann surface $\Sigma$ must be an A-twisted sigma model with target $\flatmoduli{C}$, and action %(\ref{action:DWsigma}) is
\begin{equation}
\boxed{\begin{aligned}
	S_{\text{DW}}'&={1\over e^2}\int_{\Sigma} d^2z\ \left(G^{\textrm{flat}}_{I\ov J}\cbrac{\half\partial_{z}X^{I}\partial_{\ov{z}}{X}^{\ov J}
		+\half\partial_{\ov{z}}X^{I}\partial_z{X}^{\ov J}
		+{{\rho}_{z}}^{\ov J}\nabla_{\ov{z}}\chi^{I}
		+{\rho_{\ov{z}}}^{I}\nabla_z{\chi}^{\ov J}}\right.\\
	&\qquad\qquad\qquad\left.\phantom{\half}
	-R_{I\ov JK\ov L}{{\rho}_{\ov z}}^{I}{\rho_{z}}^{\ov J}{\chi}^K\chi^{\ov L}\right)
	+{i\theta}\int_{\Sigma}\ X^*\omega_{\text{flat}}
	\end{aligned}}
\end{equation}

In \S\ref{sec:3}, we provide a review of SW theory, and its relation to monopole Floer homology when the underlying four-manifold $M_4$ has boundary $Y_3$. In particular, if the moduli space of monopoles is zero-dimensional, the partition function on $M_4$ is expressed as a sum of monopole Floer homology classes $\Psi_{\text{inst}}$ %(\ref{ob:SWfloer}). 
\begin{equation}
\boxed{Z_{M_4}=\vev{1}_{\Psi(\Phi_{Y_3})} = \sum_i  \Psi_{\textrm{mono}}(\Phi^i_{Y_3})}
\end{equation}

We will also take $M_4=\Sigma\times C$, and shrink $C$, to obtain a sigma model on $\Sigma$ with target $\CM(C)$. This space is characterized by %(\ref{eqn:vortex})
\begin{equation}
\boxed{\begin{aligned}
	&F_{w\ov{w}}={i\over 4}\cbrac{1-\vbrac{\varphi}^2}\\
	&D_{\ov{w}}\varphi=0
	\end{aligned}
}
\end{equation}
which corresponds to the moduli space of vortices on $C$, which we will denote by $\vortexmoduli{q}{C}$. Furthermore, upon shrinking $C$, a $\CQ$-exact topological term $S_{\text{top}}$, which can be added inconsequentially, leads to the pullback, i.e. %(\ref{charge:degreepullbackSW})
\begin{equation}
\boxed{S_{\text{top}}={1\over 8\pi^2}\int_{M_4}\ \CF\wedge \CF=\int_{\Sigma}\ X^*\omega_{\text{vort}}}
\end{equation}
Hence, the ensuing 2d model on $\Sigma$ must also be an A-model with target $\vortexmoduli{q}{C}$, and action%(\ref{action:SWsigma}) is
\begin{equation}
\boxed{\begin{aligned}
	S_{\text{SW}}'&={1\over e^2}\int_{\Sigma} d^2z\ \left(G^{\text{vort}}_{I\ov J}\cbrac{
		%	\quar\partial_{z}X^{I}\partial_{\ov{z}}{X}^{\ov J}
		%	+\quar\partial_{\ov{z}}X^{I}\partial_z{X}^{\ov J}
		\half\partial_{z}X^{I}\partial_{\ov{z}}{X}^{\ov J}
		+\half\partial_{\ov{z}}X^{I}\partial_z{X}^{\ov J}
		+{{\rho}_{z}}^{\ov J}\nabla_{\ov{z}}\chi^{I}
		+{\rho_{\ov{z}}}^{I}\nabla_z{\chi}^{\ov J}}\right.\\
	&\qquad\qquad\qquad\left.\phantom{\half}
	-R_{I\ov JK\ov L}{{\rho}_{\ov z}}^{I}{\rho_{z}}^{\ov J}{\chi}^K\chi^{\ov L}\right)
	+{i\theta}\int_{\Sigma}\ X^*\omega_{\text{vort}}
	%-{2\pi^2\over e^2}\int_{\Sigma}\ X^*\omega_{\text{vort}}
	\end{aligned}}
\end{equation}

In \S\ref{sec:4}, we provide physical proofs of various mathematical conjectures and theorems, by exploiting the fact that physical states of a TQFT are insensitive to topological deformations. We first prove the Atiyah-Floer conjecture, which relates instanton and Lagrangian intersection Floer homologies. This takes the form% (\ref{math:atiyahfloer})
\begin{equation}
\label{sum:4-1}
\boxed{\HFinst{Y_3}\cong\HFlagr{\flatmoduli{\Sigma},L_0,L_1}}
\end{equation}
This is proved by studying DW theory on the four-manifold $M_4=\real^+\times Y_3\cong \real^+\times I\times_f \Sigma$ (LHS of (\ref{sum:4-1})), and identifying its states with those of the A-model on $\real^+\times I$ with target $\flatmoduli{\Sigma}$ (RHS of (\ref{sum:4-1})).

The second mathematical claim is Mu\~noz's theorem, which relates instanton and symplectic Floer cohomologies, whereby the result of \cite{sadov1995equivalence} further relates symplectic Floer cohomology with quantum cohomology. It takes the form
\begin{equation}
\label{sum:4-2}
\boxed{\Qch{\flatmoduli{\Sigma}}\cong\cHFsymp{\flatmoduli{\Sigma}}\cong\cHFinst{\Sigma\times S^1}}
\end{equation}
This is proved by studying DW theory on the four-manifold $M_4=\Sigma\times S^1\times \real^+$ (RHS of (\ref{sum:4-2})), and identifying that with the A-model on $\real^+\times S^1$ with target $\flatmoduli{\Sigma}$ (centre of (\ref{sum:4-2})). The first equality follows from \cite{sadov1995equivalence}, as mentioned. 

The same analysis is carried out for SW theory, in which we consider monopole analogs of (\ref{sum:4-1}) and (\ref{sum:4-2}). In an analogous manner, we prove the monopole Atiyah-Floer conjecture, which takes the form% (\ref{math:atiyahfloermonopole})
\begin{equation}
\label{sum:4-3}
\boxed{\HFmono{q,Y_3}\cong\HFheeg{\vortexmoduli{q}{\Sigma},L_0,L_1}}
\end{equation}
Our analysis of SW theory also allows us to deduce the {\it mathematically novel} monopole analog of Mu\~noz's theorem, which takes the form %(\ref{math:munozmonopole})
\begin{equation}
\label{sum:4-4}
\boxed{\Qch{\vortexmoduli{q}{\Sigma}}\cong\cHFsymp{\vortexmoduli{q}{\Sigma}}\cong\cHFmono{q,\Sigma\times S^1}}
\end{equation}

These results can be generalized to higher rank gauge groups $G$, because the physical analysis is either independent of the choice of $G$, or simply involves a straightforward extension. In doing so, we can obtain higher rank generalizations of (\ref{sum:4-1})--(\ref{sum:4-4}).

In \S\ref{sec:5}, we consider DW theory on $M_4=\Sigma\times D\cong \Sigma\times S^1\times \real^+$, with gauge group $G=SU(2)$. Instanton Floer homology can be defined on $Y_3=\Sigma\times S^1$. Moreover, shrinking $D$ allows us to identify the target of the A-model on $\Sigma$ with the based loop group $\Omega G$. Such an A-model is known to possess affine symmetry \cite{ashwinkumar2018little}. The corresponding A-model states form modules of an affine Lie algebra $\mathfrak{g}_{\text{aff}}$, which span the space of $\frak g_{\textrm {aff}}$-modules on $\Sigma$   that we denote by $\mathfrak{G}_{\text{mod}}(\Sigma)$. We can identify the corresponding partition functions, and hence establish the {\it mathematically novel} isomorphism %in (\ref{math:affine-iso-HFi}), which takes the form
\begin{equation}
\label{sum:5-1}
\boxed{\HFinst{\Sigma\times S^1}\cong\mathfrak{G}_{\text{mod}}(\Sigma)}
\end{equation}

Next, we study DW theory on $M_4=\Sigma\times_f D\cong M_{g,p}\times \real^+$, where $M_{g,p}$ is a Seifert manifold, $g$ is the genus of $\Sigma$, and $p$ is the Chern number of the $S^1$ bundle which characterizes the nontriviality of the fibration. As such, we see that $\Sigma\times S^1$ is a trivially-fibered Seifert manifold -- i.e. $\Sigma\times S^1=M_{g,0}$. By inserting $p$ copies of the fibering operator $\CP$ \cite{blau2006chern} in the partition function over $M_{g,0}$, which has the effect of shifting the Chern number by $p$, the partition function over $M_{g,p}$ is obtained. We may then generalize (\ref{sum:5-1}) to the case where $\Sigma\times S^1$ is replaced by $M_{g,p}$. 
Denoting $\mathfrak{G}_{\text{mod},p}(\Sigma)$ as the space where each basis component is now acted upon by $p$ copies of a suitable representation of $\cal P$, we similarly show that there is a {\it mathematically novel} isomorphism %(\ref{math:affine-iso-HFi-p})
\begin{equation}
\label{sum:5-2}
\boxed{\HFinst{M_{g,p}}\cong\mathfrak{G}_{\text{mod},p}(\Sigma)}
\end{equation}

It is straightforward to see, from \eqref{sum:4-2} and \eqref{sum:5-1}, that there is yet another \textit{mathematically novel} identity of the form
\begin{equation}
\label{sum:5-3}
\boxed{\Qch{\flatmoduli{\Sigma}}\cong \mathfrak{G}_{\text{mod}}(\Sigma)}
\end{equation}

In \S\ref{sec:6}, as a preliminary step, we first study an A-model on $D\cong \real^+\times S^1$ with target $\flatmoduli{\Sigma}$. Since the theory is topological, we may further shrink $S^1$ so that we get a 1d sigma model which turns out to be a quantum mechanical model on $\flatmoduli{\Sigma}$ with action %(\ref{action:QM})
\begin{equation}
\label{sum:6.1}
\boxed{	S_{\text{QM}}={1\over \hbar}\int d\tau\ 
	\half\dot{X}^I\dot{X}_I}
\end{equation}
Thus, states of the A-model on $D$ with target $\flatmoduli{\Sigma}$ are identified with quantum mechanical states on $\flatmoduli{\Sigma}$. (\ref{sum:6.1}) also means we can write the commutator relations for the collective coordinates $X$ as %(\ref{eqn:commutatorspi})
\begin{equation}
\boxed{[\hat{X}^I,\hat{P}^J]=\hbar\delta^{IJ}}
\end{equation}
which amounts to quantizing $\flatmoduli{\Sigma}$. 

Next, we make use of DW theory on $M_4=\Sigma\times D$, to obtain, upon shrinking $D$, an A-model on $\Sigma$ with target $\Omega G$, and, upon shrinking $\Sigma$, an A-model on $D$ with target $\flatmoduli{\Sigma}$. In doing so, we can derive Falting's definition of the Verlinde formula \cite{faltings1994proof,beauville1994conformal} %(\ref{math:verlindeunderlying})
\begin{equation}
\label{sum:6.3}
\boxed{{V}_{\ell} (\Sigma)\cong H^0(\flatmoduli{\Sigma},\CL^\ell)}
\end{equation}
where ${V_\ell}(\Sigma)$ is the space of zero-point conformal blocks of $\mathfrak{g}_{\text{aff}}$ at level $\ell$ on $\Sigma$. The LHS of (\ref{sum:6.3}) is obtained from the A-model on $\Sigma$ with target $\Omega G$, and the RHS is obtained from the quantum mechanical model on $\flatmoduli{\Sigma}$, where $H^0$ is the space of holomorphic sections of the determinant line bundle $\cal L$. 

We also derive Pauly's definition of the Verlinde formula \cite{pauly1996espaces}, which considers extra operator insertions in the theory. This derivation will proceed similar to the case with \textit{no} operator insertions, because we can exploit the position-independence of operator insertions in an A-model. We are then able to derive the isomorphism %(\ref{math:verlindepauly-physics}), which takes the form
\begin{equation}
\boxed{{V_\ell}(\Sigma, \vec{p})\cong H^0(\CM_{\text{para}}\cbrac{\Sigma,\vec{p}},\CL^\ell)}
\end{equation}
where $\vec{p}=(p_1,\cdots,p_n)$ are the operator insertion points on $\Sigma$, and ${V_\ell}(\Sigma, \vec{p})$ is the space of $n$-point conformal blocks of $\mathfrak{g}_{\text{aff}}$ at level $\ell$ on $\Sigma$.

\bigskip\noindent\textit{Conventions}

The labeling conventions for the indices used in this paper are as follows:

\renewcommand{\arraystretch}{1.5}
\begin{center}
	\begin{tabular}[p]{c|c|c}
&Spin&index\\ \hline
4d (on $M_4$)&$1/2$&$\alpha,\beta,\cdots=1,2$\\\hline
4d (on $M_4$)&1&$\mu,\nu,\cdots=1,2,3,4$\\\hline
%3d&1&$i,j,\cdots=1,2,3$\\\hline
2d (on $\Sigma$)&1&$A,B,\cdots=1,2$\\\hline
2d (on $C$)&1&$a,b,\cdots=3,4$\\\hline
$N$-d (on moduli space $\CM$)&1&$I,J,\cdots=1,\cdots,N$
\end{tabular}
\end{center}
We shall also denote components of (left-) right-handed spinors by (un)dotted indices ($\alpha$)$\dot{\alpha}$.

\bigskip\noindent\textit{Acknowledgments}

We would like to thank V. Mu\~noz for helpful discussions.
We would also like to thank M. Marino for pointing out his earlier work \cite{lozano2001donaldson}, in which computational proofs were provided for Mu\~noz's theorem, and the Verlinde formula. These proofs involved DW theory on a specific product ruled surface -- i.e. DW theory on $\Sigma\times S^2$. This differs from our work, in which a more general conceptual approach is taken.

This work is supported in part by the NUS FRC Tier 1 grant R-144-000-377-114, and the MOE Tier 2 grant R-144-000-396-112.

\section{Instanton Floer Homology and a 2d A-model from 4d \susy{2} TQFT\label{sec:2}}

\subsection{Review of Donaldson-Witten Theory and its Relation to Instanton Floer Homology\label{sec:DW}}

In this subsection, we shall review some pertinent details about DW theory, and how its invariants can be associated with instanton Floer homology when the underlying four-manifold has a boundary.

\subsubsection*{\textit{Donaldson-Witten Theory}}

%Introduction and action
Let us consider an arbitrary four-manifold $M_4$, on which we define a pure \susy{2} theory with gauge group $G$.
%\footnote
%{
%	Note that $G$ was taken to be $\SO{3}$ or $\SU{2}$ in the original DW theory, so that the Donaldson invariants could be retrieved. The gauge group can be generalized to any simple Lie group $G$, for which the corresponding topological invariants may be computed.
%} 
Mathematically, this is the same as defining a $G$ principal bundle $E\to M_4$, on which a gauge theory may be defined. Let us focus on the simple case where $G=SU(2)$, bearing in mind that the following arguments can also be applied to a higher-rank gauge group.

Upon topological twisting, some of the supercharges, which were originally spinors, become scalars. For brevity, we shall consider one such nilpotent scalar supercharge $\qcharge$,\footnote
{
	More accurately, $\qcharge$ is a linear combination of two scalar supercharges obtained from topological twisting.
}
where there is a $U(1)_R$ R-symmetry. The action can be written as \cite{witten1988topological,labastida2005topological}
\begin{equation}
\label{S}
\begin{aligned}
	S_{\text{DW}}&={1\over e^2}\int_{M_4}d^4x\  \sqrt{G_{M_4}}\tr{\ccbrac{\qcharge,V_{\text{DW}}}}+{i\theta\over 8\pi^2}\int_{M_4}\tr\cbrac{F\wedge F}\\
%	&={1\over e^2}\int_{M_4}d^4x\  \sqrt{G_{M_4}}\tr\left({1\over 4}F^{\mu\nu}F_{\mu\nu}
%	+D_\mu\phi D^\mu\phi^\dagger
%	+i\chi_{\alpha\dot{\alpha}} D^{\dot{\alpha}\alpha}\eta
%	+i{\chi^{\dot{\beta}}}_{\alpha}D^{\dot{\alpha}\alpha}\lambda_{\dot{\alpha}\dot{\beta}}\right.\\
%	&\qquad\qquad\qquad\qquad\left.
%	-{i\over \sqrt{2}}\lambda^{\dot{\alpha}\dot{\beta}}[\phi,\lambda_{\dot{\alpha}\dot{\beta}}]
%	-{i\over \sqrt{2}}\chi_{\alpha\dot{\alpha}}[\chi^{\alpha\dot{\alpha}},\phi^\dagger]
%	-{i\sqrt{2}}\eta[\phi,\eta]
%	+{1\over 2}[\phi,\phi^\dagger]^2\right)\\
%	&\qquad+{i\theta\over 8\pi^2}\int_{M_4}\tr\cbrac{F\wedge F},
	&={1\over e^2}\int_{M_4}d^4x\  \sqrt{G_{M_4}}\tr\left(-{1\over 4}F^{\mu\nu}F_{\mu\nu}
	+D_\mu\phi D^\mu\phi^\dagger
	-i\chi_{\mu} D^{\mu}\eta
	-{i}\lambda_{\dot{\alpha}\dot{\beta}}\cbrac{\ov\sigma_{\mu\nu}}^{\dot{\alpha}\dot{\beta}}D^\mu\chi^\nu\right.\\
	&\qquad\qquad\qquad\qquad\left.
	+{1\over \sqrt{2}}\lambda^{\dot{\alpha}\dot{\beta}}[\lambda_{\dot{\alpha}\dot{\beta}},\phi]
	-{1\over 2\sqrt{2}}\chi_{\mu}[\chi^{\mu},\phi^\dagger]
	+{i\sqrt{2}}\eta[\phi,\eta]
	-{i\over 2}[\phi,\phi^\dagger]^2\right)\\
	&\qquad+{i\theta\over 8\pi^2}\int_{M_4}\tr\cbrac{F\wedge F},
\end{aligned}
\end{equation}
where $V_{\text{DW}}$ is a gauge-invariant fermionic operator %(see (2.39) of \cite{witten1988topological} or (5.31) of \cite{labastida2005topological} for the full expression for $V_{\text{DW}}$.) 
carrying an R-charge of $-1$ and scaling dimension $0$, and the ``electric" and ``magnetic" coupling constants $e$ and $\theta$ make up the complex gauge coupling constant $\tau={4\pi i\over e^2}+{\theta\over 2\pi}$. $G_{M_4}$ is the metric on $M_4$. 

Bosonic degrees of freedom are described by a gauge field $A_\mu$ and a complex scalar field $\phi$. Fermionic degrees of freedom are described by a 0-form $\eta$, a 1-form $\chi_\mu$, and a 2-form $\lambda_{\mu\nu}$. These fields are all in the adjoint representation of the gauge group $SU(2)$. The covariant derivative is defined as $D_\mu=\partial_\mu-i[A_\mu,\cdot]$. Note that because of the nilpotency of $\qcharge$, this action is $\qcharge$-invariant.  

%susy transformations
For completeness, we shall also state the supersymmetry transformations here. They are
\begin{equation}
\label{susy:twistedgeneral}
\begin{aligned}
	\delta A_\mu&=\zeta\chi_\mu\\
	\delta\phi&=0\\
	\delta\phi^\dagger&=2\sqrt{2}i\zeta\eta\\
	\delta\eta&=i\zeta[\phi,\phi^\dagger]\\
	\delta\chi_\mu&=2\sqrt{2}\zeta D_\mu\phi\\
	\delta\lambda_{\dot{\alpha}\dot{\beta}}&=i\zeta F^+_{\dot{\alpha}\dot{\beta}},
\end{aligned}
\end{equation}
where $\zeta$ is an arbitrary Grassmannian parameter.

%correlation functions
\bigskip\noindent{\it Observables of the Donaldson-Witten Theory}

Now consider the set of $n$ $\qcharge$-invariant operators $\mathcal{O}_r$, where $r=1,\dots,n$, and their correlation functions that take the form
\begin{equation}
\label{CF}
	\vev{\mathcal{O}_1\dots\mathcal{O}_n} = \int \CD \Phi\ \mathcal{O}_1 \dots \mathcal{O}_n  \e{- {S}},
\end{equation}
where $\CD \Phi$ denotes the total path-integral measure over all fields, and $S$ is a generic $\qcharge$-exact action.\footnote
{
	To describe DW theory, we may simply take $S=S_{\text{DW}}$. Likewise in \S\ref{sec:SW}, we take $S=S_{\text{SW}}$ to obtain SW theory.
} Recall the important fact that for any metric $G_{\mu\nu}$, the stress tensor $T_{\mu \nu}\propto{\delta S\over \delta G_{\mu\nu}}$ is $\qcharge$-exact in a TQFT -- i.e. we may write $T_{\mu \nu} = \{ \qcharge, \Lambda_{\mu\nu}\}$ for some fermionic operator $\Lambda_{\mu\nu}$. Hence, varying the correlation function (\ref{CF}) with respect to the metric yields ${\delta\over\delta G_{\mu\nu}}\langle \mathcal{O}_1 \dots \mathcal{O}_n \rangle \propto \langle \mathcal{O}_1 \dots \mathcal{O}_n\ T_{\mu \nu} \rangle =  \langle \mathcal{O}_1 \dots \mathcal{O}_n \cdot \ \{ \qcharge, \Lambda_{\mu\nu}\} \rangle =  \langle \{ \qcharge ,\mathcal{O}_1 \dots \mathcal{O}_n \, \Lambda_{\mu\nu}\} \rangle = 0$, which tells us that physical observables of a TQFT are independent of the spacetime geometry. Here we have also made use of the fact that $\langle \{ \qcharge, \dots \} \rangle = 0$ since $\qcharge$ generates a supersymmetry of the theory.

Furthermore, varying the correlation function with respect to the gauge coupling $e$ yields
$
	{\delta \over \delta e} \langle \mathcal{O}_1 \dots \mathcal{O}_n \rangle  = {2 \over e^3} \langle \mathcal{O}_1 \dots \mathcal{O}_n  \{ \qcharge , V \}  \rangle  = {2 \over e^3} \langle \{ \qcharge, \mathcal{O}_1 \dots \mathcal{O}_n \, V \}  \rangle = 0.
$
In other words, a correlation function of $\qcharge$-invariant operators is also independent of the gauge coupling $e$, and so it can be taken to be any value without any consequences. This then means that we can set $e\to0$, which is the same as going to the semiclassical limit of the gauge theory. 

%mode expansion
\bigskip\noindent{\it Computation of Observables}

Here, we can carry out a Fourier expansion of $\Phi$ about its classical values, which we denote by $\Phi_0$ -- the zero modes. By a classical configuration, we mean that the $\Phi_0$'s minimize the action (\ref{S}) to zero. This is needed to prevent the path integral from vanishing, since the exponent in the integrand of (\ref{CF}) will otherwise blow up in the limit $e\to 0$. The Fourier expansion yields
\begin{equation}
\label{eqn:fourier}
	\Phi=\Phi_0+\sum_{s> 0}{\Phi_s},
\end{equation}
where the $\Phi_s$'s are small and may be regarded as fluctuations about $\Phi_0$'s which describe quantum corrections. Since we are taking the semiclassical limit, we need only consider up to quadratic fluctuations about $\Phi_0$'s. The integration measure may now be rewritten as
\begin{equation}
	\CD\Phi=d\Phi_0\prod_{s>0}^{}d\Phi_s,
\end{equation}
so that the correlation function (\ref{CF}) may be written as
\begin{equation}
\label{ob:DWn-point-int}
	\vev{\mathcal{O}_1\cdots\mathcal{O}_n}
	=\cbrac{\int d\Phi_0\ \mathcal{O}_1 \dots \mathcal{O}_n\e{- {S_{\text{int}}}}}\ 
	\cbrac{\int\prod_{s>0}^{}d\Phi_s\  \e{- {S_{\text{KE}}}}},
\end{equation}
where, respectively, $S_{\text{int}}$ and $S_{\text{KE}}$ are the interacting and kinetic parts of (\ref{S}). The operators $\mathcal{O}_r$ can be expressed purely in terms of zero modes, since we can place them far apart from each other so that they do not interact. This is valid, because operator insertions are independent of position in a TQFT. Let us further insist that the product $\CO_1\cdots\CO_n$ saturate the $U(1)_R$ charge. Then, fermi fields in $S_{\text{int}}$ will contribute nothing to the integral because of the nature of Grassmannian integrals. Finally, as shown later, the bosonic part of $S_{\text{int}}$ can be set to zero due to the BPS equations. 

In doing so, we have factorized the correlation function into two parts -- one that is dependent only on the zero modes, and another one dependent only on the fluctuations. This means that (\ref{ob:DWn-point-int}) can be written as
\begin{equation}
\label{ob:DWn-point}
\vev{\mathcal{O}_1\cdots\mathcal{O}_n}
=\cbrac{\int d\Phi_0\ \mathcal{O}_1 \dots \mathcal{O}_n}\ 
\cbrac{\int\prod_{s>0}^{}d\Phi_s\  \e{- {S_{\text{KE}}}}}.
\end{equation}

%fluctuations
Let us first look at the fluctuations $\Phi_s$'s. The kinetic terms of the action take the form
\begin{equation}
\begin{aligned}
	S_{\text{KE}}&={1\over e^2}\int_{M_4}d^4x\ \sqrt{G_{M_4}}\tr\cbrac{-{1\over 4}F^{\mu\nu}F_{\mu\nu}
		+D_\mu\phi D^\mu\phi^\dagger
		-i\chi_{\mu} D^{\mu}\eta
		-{i}\lambda_{\dot{\alpha}\dot{\beta}}\cbrac{\ov\sigma_{\mu\nu}}^{\dot{\alpha}\dot{\beta}}D^\mu\chi^\nu}\\
	&=\mathcal{B}\Delta_B \mathcal{B}+i\mathcal{F}D_F\mathcal{F},
\end{aligned}
\end{equation}
where $\Delta_B$ and $iD_F$ are bosonic and fermionic elliptic operators respectively, while $\mathcal{B}$ and $\mathcal{F}$ are the bosonic and fermionic field content respectively, collected in column vectors. These operators can be diagonalized, such that 
\begin{align}
	&\Delta_B\mathcal{B}_s=(\Varepsilon_s)^2\mathcal{B}_s\\
	&iD_F\mathcal{F}_s=\Varepsilon_s\mathcal{F}_s.
\end{align}
Note that the eigenvalues $\Varepsilon_s$ take the same values for both bosons and fermions because of supersymmetry. The contribution from the non-zero modes to the correlation function (\ref{CF}) is then
\begin{equation}
\label{eqn:nonzeromodes}
\begin{aligned}
	\int\prod_{s>0}^{}d\Phi_s\ \e{- {(\Varepsilon_s)^2(\mathcal{B}_s)^2+\Varepsilon_s(\mathcal{F}_s)^2}}&=\prod_{s>0}^{}{\Varepsilon_s\over \sqrt{(\Varepsilon_s)^2}}\\
	&=\pm 1,
\end{aligned}
\end{equation}
where the $\pm$ sign arises because of the square roots. Hence, we may define correlation functions up to a sign.

%zero modes
Next, let us look at the zero modes. We note that the BPS equations may be obtained by setting supersymmetry transformations of the fermi fields, in (\ref{susy:twistedgeneral}), to zero. The BPS equations are then
\begin{subequations}
	\begin{eqnarray}
		\label{con:BPSa}
		&[\phi,\phi^\dagger]=0,\\
		\label{con:BPSb}
		&D_\mu\phi=0,\\
		\label{con:BPSc}
		&F_{\mu\nu}+\half\epsilon_{\mu\nu\rho\lambda}F^{\rho\lambda}=0.
	\end{eqnarray}
\end{subequations}
The zero modes $\Phi_0$'s, being classical configurations, must satisfy the BPS equations in (\ref{con:BPSa}), (\ref{con:BPSb}) and (\ref{con:BPSc}). In particular, (\ref{con:BPSc}) implies that the field strength of the zero mode gauge field $A_0$ must be anti self-dual, and hence they must produce anti instantons. To look for nearby zero modes, we study fluctuations about an anti instanton, which we further assume to be isolated. From here on, we shall deal only with anti instantons, and as such we will simply refer to them as instantons for brevity.

The constraints $D_\mu \phi_0 =0$ and $[\phi_0, {\phi_0}^\dagger] =0$ are satisfied by the trivial solution $\phi_0 =0$. This is that same as insisting that $\phi$ has no zero modes. Additionally, this also means that the zero modes $A_0$ correspond to irreducible connections.
\footnote{
	A connection $A$ is irreducible if the stabilizer of $G$ is equal to its center $Z(G)$. Elements of the stabilizer $\text{Stabl}(G)$ are covariantly constant -- i.e. $\gamma\in\text{Stabl}(G)$ take the form $D_A\gamma=0$. Elements of $Z(G)$ commute with all other elements in $G$ -- i.e. $\widetilde{\gamma}\in Z(G)$ satisfies $[\widetilde{\gamma},f]=0$ for $f\in G$. Here, $\phi_0$, which is a $\gf$-valued function, satisfies both conditions. On one hand, we have $D_\mu \phi_0 =0$ from (\ref{con:BPSb}), which means that $\phi_0\in\text{Lie}(\text{Stabl}(G))$. On the other hand, (\ref{con:BPSa}) and the assumption $\phi=0$ implies that $[\phi,T^a]=0$, where $T^a$ are generators of $G$. This means that $\phi\in \text{Lie}(Z(G))$. Hence, the connections $A$ considered here are indeed irreducible.
} 
This further tells us that the \textit{only} bosonic zero modes come from the gauge field $A_0$, and that they correspond to irreducible, anti self-dual connections. Hence, bosonic zero modes are characterized by the relation (\ref{con:BPSc}) which may be written as the anti self-duality condition
\begin{equation}
\label{con:antiselfdual}
	F=-*F,
\end{equation}
where $*$ is the Hodge star operation. Recall that we have defined the bundle $E$ over $M_4$ with gauge curvature $F$. 

Further taking the modulo of gauge transformations that leave (\ref{con:BPSc}) invariant, the fluctuations $A_s$ that appear in the path-integral measure will correspond to the collective coordinates of the moduli space of instantons with the instanton number $k$ defined by
\begin{equation}
\label{eqn:FF}
	k={1\over 8\pi^2}\int_{M_4} \tr \cbrac{F\wedge F}.
\end{equation}
We shall denote this moduli space by $\instantonmoduli{k}{M_4}$, and correspondingly its (virtual) dimension by $N_{\text{I}}=\text{dim}\cbrac{\instantonmoduli{k}{M_4}}$. 
We shall now analyze the zero modes of the fermions $\eta, \chi_\mu,\lambda_{\mu\nu}$. Since we have restricted ourselves to gauge connections $A$ that are irreducible, and moreover, since they are also regular, it can be argued that $\eta$ and $\lambda$ do not have any zero modes \cite{moore1997integration}. Thus, the only fermionic expansion coefficients that contribute to the path-integral measure come from $\chi$. 

The number of bosonic zero modes is, according to our analysis above, given by the dimension ${N_{\text{I}}}$ of $\instantonmoduli{k}{M_4}$. As for the zero modes of $\chi$,  the dimension of the kernel of the Dirac operator $iD_F$ which acts on $\chi$ in the  Lagrangian is equal to the index of $iD_F$; in other words, the number of zero modes of $\chi$ is given by ${\rm dim} ({\rm Ker} (iD_F)) = {\rm ind}(iD_F)$. This index also counts the number of infinitesimal connections $\delta A$ where gauge-inequivalent classes of $A + \delta A$ satisfy the anti self-duality condition (\ref{con:BPSc}). Therefore, the number of zero modes of $\chi$ will also be given by the dimension $N_{\text{I}}$ of $\instantonmoduli{k}{M_4}$. Altogether, this means that after integrating out the non zero modes, we can write the remaining part of the measure in the expansion coefficients ${A_0}^i$ and ${\chi_0}^i$ of the zero modes of $A$ and $\chi$ as
\begin{equation}
	\prod_{i=1}^{N_{\text{I}}} d{A_0}^i d{\chi_0}^i.
	\label{measure} 
\end{equation}
Notice that the ${N_{\text{I}}}$ distinct $d{\chi_0}^i$'s anti-commute. Hence, (\ref{measure}) can be interpreted as a natural measure for the integration of a differential form on $\instantonmoduli{k}{M_4}$. 

%Conditions on the Q-invariant operators, and relationship between U(1)R charge and s
%Next, recall that in order for a topological twist on a \susy{2} super Yang-Mills theory to be carried out, we take the diagonal subgroup of the rotation group (either left- or right-handed rotations) $\SU{2}$ and the R-symmetry group $\SU{2}_R$, which shifts spins on the spinor supercharges by $\pm \half$. However, the full classical R-symmetry
%\footnote
%{
%	Quantum mechanically $\U{1}_R$ is broken down to a discrete subgroup $\integer_{4N}$ for the pure $\SU{N}$ \susy{2} Yang-Mills theory. Due to the presence of instantons, the measure is only invariant under the discrete group $\integer_{4N}$. This will not an issue here, since we are currently interested in the classical (zero) modes of the theory.
%} 
%is really $\SU{2}_R\times\U{1}_R$. 
Next, it is known that the $\U{1}_R$ charges
\footnote
{
	Quantum mechanically $\U{1}_R$ is broken down to a discrete subgroup $\integer_{4N}$ for the pure $\SU{N}$ \susy{2} Yang-Mills theory. Due to the presence of instantons, the measure is only invariant under the discrete group $\integer_{4N}$. This will not an issue here, since we are currently interested in the classical (zero) modes of the theory.
} 
for the physical fields $(A_\mu,\phi,\eta,\chi_\mu,\lambda_{\mu\nu})$ are given by $(0,2,-1,1,-1)$. Hence, the only non-zero contributions to the $\U{1}_R$ charge of the measure come from the fermions $\eta,\chi_\mu,\lambda_{\mu\nu}$. Using an index theorem, it can be shown that the number of $\chi_0$'s minus the combined number of $\eta_0$ and $\lambda_0$ is also ${N_{\text{I}}}$.
%, the virtual dimension of the moduli space of instantons. 
For a compact semi-simple gauge group $G$, we have \cite{atiyah1978self}
\begin{equation}
\label{eqn:diminstantonmod}
	{N_{\text{I}}}=p_1[E]-\half\text{dim}G\ \cbrac{\chi-\tau},
\end{equation}
where $p_1[E]$ is the first Pontrjagin class of the bundle $E$, while $\chi$ is the Euler characteristic of $M_4$, and $\tau$ is the signature of $M_4$. In particular when $G=\SU{2}$, this becomes
\begin{equation}
	{N_{\text{I}}}=8k-{3\over 2}\cbrac{\chi-\tau},
\end{equation}
where $k$ is the instanton number like before.

As mentioned in an earlier part, the integer ${N_{\text{I}}}$ also coincides with the number of $\chi$ zero modes. This means that the number of $\chi_0$ is determined by the $\U{1}_R$ charge of the integration measure. In order for the correlation function to be non-vanishing, we need the product of operators $\mathcal{O}_1\cdots\mathcal{O}_n$ to carry a $\U{1}_R$ charge equal to ${N_{\text{I}}}$, because of the nature of Grassmannian integrals. For an integration measure $d{\chi_0}^1\cdots d{\chi_0}^{N_{\text{I}}}$, we need the integrand to be of the form $\chi_{p_1}\cdots\chi_{p_{N_{\text{I}}}}$ -- i.e. the integrand must contain some arbitrary product of the expansion coefficients for which each expansion coefficient appears \textit{exactly once}.% -- in order for the integral to be non-vanishing.

To elaborate on the point made about the measure in (\ref{measure}), we may also consider a linear combination of these products, which is totally anti-symmetric -- i.e.
\begin{equation}
\label{eqn:n-form}
	\mathcal{O}_1\cdots\mathcal{O}_n=\Omega^{p_1\cdots p_{N_{\text{I}}}}\chi_{p_1}\cdots\chi_{p_{N_{\text{I}}}}.
\end{equation}
In doing so, we may write the $n$-point correlation function (\ref{CF}) in terms of (\ref{eqn:n-form}) as
\begin{equation}
\begin{aligned}
	\int\prod_{i=1}^{{N_{\text{I}}}}d {A_0}^i d{\chi_0}^i\ \Omega^{p_1\cdots p_{N_{\text{I}}}}\chi_{p_1}\cdots\chi_{p_{N_{\text{I}}}}&=\int d{A_0}^{p_1}\cdots d{A_0}^{p_{N_{\text{I}}}}\ {1\over {N_{\text{I}}}!}\Omega^{p_1\cdots p_{N_{\text{I}}}}\\
	&=\int_{\instantonmoduli{k}{M_4}}\ {1\over {N_{\text{I}}}!}\Omega^{s_1\cdots i_{N_{\text{I}}}}d{A_0}^{i_1}\wedge\cdots\wedge d{A_0}^{i_{N_{\text{I}}}}\\
	&=\int_{\instantonmoduli{k}{M_4}}\ \Omega,
\end{aligned}
\end{equation}
where we have made use of the fact that $\int d\chi_1\cdots d\chi_{N_{\text{I}}}\ \cbrac{\chi_{p_1}\cdots \chi_{p_{N_{\text{I}}}}}={1\over {N_{\text{I}}}!}\delta^1_{\left[\phantom{\over}p_1\right.}\cdots\delta^{N_{\text{I}}}_{\left.p_{N_{\text{I}}}\right]}$ in the first equality. Hence, the correlation function (\ref{CF}) can be viewed as an integration of the top form, $\Omega$, over $\instantonmoduli{k}{M_4}$. This just defines the Donaldson invariants.
%Thus, the $\U{1}_R$ charge of the fermionic measure also determines the degree of the top form $\Omega$, defined on $\instantonmoduli{k}{M_4}$.

%Note that insofar, the observables we considered are \textit{not} locally defined. Instead, they correspond to global data -- while the original theory was local.

\subsubsection*{\textit{Relation between Donaldson-Witten Theory and Instanton Floer Homology}\label{sec:instantonfloer}}

We shall now review various formulas presented by Donaldson and Atiyah in \cite{atiyah1988new,donaldson2002floer}, that relate Donaldson and Floer theory on a four-manifold with boundary.

To this end, consider DW theory on $M_4 = Y_3 \times \real^+$, where $Y_3$ can be interpreted as the boundary of $M_4$, and the half real-line $\real^+$ can be interpreted as the time direction. We may then include the other nilpotent scalar supercharge $\ov{\qcharge}$, so that the Hamiltonian $H$ may be defined via the supersymmetry algebra 
\begin{equation}
\label{susy:H}
	\ccbrac{\qcharge,\ov{\qcharge}}=2H.
\end{equation}
Furthermore, the operators $\qcharge,\ov{\qcharge}$ and $H$ all commute with each other, which means that energy eigenstates are in one-to-one correspondence with the states of either $\qcharge$- or $\ov{\qcharge}$- cohomologies.

Next, we recall that physical operators $\CO$ must be $\qcharge$-invariant. At the same time, $\CO$ cannot be allowed to be $\qcharge$-exact -- i.e. $\CO\neq\ccbrac{\qcharge,\widetilde{\CO}}$, for some other operator $\widetilde{\CO}$. Otherwise, correlation functions involving such a $\CO$ will be vanishing, since $\abrac{\ccbrac{\qcharge,\cdots}}=0$. Using the state-operator correspondence, we then have $\qcharge$-closed states $\ket{\CO}$ which cannot be written as $\qcharge\ket{\widetilde{\CO}}$ -- these states $\ket{\CO}$ then correspond to classes of the $\qcharge$-cohomology of DW theory. 

Note also, that these states $\ket{\CO}$ must transform in the same manner under $\ov{\qcharge}$, because $\qcharge$ and $\ov{\qcharge}$ are mapped into each other under a time-reversal symmetry transformation (See equation (4.9) of \cite{witten1988topological}). As such, the states $\ket{\CO}$ are also identified with classes of the $\ov{\qcharge}$ cohomology group. Collectively, we have $\qcharge\ket{\CO}=0=\ov{\qcharge}\ket{\CO}$. (\ref{susy:H}) then implies that these states have zero energy eigenvalues. Hence, zero energy ground states also correspond to classes of the $\qcharge$-cohomology group.

Physically, an instanton allows for quantum tunneling between two ground states. We want, however, unique ground states that cannot be accessed from other states via tunneling. 
The topological term can be interpreted as a supersymmetric 1d sigma model on the space of $SU(2)$ connections on $Y_3$ with potential $h_{Y_3} = \half \int_{Y_3} {\rm Tr} \, (A_{Y_3} \wedge dA_{Y_3} + {2\over 3} A_{Y_3} \wedge A_{Y_3} \wedge A_{Y_3})$ -- the Chern-Simons functional of $A_{Y_3}$. Ground states are obtained by extremizing the potential $h_{Y_3}$ -- which turn out to be flat connections on $Y_3$. This describes the $\qcharge$-cohomology of the 1d sigma model, which can be identified with the instanton Floer homology $\HFinst{Y_3}$.

%This 1d supersymmetric sigma model can be viewed as supersymmetric quantum mechanics, if $\real^+$ is taken to be the temporal direction. As such, the Hilbert space of states may be identified as $\mathscr{H}=\mAY / {\cal G}_{Y_3}$.

According to the general ideas of quantum field theory, when the theory is formulated on such an $M_4$, one must specify the boundary values of the path-integral fields along ${Y_3}$. Let us denote $\Phi_{Y_3}$ to be the restriction of these fields to ${Y_3}$; then, in the Hilbert space of states $\mathscr H$, specifying a set of boundary values for the fields on ${Y_3}$ is tantamount to selecting a functional $\Psi(\Phi_{Y_3}) \in \mathscr H$. Since the $\qcharge$-cohomology of the sigma model is annihilated by $H$ -- i.e. it is time-invariant --  one can take an arbitrary time-slice in $M_4$ and study the quantum theory formulated on ${Y_3}$ instead; in this way, $\Psi(\Phi_{Y_3}) \in \mathscr H$ can be interpreted as a state in the Hilbert space $\mathscr H$ of the quantum theory on ${Y_3}$. As a result, via a state-operator mapping of the TQFT, we may use $\Psi$ to impose boundary conditions on the fields. The correlation function will then be given by 
\begin{equation}
	\langle \mo_1 \dots \mo_n \rangle_{\Psi(\Phi_{Y_3})} = \int \CD \Phi \  \e{- {S}} \ \mo_1 \dots \mo_n \cdot \Psi(\Phi_{Y_3}).
\label{CF_b}
\end{equation}

Since the theory ought to remain topological in the presence of a boundary ${Y_3}$, it must be that $\mbrac{\qcharge, \mo_r} = 0 = \mbrac{\qcharge, \Psi}$. Moreover, if $\Psi = \mbrac{\qcharge, \dots}$, the fact that $\mbrac{\qcharge, \mo_r} = 0$ implies that (\ref{CF_b}) will also be zero. Thus, (\ref{CF_b}) depends \textit{only} on $\Psi$ via its interpretation as a $\qcharge$-cohomology class, and since $\Psi$ is associated with the quantum theory on ${Y_3}$, we can identify $\Psi$ as a class in the instanton Floer homology $\HFinst{Y_3}$. (\ref{CF_b}) will then represent a Donaldson invariant with values in $\HFinst{Y_3}$ -- a homology group from instantons. This is the relation between Donaldson and Floer theory on $M_4$ as described by Donaldson in \cite{donaldson2002floer}. The underlying chain complex of $\HFinst{Y_3}$ is graded by the relative Morse index between a pair of critical points of the Chern-Simons functional. Since the coboundary operator corresponding to $\CQ$ counts the number of instanton tunneling solutions between a pair of critical points, the grading of $\HFinst{Y_3}$ coincides with the instanton number.
% flow lines between these critical points are counted by instantons in DW theory, the grading of $\HFinst{Y_3}$ coincides with the instanton number.

In particular, if the virtual dimension $N_{\text{I}}$ (given by an open version of (\ref{eqn:diminstantonmod})) vanishes,\footnote
{
	As it will be explained in \S\ref{sec:atiyah-floer}, we may write $M_4=\real^+\times Y_3\cong\real^+\times I\times \Sigma$, where $I$ is an interval and $\Sigma$ is a compact Riemann surface with genus $g$. The open version of the index theorem differs from (\ref{eqn:diminstantonmod}) by the eta invariant, which also has no dependence on $g$. Since we can write $M_4$ as a product manifold, its Euler characteristic can be written as $\chi(M_4)=\chi(\real^+)\cdot\chi(Y_3)$. Furthermore, since $\real^+$ is topologically trivial -- i.e. $\chi(\real^+)=0$ -- the Euler characteristic of $M_4$ vanishes. Hence, setting $N_{\text{I}}=0$ places no constraints on the value of $g$.
} which means that the operators $\mo_1 \dots \mo_n$ \textit{must} be replaced with the identity operator $1$, (\ref{CF_b}) becomes
\begin{equation}
\label{ob:DWfloer}
	\boxed{Z_{M_4}=\vev{1}_{\Psi(\Phi_{Y_3})} = \sum_i  \Psi_{\textrm{inst}}(\Phi^i_{Y_3})}
\end{equation}
In other words, the partition function on $M_4$ is a sum of instanton Floer homology classes $\Psi_{\textrm{inst}}(\Phi^i_{Y_3})$, where $i$ denotes the $i^{\textrm{th}}$ (flat) gauge connection on $Y_3$ that descends from an instanton solution on $M_4$. 

%That is the Atiyah–Patodi–Singer index theorem.

We will be dealing with observables of the form (\ref{ob:DWfloer}), unless otherwise stated.
	
\subsection{Donaldson-Witten Theory and a 2d A-model\label{sec:DWadiabatic}}

In \S\ref{sec:DW}, we see that the partition function of DW theory sums classes of instanton Floer homology associated with the boundary of $M_4$. In this paper, we would like to relate Floer homologies to quantum cohomology, which is defined by 2d topological sigma models \cite{witten1988topologicalsig}. 

To do that, we take $M_4=\Sigma\times C$, where $\Sigma$ and $C$ are both smooth and compact Riemann surfaces with genera $g$ and $h$, respectively. Unless stated, we shall assume that $g,h>1$, so that the connections on the reduced Riemann surface will be irreducible. We may shrink either Riemann surface to obtain an effective 2d theory on the remaining Riemann surface. In this subsection, we will explain that this effective 2d theory is the relevant 2d topological sigma model. To this end, we shall rederive the result obtained by Bershadsky et al. \cite{bershadsky1995topological}, and flesh out the points relevant to our paper.

\subsubsection*{\textit{The Adiabatic Limit and a 2d A-model with Target the Moduli Space of Flat Connections}}

\bigskip\noindent{\it The Adiabatic Limit}

Since we have a product manifold, the metric may be written in a block diagonal form
\begin{equation}
\label{met:blockdiagonalgen}
ds^2=\cbrac{G_{\Sigma}}_{AB}dx^Adx^B+\cbrac{G_{C}}_{ab}dx^adx^b,
\end{equation}
and hence, we may also decompose the $G$-connections on $M_4$ as $A_\mu dx^\mu=\cbrac{A_{\Sigma}}_A dx^A+\cbrac{A_{C}}_a dx^a$, where $A_{\Sigma}$ and $A_{C}$ are connections on $\Sigma$ and $C$, respectively.

To shrink $C$, we may deform this metric by multiplying in a scaling factor $\varepsilon$, such that
\begin{equation}
ds^2\to ds'^2=\cbrac{G_{\Sigma}}_{AB}dx^Adx^B+\varepsilon\cbrac{G_{C}}_{ab}dx^adx^b,
\end{equation}
and then set $\varepsilon\to0$.

This is known as the \textit{adiabatic limit}, which is the same as physically shrinking $C$, leaving behind an 2d effective gauge theory. Equivalently, we may view this as the limit in which we take $\Sigma$ to be much bigger than $C$. This is why we have denoted indices on the small Riemann surface $C$ as $a,b,\cdots$, and those on its large counterpart $\Sigma$ as $A,B,\cdots$.

Consider now a generic pure Yang-Mills theory, which may not necessarily be supersymmetric, for which the kinetic action of the gauge fields may be written as
\begin{equation}
S_{\text{KE}}={1\over 4e^2} \int_{M_4}dx^4\sqrt{G_{M_4}}\tr\cbrac{F^{\mu\nu}F_{\mu\nu}}.
\end{equation}
This can be decomposed as
\begin{equation}
\label{eqn:gaugeKEgen}
S_{\text{KE}}={1\over 4e^2} \int_{M_4}dx^4\sqrt{G_{M_4}}\tr\cbrac{F^{\mu\nu}F_{\mu\nu}}
={1\over 4e^2}\int_{\Sigma\times C}dx^4\sqrt{G_{\Sigma}}\sqrt{G_{C}}\tr\cbrac{F^{AB}F_{AB}+F^{ab}F_{ab}+2F^{Aa}F_{Aa}}.
\end{equation}
Upon applying the $\varepsilon$ deformation, (\ref{eqn:gaugeKEgen}) becomes
\begin{equation}
\label{eqn:deformedgaugeKEgen}
{1\over 4e^2}\int_{\Sigma\times \varepsilon C}dx^4\sqrt{G_{\Sigma}}\sqrt{G_{C}}\tr\cbrac{\varepsilon F^{AB}F_{AB}+\varepsilon^{-1}F^{ab}F_{ab}+2F^{Aa}F_{Aa}},
\end{equation}
having noted that $F^{ab}\to\varepsilon^{-2}F^{ab}$, $F^{Aa}\to\varepsilon^{-1}F^{Aa}$ and $\sqrt{G_{C}}\to\varepsilon\sqrt{G_{C}}$.

Setting $\varepsilon\to 0$ amounts to the first term vanishing. Furthermore, to prevent the second term from blowing up, we are forced to impose the \textit{flatness condition}
\begin{equation}
\label{con:flatnessgen}
\boxed{F_{ab}=0}
\end{equation}
This means that there are flat connections living on the small Riemann surface $C$, and these are the bosonic zero modes on $C$. Further taking variations around $A_C$ in the form of gauge transformations $A_C\to A_C+\delta A_C$, we may then look for nearby gauge-inequivalent solutions. Equivalently, (\ref{con:flatnessgen}) can be linearized to give
\begin{equation}
\label{lin:flat}
D_a \delta A_b=D_b \delta A_a,
\end{equation}
which provides solutions correspond to nearby zero modes. So, these solutions span the moduli space of flat connections on $C$, which we shall denote by $\flatmoduli{C}$.

It is seen that only the last term of (\ref{eqn:deformedgaugeKEgen}) -- i.e. the mixed term -- survives when we shrink $C$. As we will see, this surviving term becomes the action for a 2d sigma model on $\Sigma$.

Furthermore, since we have shrunken the $x^3$- and $x^4$- directions away, the ensuing 2d action on $\Sigma$ should no longer have any dependence on these coordinates. At the same time, information about the topology of the original four-manifold $\Sigma\times C$ must be retained even if it is shrunken away, since we began with a 4d TQFT. This information is encapsulated fully, through the maps $X:\Sigma\to\flatmoduli{C}$. These maps are collective coordinates on $\flatmoduli{C}$, at least for the zero instanton sector.

\bigskip\noindent\textit{The Moduli Space of Flat Connections}

Like \cite{bershadsky1995topological}, we may write the variations $\delta A_C$ in terms of \textit{basis} cotangent vectors $\alpha_{IC}$ on $\flatmoduli{C}$, up to a gauge transformation. Here, the indices $I=1,\cdots,\dim(\flatmoduli{C})$ describe the collective coordinates on $\flatmoduli{C}$. Explicitly, these variations may be written as
\begin{equation}
{\partial A_C\over \partial X_I}=\alpha_{IC}-D_C E_I,
\end{equation}
where the gauge parameter $E_I$ can now be identified with connections on $\flatmoduli{C}$. The target space is curved in general.

It is well known that the moduli space of flat connections on a Riemann surface is a K\"ahler manifold. We may define its symplectic form and metric as \cite{bershadsky1995topological}
\begin{align}
\label{eqn:kahlersymplectic}
&\omega^{\text{flat}}_{IJ}=\int_{C}d^2w\ \tr\cbrac{\alpha_{Iw}\alpha_{J\ov{w}}-\alpha_{I\ov{w}}\alpha_{Jw}},\\
\label{eqn:kahlermetric}
&G^{\text{flat}}_{IJ}=\int_{C}d^2w\ \tr\cbrac{\alpha_{Iw}\alpha_{J\ov{w}}+\alpha_{I\ov{w}}\alpha_{Jw}},
\end{align}
where we have switched to complex coordinates on $M_4$, defined by
\begin{equation}
\label{eqn:complexcoords}
\begin{aligned}
&z=x^1+ix^2,&&w=x^3+ix^4,\\
&\ov{z}=x^1-ix^2,&&\ov{w}=x^3-ix^4,
\end{aligned}
\end{equation}
and
\begin{equation}
\begin{aligned}
&A_z={1\over 2}\cbrac{A_1-iA_2},&&A_w={1\over 2}\cbrac{A_3-iA_4},\\
&A_{\ov{z}}={1\over 2}\cbrac{A_1+iA_2},&&A_{\ov{w}}={1\over 2}\cbrac{A_3+iA_4}.
\end{aligned}
\end{equation}

\bigskip\noindent{\it The Sigma Model Action}

Furthermore, it is known that for flat connections on $C$, we can write the mixed components of the field strength \cite{bershadsky1995topological,popov2015loop} as
\begin{equation}
\label{eqn:flatmixedFgen}
F_{\Sigma C}=\cbrac{\partial_\Sigma X^I}\alpha_{I C},
\end{equation}
where $X^I=X^I(z,\ov{z})$ are the \textit{real} local coordinates on $\flatmoduli{C}$. Here, we have chosen the gauge-fixing condition \cite{bershadsky1995topological} $D_C\alpha_{I C}=0$.

Making use of (\ref{eqn:flatmixedFgen}), we can then write the surviving term in (\ref{eqn:deformedgaugeKEgen}) as
\begin{equation}
\label{eqn:sigma0real}
\begin{aligned}
S_{\text{KE}}'&={1\over e^2}\int_{\Sigma}d^2z\ \cbrac{\int_C d^2w\ \tr\cbrac{\alpha_{Iw}\alpha_{J\ov{w}}+\alpha_{I\ov{w}}\alpha_{Jw}}}\partial_zX^I\partial_{\ov{z}}X^J\\
&={1\over e^2}\int_{\Sigma}d^2z\ G^{\text{flat}}_{IJ}\partial_zX^I\partial_{\ov{z}}X^J.
\end{aligned}
\end{equation}
This is the standard action for a 2d sigma model on $\Sigma$, with constant maps to the target space $\flatmoduli{C}$, which resulted from the 4d Yang-Mills action in the \textit{zero} instanton sector.

Since $\flatmoduli{C}$ is complex, we can write the action (\ref{eqn:sigma0real}) in terms of complex coordinates $X_I,{X}_{\ov J}$, where the barred indices denote complex conjugation. The action may be rewritten as 
\begin{equation}
\label{eqn:sigma0complex}
S_{\text{KE}}'={1\over 2e^2}\int_{\Sigma}d^2z\ G^{\text{flat}}_{I\ov{J}}\cbrac{\partial_z X^I\partial_{\ov{z}}{X}^{\ov J}+\partial_{\ov{z}} X^I\partial_z { X}^{\ov J}}.
\end{equation}

%It comes at no surprise that the gauge fields in the DW theory would become coordinates on the corresponding moduli space of flat connections on $C$. After all, fluctuations of the gauge field $A$ did become the coordinates on $\instantonmoduli{k}{M_4}$ in the DW theory.

As an aside, we may link this to our previous discussion about instanton Floer homology. There, we mentioned that the critical points of the Chern-Simons functional correspond to flat connections. This can be viewed alternatively as shrinking $Y_3$, which amounts to deforming the kinetic term $F\wedge*F$ in the same manner as we did in (\ref{eqn:deformedgaugeKEgen}). This then leads to the same conclusion that the zero modes (critical points) are flat connections on $Y_3$, a result analogous to that in (\ref{con:flatnessgen}). %The difference, however, is that the maps $\real^+\to\flatmoduli{Y_3}$ are no longer necessarily holomorphic. In fact, it was shown in \cite{deser2015sigma} that Yang-Mills instanton equations only reduce to holomorphicity conditions if the reduced manifold is even-dimensional.

\bigskip\noindent{\it Pullback of the Symplectic Form}

There is a topological term in the action (\ref{S}), which takes the form $\int_{M_4}\tr\cbrac{F\wedge F}$. Let us now examine the instanton number (\ref{eqn:FF}), which may be written in component form as
\begin{equation}
\label{eqn:instantonnumbergen}
k={1\over 16\pi^2}\int_{M_4} d^4x \tr\cbrac{{1\over 2}\epsilon^{\mu\nu\rho\lambda}F_{\mu\nu}F_{\rho\lambda}}.
\end{equation}
Taking the four manifold to be of the form $M_4=\Sigma\times C$, and shrinking $C$, (\ref{eqn:instantonnumbergen}) becomes
\begin{equation}
\label{int:1}
\begin{aligned}
k&={1\over 8\pi^2}\int_{\Sigma\times \varepsilon C} d^4x \tr\cbrac{\epsilon^{AaBb}F_{Aa}F_{Bb}}\\
&={1\over 4\pi^2}\int_{\Sigma\times \varepsilon C} d^4x \tr\cbrac{F_{zw}F_{\ov{z}\ov{w}}-F_{z\ov{w}}F_{\ov{z}w}},
\end{aligned}
\end{equation}
where the flatness condition (\ref{con:flatnessgen}) has been applied. Let us switch gears for a moment, and use real coordinates $X$, like (\ref{eqn:sigma0real}). Further using (\ref{eqn:flatmixedFgen}), (\ref{int:1}) then becomes
\begin{equation}
\label{charge:degree}
\begin{aligned}
k&={1\over 2\pi^2}\int_{\Sigma}d^2z(\partial_z X^I\partial_{\ov{z}}X^{J})\cbrac{\int_{C}d^2w\ \tr\cbrac{\alpha_{I w}\alpha_{J \ov{w}}-\alpha_{I \ov{w}}\alpha_{J w}}}\\
&={1\over 2\pi^2}\int_{\Sigma}d^2z\ \omega^{\text{flat}}_{IJ}(\partial_z X^I\partial_{\ov{z}}X^{J}),
\end{aligned}
\end{equation}
where the last equality makes use of the definition of the K\"ahler symplectic form $\omega^{\text{flat}}_{IJ}$ (\ref{eqn:kahlersymplectic}). This expression defines the degree \cite{hori2003mirror,salamon1998notes} of generic maps $X$. Hence, the instanton number $k$ is identified with the degree of maps $X:\Sigma\to\flatmoduli{C}$, and we can then say that there is an isomorphism between the moduli space of $k$ instantons and the moduli space of degree $k$ maps from $\Sigma$ to $\flatmoduli{C}$, which we shall denote as $\mapsmoduli{k}{\Sigma\to\flatmoduli{C}}$.

%This is then a generalization of \cite{bershadsky1995topological}, in which only the \textit{zero} instanton sector was considered. There, $X$ are constant maps -- or maps of zero degree -- and hence describe the usual coordinates on $\flatmoduli{C}$. We have shown explicitly, that in the non-zero instanton sector (with instanton number $k$), the relevant moduli space is that of degree $k$ maps instead.

Note that (\ref{charge:degree}) can be rewritten as a pullback of the K\"ahler form $\omega_{\text{flat}}$. Specifically, one can write
\begin{equation}
\label{charge:degreepullback}
\boxed{k={1\over 8\pi^2}\int_{M_4}\ \tr \cbrac{F\wedge F}=\int_{\Sigma}\ X^*\omega_{\text{flat}}}
\end{equation}
We shall use the expression in (\ref{charge:degreepullback}) for compactness.

\bigskip\noindent{\it Adiabatic Limit of Donaldson-Witten Theory}

Before we analyze DW theory on $M_4=\Sigma\times C$, let us first look at a pure, untwisted \susy{2} super Yang-Mills theory on a flat four-manifold $M_4=\real^4=\real^2\times\real^2$, which contains $8$ supercharges. Following which, we may replace one of the 2d planes $\real^2$ by a compact Riemann surface which we call $C$, so that the four-manifold now becomes $M_4=\real^2\times C$. In doing so, supersymmetry is broken by \textit{half} -- i.e. only $4$ supercharges survive. Further reducing on $C$ then gives rise to a 2d theory on $\real^2$ with $4$ supercharges. This is a 2d theory with \susy{(2,2)} supersymmetry. 

A topological twist can be carried out, so that the remaining flat two-manifold $\real^2$ can be replaced by a generic curved Riemann surface $\Sigma$. Consequently, this gives us a 4d twisted \susy{2} pure gauge theory on $M_4=\Sigma\times C$ -- i.e. DW theory -- which descends to a 2d twisted \susy{(2,2)} theory on $\Sigma$.

Note that as a consequence of (\ref{con:flatnessgen}) and (\ref{eqn:sigma0complex}), shrinking $C$ leads to a 2d sigma model on $\Sigma$ with target $\flatmoduli{C}$. This means that the 2d \susy{(2,2)} theory is a sigma model on target $\flatmoduli{C}$ with worldsheet $\Sigma$, whose fermi fields can be determined from supersymmetry.

Let us briefly explain how the fermions in the A-model are determined from DW theory. These worldsheet fermions descend from 4d DW fermions -- i.e. the 1-form $\chi_C$ and 2-form $\lambda_{C\Sigma}$, defined over $M_4$ -- which were the only non-auxiliary fermi fields obtained upon shrinking $C$. These fermions are really cotangent vectors on the ensuing target space $\flatmoduli{C}$, and hence can be written in terms of basis cotangent vectors $\alpha$ as
\begin{align}
&\chi_w={\chi}^{\ov I}\alpha_{\ov{I}w}\\
&\chi_{\ov{w}}=\chi^{I}\alpha_{I\ov{w}}\\
&\lambda_{wz}={{\rho}_{z}}^{\ov I}\alpha_{\ov{I}w}\\
&\lambda_{\ov{w}\ov{z}}={\rho_{\ov{z}}}^{I}\alpha_{I\ov{w}}.
\end{align}

From the above, one can see that $\chi_C$ can also be interpreted as a 1-form on $C$, and a 0-form on $\Sigma$. Likewise, $\lambda_{C\Sigma}$ can also be interpreted as a 1-form on $C$, and a 1-form on $\Sigma$.

Next, consider the 4d supersymmetry transformation $\delta\lambda_{\mu\nu}=i\zeta F^+_{\mu\nu}$ (see \eqref{susy:twistedgeneral}). After shrinking $C$, the only surviving 2-form fermi terms are $\lambda_{wz}={{\rho}_{z}}^{\ov I}\alpha_{\ov{I}w}$ and $\lambda_{\ov{w}\ov{z}}={\rho_{\ov{z}}}^{I}\alpha_{I\ov{w}}$. Further recalling \eqref{eqn:flatmixedFgen}, and using complex coordinates on $\flatmoduli{C}$, the ensuing 2d supersymmetry transformations\footnote
{
	It can be seen from \cite{bershadsky1995topological} that the only non-zero basis cotangent vectors are $\alpha_{I \ov{w}}$ and $\alpha_{\ov I {w}}$. We have made use of this fact to obtain \eqref{susy:DWA-model}.
} 
can then be written as
\begin{equation}
\label{susy:DWA-model}
\begin{aligned}
	&\delta{{\rho}_{z}}^{\ov I}=-{i\over 2}\zeta\cbrac{\p_{\ov z}X^I}\\
	&\delta{\rho_{\ov{z}}}^{I}=-{i\over 2}\zeta\cbrac{\p_{z}X^{\ov I}}.
\end{aligned}
\end{equation}
Hence, the corresponding BPS equations of the 2d sigma model are
\begin{equation}
\begin{aligned}
&\p_{\ov z}X^I=0\\
&\p_{z}X^{\ov I}=0,
\end{aligned}
\end{equation}
which means that the maps $X^I:\Sigma\to\flatmoduli{C}$ are \textit{holomorphic maps}.

%Finally, that we have the pullback in (\ref{charge:degreepullback}) means that the \susy{(2,2)} sigma model on $\Sigma$ must correspond to the \textit{A-twisted} sigma model. Since it is an A-model, the maps $X:\Sigma\to\flatmoduli{C}$ must be \textit{holomorphic} maps of degree $k$; these maps are also known as 2d holomorphic instantons.

Finally, that we have the pullback in (\ref{charge:degreepullback}) means that the \susy{(2,2)} sigma model on $\Sigma$ must correspond to the \textit{A-twisted} sigma model, which has the action
\begin{equation}
\boxed{\begin{aligned}
\label{action:DWsigma}
S_{\text{DW}}'&={1\over e^2}\int_{\Sigma} d^2z\ \left(G^{\text{flat}}_{I\ov J}\cbrac{\half\partial_{z}X^{I}\partial_{\ov{z}}{X}^{\ov J}
	+\half\partial_{\ov{z}}X^{I}\partial_z{X}^{\ov J}
	+{{\rho}_{z}}^{\ov J}\nabla_{\ov{z}}\chi^{I}
	+{\rho_{\ov{z}}}^{I}\nabla_z{\chi}^{\ov J}}\right.\\
&\qquad\qquad\qquad\left.\phantom{\half}
-R_{I\ov JK\ov L}{{\rho}_{\ov z}}^{I}{\rho_{z}}^{\ov J}{\chi}^K\chi^{\ov L}\right)
+{i\theta}\int_{\Sigma}\ X^*\omega_{\text{flat}}
\end{aligned}}
\end{equation}
where $R_{I\ov JK\ov L}$ is the Riemann curvature tensor on $\flatmoduli{C}$, and $\nabla_{\ov z}\chi^{I}=\partial_{\ov z}\chi^{I}+\chi^{J}\Gamma^{I}_{JK}\partial_{\ov z}X^{K}$. Here, $\Gamma^{I}_{JK}$ are the Christoffel symbols on $\flatmoduli{C}$. The maps $X^I:\Sigma\to\flatmoduli{C}$ are \textit{holomorphic} maps of degree $k$; these maps are also known as 2d holomorphic instantons.

The reader may refer to appendix A of \cite{bershadsky1995topological} for more details on the derivation of the fermionic part of (\ref{action:DWsigma}). 

This action can further be rewritten \cite{ashwinkumar2018little} as the sum of a $\qcharge$-invariant term (in the perturbative regime), and a metric-independent one (topological) with mixed gauge coupling. Hence, $\qcharge$-cohomology can also be defined on the topological sigma model on $\Sigma$. The crux here is that the $\qcharge$-cohomology of the 4d DW theory is isomorphic to the $\qcharge$-cohomology of the 2d A-twisted sigma model on $\Sigma$ with target $\flatmoduli{C}$, since the scalar $\qcharge$ is scale-invariant. Therefore, states of DW theory are identified with states of the 2d A-model.

\section{Monopole Floer Homology and a 2d A-model from 4d \susy{2} TQFT\label{sec:3}}

\subsection{Review of Seiberg-Witten Theory and its Relation to Monopole Floer Homology\label{sec:SW}}

In the low energy limit of DW theory, the gauge group symmetry $G$ is broken to its maximal torus $\maxtorus=\U{1}^R$, where $R$ is the rank of $G$. In particular, $G=\SU{2}$ is broken to $\U{1}$ so that we now have a \susy{2} $U(1)$ gauge theory. Further coupling the pure theory to a single (spinor) monopole field $M$, one then obtains the SW theory.

Here, monopole analogs of objects defined in \S\ref{sec:DW} may also be written down. In particular, monopole Floer homology can be defined instead of instanton Floer homology considered in \S\ref{sec:DW}. In this paper, we shall describe SW theory independently of DW theory, and take $G=U(1)$.

\subsubsection*{\textit{Seiberg-Witten Theory}}

For the purpose of this paper, we will work exclusively with the topologically twisted version
\footnote
{
	The twisted theory differs from the untwisted one in that the monopole fields, which were originally made up of a $\SU{2}_R$ doublet of complex scalar fields, become spinors upon topological twisting. 
} 
of the SW theory. Henceforth, we shall refer to this twisted theory simply as the SW theory. Like DW theory, this is a \susy{2} gauge theory. There are two main differences however. Firstly, the gauge group is now just $\U{1}$,
\footnote
{
More generally, it is an abelian group given by $\U{1}^R$, where $R\in\integer^+$. There is also a nonabelian version of SW theory, which produces nonabelian monopoles instead \cite{labastida2005topological}.
} 
which immensely simplifies calculations. Secondly, the SW theory is made up of a \susy{2} $\U{1}$ pure gauge theory coupled to a massless monopole hypermultiplet.

Being a TQFT, the action is $\qcharge$-exact up to a topological term (See \cite{moore1997integration} for example). We can write the action in \cite{labastida2005topological}, after some appropriate rescalings, as

\begin{equation}
\label{action:SW}
\begin{aligned}
S_{\text{SW}}&={1\over e^2}\int_{M_4}d^4x\  \sqrt{G_{M_4}}{\ccbrac{\qcharge,V_{\text{SW}}}}+{i\theta\over 8\pi^2}\int_{M_4}\cbrac{F\wedge F}\\
%&={1\over e^2}\int_{M_4} d^4x\,\sqrt{G_{M_4}} \left[ 
%{1\over 4}F_{\mu\nu}F^{\mu\nu}
%+\half\cbrac{\ov{M}^{\dot{\alpha}}F^+_{\dot{\alpha}\dot{\beta}}M^{\dot{\beta}}
%-\ov{M}^{\left(\dot{\alpha}\phantom{\over}\right.}M^{\left.\dot{\beta}\phantom{\over}\right)} \ov{M}_{\left(\dot{\alpha}\phantom{\over}\right.}M_{\left.\dot{\beta}\phantom{\over}\right)} }\right.\\
%&\qquad\qquad\qquad\qquad\qquad
%+ D_\mu\phi D^\mu \ov{\phi}
%+i\chi_{\alpha\dot{\alpha}} D^{\dot{\alpha}\alpha}\eta
%+i{\chi^{\dot{\beta}}}_{\alpha}D^{\dot{\alpha}\alpha}\lambda_{\dot{\alpha}\dot{\beta}}\\
%&\qquad\qquad\qquad\qquad\qquad
%-\ov{M}^{\dot{\alpha}}D_{\dot{\alpha}\alpha}D^{\alpha\dot{\beta}}M_{\dot{\beta}}
%-i\ov{\nu}_\alpha D^{\dot{\alpha}\alpha}\mu_{\dot{\alpha}}
%-i\ov{\mu}^{\dot{\alpha}}D_{\alpha\dot{\alpha}}\nu^\alpha\\
%&\qquad\qquad\qquad\qquad\qquad
%-{i\over\sqrt{2}}\cbrac{\ov{M}^{\dot{\alpha}}\lambda_{\dot{\alpha}\dot{\beta}}\mu^{\dot{\beta}}
%+\ov{\mu}^{\dot{\alpha}}\lambda_{\dot{\alpha}\dot{\beta}}M^{\dot{\beta}}}
%+i\sqrt{2}\cbrac{\ov{M}^{\dot{\alpha}}\chi_{\alpha\dot{\alpha}}\nu^\alpha
%-\ov{\nu}_\alpha\chi^{\dot{\alpha}\alpha}M_{\dot{\alpha}}}\\
%&\qquad\qquad\qquad\qquad\qquad
%+i\sqrt{2}\cbrac{\ov{M}^{\dot{\alpha}}\eta\mu_{\dot{\alpha}}-\ov{\mu}^{\dot{\alpha}}\eta M_{\dot{\alpha}}}
%-i\sqrt{2}\ov{\nu}_{\alpha}\phi\nu^\alpha
%+i\sqrt{2}\ov{\mu}^{\dot{\alpha}}\ov{\phi}\mu_{\dot{\alpha}}
%+2\ov{M}^{\dot{\alpha}}\phi\ov{\phi}M_{\dot{\alpha}}\left.\phantom{\over}\right]\\
&={1\over e^2}\int_{M_4} d^4x\,\sqrt{G_{M_4}} \left[ 
-{1\over 4}F_{\dot{\alpha}\dot{\beta}}F^{\dot{\alpha}\dot{\beta}}
-i\ov{M}_{\left(\dot{\alpha}\phantom{\over}\right.}M_{\left.\dot{\beta}\phantom{\over}\right)}F^{\dot{\alpha}\dot{\beta}}	+\ov{M}^{\left(\dot{\alpha}\phantom{\over}\right.}M^{\left.\dot{\beta}\phantom{\over}\right)} \ov{M}_{\left(\dot{\alpha}\phantom{\over}\right.}M_{\left.\dot{\beta}\phantom{\over}\right)}\right.\\
&\qquad\qquad\qquad\qquad\qquad
+ D_\mu\phi D^\mu \ov{\phi}
-i\chi_{\mu} D^{\mu}\eta
-i\lambda_{\dot{\alpha}\dot{\beta}}\cbrac{\ov\sigma_{\mu\nu}}^{\dot{\alpha}\dot{\beta}}D^\mu\chi^\nu
-\ov{M}^{\dot{\alpha}}D_{\dot{\alpha}\alpha}D^{\alpha\dot{\beta}}M_{\dot{\beta}}\\
&\qquad\qquad\qquad\qquad\qquad
-i\ov{\nu}_\alpha D^{\dot{\alpha}\alpha}\mu_{\dot{\alpha}}
-i\ov{\mu}^{\dot{\alpha}}D_{\alpha\dot{\alpha}}\nu^\alpha
-{1\over\sqrt{2}}\cbrac{\ov{M}^{\dot{\alpha}}\chi_{\alpha\dot{\alpha}}\nu^\alpha
	-\ov{\nu}_\alpha\chi^{\dot{\alpha}\alpha}M_{\dot{\alpha}}}\\
&\qquad\qquad\qquad\qquad\qquad
-{1\over\sqrt{2}}\lambda_{\dot{\alpha}\dot{\beta}}\cbrac{(i+1)\ov{M}^{\left(\dot{\alpha}\phantom{\over}\right.}\mu^{\left.\dot{\beta}\phantom{\over}\right)}
	+(i-1)\ov{\mu}^{\left(\dot{\alpha}\phantom{\over}\right.}M^{\left.\dot{\beta}\phantom{\over}\right)}}\\
&\qquad\qquad\qquad\qquad\qquad
-\sqrt{2}\cbrac{\ov{M}^{\dot{\alpha}}\eta\mu_{\dot{\alpha}}-\ov{\mu}^{\dot{\alpha}}\eta M_{\dot{\alpha}}}
-i\sqrt{2}\ov{\nu}_{\alpha}\phi\nu^\alpha
+i\sqrt{2}\ov{\mu}^{\dot{\alpha}}\ov{\phi}\mu_{\dot{\alpha}}
-2\ov{M}^{\dot{\alpha}}\phi\ov{\phi}M_{\dot{\alpha}}\left.\phantom{\over}\right]\\
&\phantom{=}+{i\theta\over 8\pi^2}\int_{M_4}\cbrac{F\wedge F},
\end{aligned}
\end{equation}
where $V_{\text{SW}}$ is a gauge-invariant fermionic operator. Here. $M$ denotes the spinor monopole fields, and the symmetric product $\ov{M}_{\left(\dot{\alpha}\phantom{\over}\right.}M_{\left.\dot{\beta}\phantom{\over}\right)}$ may be defined in matrix form as 
\begin{equation*}
\ov{M}_{\left(\dot{\alpha}\phantom{\over}\right.}M_{\left.\dot{\beta}\phantom{\over}\right)}=
\begin{pmatrix}
-M_1\ov{M}_2&\half\cbrac{\vbrac{M_1}^2-\vbrac{M_2}^2}\\
\half\cbrac{\vbrac{M_1}^2-\vbrac{M_2}^2}&\ov{M}_1M_2
\end{pmatrix}.
\end{equation*}
The corresponding supersymmetry transformations are
\begin{equation}
\label{susy:twistedSW}
\begin{aligned}
\delta A_\mu&=\zeta\chi_\mu,\\
\delta\phi&=0,\\
\delta\phi^\dagger&=2\sqrt{2}i\zeta\eta,\\
\delta\eta&=i\zeta[\phi,\phi^\dagger],\\
\delta\chi_\mu&=2\sqrt{2}\zeta D_\mu\phi,\\
\delta\lambda_{\dot{\alpha}\dot{\beta}}&=i\zeta\cbrac{F^+_{\dot{\alpha}\dot{\beta}}+2i\ov{M}_{\left(\dot{\alpha}\right.}M_{\left.\dot{\beta}\right)}},\\
\end{aligned}
\quad\begin{aligned}
\delta M_{\dot{\alpha}}&=-\sqrt{2}\zeta \mu_{\dot{\alpha}},\\
\delta \ov{M}_{\dot{\alpha}}&=\sqrt{2}\zeta \ov{\mu}_{\dot{\alpha}},\\
\delta\mu_{\dot{\alpha}}&=2i\zeta\phi M_{\dot{\alpha}},\\
\delta\ov{\mu}_{\dot{\alpha}}&=2i\zeta \ov{M}_{\dot{\alpha}}\ov{\phi},\\
\delta\nu^\alpha&=-i\sqrt{2}\zeta D^{\dot{\alpha}\alpha}M_{\dot{\alpha}},\\
\delta\ov{\nu}_\alpha&=-i\sqrt{2}\zeta D_{\alpha\dot{\alpha}}\ov{M}^{\dot{\alpha}}.
\end{aligned}
\end{equation}
Notice that all fields of the \susy{2} vector multiplet transform in the same way as (\ref{susy:twistedgeneral}), except for the fermionic 2-form $\lambda_{\mu\nu}$, whose susy transformation is modified by the presence of monopoles. It should be noted that if the hypermultiplet fields are set to zero, a pure $U(1)$ theory will be obtained, with the action and supersymmetry transformations taking the form of (\ref{S}) and (\ref{susy:twistedgeneral}).

\bigskip\noindent\textit{Observables of the Seiberg-Witten Theory}

There are many similarities between observables of the DW theory and those of the SW theory. 
%\footnote
%{
%	Likewise, we assume that the DW and SW theories are not equivalent in general. For instance, we can be comparing two such theories with gauge groups of different ranks.
%} 
%Let us first describe the direct analogies we can draw from the DW theory, before proceeding to describe the differences.
Like the DW theory, we may also consider the set of $n$ $\qcharge$-invariant operators $\CO_r$ where $r=1,\cdots, n$.  The correlation functions take the same form as (\ref{CF}), and the same arguments apply, since we may also write the SW action in a $\qcharge$-exact form.

Furthermore, due to supersymmetry, the theory is symmetric in the number of bosonic and fermionic degrees of freedom despite the extra field content. Hence, the contribution from the non-zero modes to the correlation function is still a factor of $\pm 1$. Correlation functions may then be defined up to a sign.

In the zero mode sector, however, we start to see some differences. The BPS equations are now
\begin{subequations}
	\begin{eqnarray}
	\label{con:SWBPSa}
	&D_\mu\phi=0,\\
	\label{con:SWBPSb}
	&F^+_{\dot{\alpha}\dot{\beta}}+2i\ov{M}_{\left(\dot{\alpha}\right.}M_{\left.\dot{\beta}\right)}=0\\
	\label{con:SWBPSc}
	&\phi M_{\dot{\alpha}}=0\\
	\label{con:SWBPSd}
	&\ov{M}_{\dot{\alpha}}\ov{\phi}=0\\
	\label{con:SWBPSe}
	&D^{\dot{\alpha}\alpha}M_{\dot{\alpha}}=0\\
	\label{con:SWBPSf}
	&D_{\alpha\dot{\alpha}}\ov{M}^{\dot{\alpha}}=0.
	\end{eqnarray}
\end{subequations}
We shall take the trivial solution $\phi_0=0$ like before. Then, (\ref{con:SWBPSb}), (\ref{con:SWBPSe}) and (\ref{con:SWBPSf}) are, collectively, the well-known SW equations, which take the form
\begin{subequations}
	\label{eqn:SW}
	\begin{eqnarray}
	\label{eqn:SWa}
	&F^+_{\dot{\alpha}\dot{\beta}}=-2i\ov{M}_{\left(\dot{\alpha}\right.}M_{\left.\dot{\beta}\right)}\\
    &D^{\dot{\alpha}\alpha}M_{\dot{\alpha}}=0\\
    &D_{\alpha\dot{\alpha}}\ov{M}^{\dot{\alpha}}=0
%	&\cbrac{\ov{\sigma}^\mu}^{\dot{\alpha}\alpha}D_\mu M_{\dot{\alpha}}=0\\
%	&\cbrac{\sigma^\mu}_{\alpha\dot{\alpha}}D_\mu\ov{M}^{\dot{\alpha}}=0.
	\end{eqnarray}
\end{subequations}
for which the solutions are $U(1)$ monopoles. So, zero modes of SW theory are monopoles. Gauge-inequivalent solutions to (\ref{eqn:SW}) then span the moduli space of charge $q$ monopoles, which we shall denote by $\monopolemoduli{q}{M_4}$. Like instantons, the monopole charge is a topological invariant, and is defined by the first Chern number \cite{elbistan2017weyl,shnir2006magnetic} of the $U(1)$ line bundle over $M_4$. Over a closed surface ${\mathscr S}$, this may be written as
\begin{equation}
\label{charge:monopole}
q = {1\over 2\pi}\int_{\mathscr S}\, F\in\integer.
\end{equation}
In general, we may choose any ${\mathscr S}\subset M_4$, as long as we assume that ${\mathscr S}$ wraps around the monopole. 

Let us define a $U(1)$ line bundle $L$, and a spin bundle $S=S^+\oplus S^-$. A section of $L$ is the gauge field $A$, while a section of $S^\pm$ is either a left- or right-handed spinor, depending on the sign. We take $M$ to be a section of $S^+$. Hence, the topological data of the SW equations is determined by the bundles $L$ and $S^+$. 

$U(1)_R$ charges for the fields $(A_\mu,\phi,\eta,\chi_\mu,\lambda_{\mu\nu},M^{\dot{\alpha}},\mu^{\dot{\alpha}},\nu^{\alpha})$ are $(0,2,-1,1,-1,0,-1,-1)$, respectively. The hypermultiplet fermions $\mu^{\dot{\alpha}}$ and $\nu^\alpha$, and their conjugates have opposite R-charges. Hence, the only non-vanishing contributions to the $U(1)_R$ charge of the measure come from the \susy{2} vector multiplet fermions $\eta,\chi_\mu,\lambda_{\mu\nu}$ like before. The virtual dimension of the moduli space of charge $q$ monopoles, $N_{\text{M}}=\dim\cbrac{\monopolemoduli{q}{M_4}}$, is \cite{witten1994monopoles}
\begin{equation}
\label{NM}
N_{\text{M}}=-{2\chi+3\tau\over 4}+c_1(L)^2,
\end{equation}
where $\chi$ is the Euler characteristic of $M_4$, and $\tau$ is the signature of $M_4$. Like before, operator insertions must absorb the fermionic zero modes, or else correlation functions will be vanishing due to the nature of Grassmannian integrals.

\subsubsection*{\textit{Relation Between Seiberg-Witten Theory and Monopole Floer Homology}}

We shall take $M_4=\real^+\times Y_3$, where $\real^+$ is identified with the time direction. Like DW theory, we can rewrite the SW theory as a supersymmetric 1d sigma model on the space of $U(1)$ connections and monopole fields on $Y_3$. Instead of having a Chern-Simons potential like the DW case, we now have a Chern-Simons-Dirac potential \cite{marcolli1999seiberg}, which takes the form $h_{Y_3}=\half \int_{Y_3}\cbrac{A_{Y_3}\wedge d A_{Y_3}+(M,D_{Y_3}M)}$, where the second term $(M,D_{Y_3}M)$ -- an inner product -- describes the dynamics of the monopole (spinor) fields in the presence of gauge fields on $Y_3$. 

The critical points of the Chern-Simons-Dirac potential are in one-to-one correspondence with the ground states and hence $\qcharge$-cohomology of the 1d sigma model, and they correspond to the monopole Floer homology $\HFmono{q,Y_3}$, where $q$ is the monopole charge (\ref{charge:monopole}). The underlying chain complex of $\HFmono{q,Y_3}$ is graded by the relative Morse index between a pair of critical points of the Chern-Simons-Dirac functional. Since the coboundary operator corresponding to $\CQ$ counts the number of solutions of the SW equations between a pair of critical points, the grading of $\HFmono{q,Y_3}$ coincides with the number of solutions of the SW equations.

If the virtual dimension $N_{\text{M}}$ (given by an open version of (\ref{NM})) vanishes,\footnote
{
	As it will be explained in \S\ref{sec:atiyah-floer}, we may write $M_4=\real^+\times Y_3\cong\real^+\times I\times \Sigma$, where $I$ is an interval and $\Sigma$ is a compact Riemann surface with genus $g$. The open version of the index theorem differs from (\ref{NM}) by the eta invariant, which also has no dependence on $g$. Since we can write $M_4$ as a product manifold, its Euler characteristic can be written as $\chi(M_4)=\chi(\real^+)\cdot\chi(Y_3)$. Furthermore, since $\real^+$ is topologically trivial -- i.e. $\chi(\real^+)=0$ -- the Euler characteristic of $M_4$ vanishes. Hence, setting $N_{\text{M}}=0$ places no constraints on the value of $g$.
} the only SW observables are partition functions of the form (\ref{ob:DWfloer}):
\begin{equation}
\label{ob:SWfloer}
\boxed{Z_{M_4}=\vev{1}_{\Psi(\Phi_{Y_3})} = \sum_i  \Psi_{\textrm{mono}}(\Phi^i_{Y_3})}
\end{equation}
In other words, the partition function on $M_4$ is a sum of monopole Floer homology classes $\Psi_{\textrm{mono}}(\Phi^i_{Y_3})$, where $i$ denotes the $i^{\textrm{th}}$ gauge connection and monopole field on $Y_3$ that descends from a monopole solution on $M_4$. 

We will be dealing with observables of the form (\ref{ob:SWfloer}), unless otherwise stated.

\subsection{Seiberg-Witten Theory and a 2d A-model\label{sec:SWadiabatic}}

It was suggested in \cite{gukov2009surface} that starting with SW theory on $M_4=\real^2\times C$, a dimensional reduction on $C$ should give rise to a 2d A-model with the moduli space of charge $q$ vortices on $C$ as its target space.% Specifically, the SW equations reduce to vortex equations, which gives rise to a moduli space of vortices on $C$. 

We would now like to physically prove the suggestion in \cite{gukov2009surface}, by making use of the technique discussed in \S\ref{sec:DWadiabatic}. Furthermore, we will generalize to the case where $\real^2$ is replaced by a generically curved Riemann surface $\Sigma$. To this end, let us consider SW theory on $M_4=\Sigma\times C$.
%\footnote{
%	However tempting it might be, the twisted SW theory should not be viewed as a Donaldson twisted $U(1)$ \susy{4} theory, even though matter fields appear to be in the ``adjoint" representation of the gauge group $U(1)$. To see this, we note that the twisted SW theory \textit{does not} possess the residual $SU(2)_{R'}$ R-symmetry, which is seen in a Donaldson-twisted \susy{4} theory. Hence, instead of obtaining a hyperK\"ahler target as expected from adiabatically reducing the (twisted) \susy{4} theory \cite{bershadsky1995topological,kapustin2006electric}, the target space obtained from SW theory is to be K\"ahler instead.
%}

\subsubsection*{\textit{The Adiabatic Limit and a 2d A-model with Target Moduli Space of Charge $q$ Vortices}}

Looking at the action in (\ref{action:SW}), it can be seen that the first three terms of the $\qcharge$-exact action can be combined to give a modified gauge curvature, which can be written in bispinor representation as
\begin{equation}
\CF_{\dot{\alpha}\dot{\beta}}=F_{\dot{\alpha}\dot{\beta}}+2i\ov{M}_{\left(\dot{\alpha}\phantom{\over}\right.}M_{\left.\dot{\beta}\phantom{\over}\right)}.
\end{equation}
In the vector representation of the Lorentz group (see appendix \ref{appendix}), this can also be written as 
\begin{equation}
\label{eqn:modifiedF}
\CF_{\mu\nu}=F_{\mu\nu}-i\cbrac{\ov\sigma_{\mu\nu}}^{\dot{\alpha}\dot{\beta}}\ov{M}_{\left(\dot{\alpha}\phantom{\over}\right.}M_{\left.\dot{\beta}\phantom{\over}\right)}.
\end{equation}
%having noted that $\ov{M}^{\dot{\alpha}}\cbrac{\ov{\sigma}_{\mu\nu}}_{\dot{\alpha}\dot{\beta}}M^{\dot{\beta}}\ov{M}^{\dot{\gamma}}\cbrac{\ov{\sigma}^{\mu\nu}}_{\dot{\gamma}\dot{\delta}}M^{\dot{\delta}}
%=-2\ov{M}^{\left(\dot{\alpha}\phantom{\over}\right.}M^{\left.\dot{\beta}\phantom{\over}\right)} \ov{M}_{\left(\dot{\alpha}\phantom{\over}\right.}M_{\left.\dot{\beta}\phantom{\over}\right)}$. 
The action in (\ref{action:SW}) may then be rewritten as
\begin{equation}
\label{action:SWmodified}
\begin{aligned}
S_{\text{SW}}
&={1\over e^2}\int_{M_4} d^4x\,\sqrt{G_{M_4}} \left[ 
-{1\over 4}\CF_{\mu\nu}\CF^{\mu\nu}
+ D_\mu\phi D^\mu \ov{\phi}
-i\chi_{\mu} D^{\mu}\eta
-i\lambda_{\dot{\alpha}\dot{\beta}}\cbrac{\ov\sigma_{\mu\nu}}^{\dot{\alpha}\dot{\beta}}D^\mu\chi^\nu
\right.\\
&\qquad\qquad\qquad\qquad\qquad
-\ov{M}^{\dot{\alpha}}D_{\dot{\alpha}\alpha}D^{\alpha\dot{\beta}}M_{\dot{\beta}}
-i\ov{\nu}_\alpha D^{\dot{\alpha}\alpha}\mu_{\dot{\alpha}}
-i\ov{\mu}^{\dot{\alpha}}D_{\alpha\dot{\alpha}}\nu^\alpha\\
&\qquad\qquad\qquad\qquad\qquad
-{1\over\sqrt{2}}\lambda_{\dot{\alpha}\dot{\beta}}\cbrac{(i+1)\ov{M}^{\left(\dot{\alpha}\phantom{\over}\right.}\mu^{\left.\dot{\beta}\phantom{\over}\right)}
	+(i-1)\ov{\mu}^{\left(\dot{\alpha}\phantom{\over}\right.}M^{\left.\dot{\beta}\phantom{\over}\right)}}
-2\ov{M}^{\dot{\alpha}}\phi\ov{\phi}M_{\dot{\alpha}}\\
&\qquad\qquad\qquad\qquad\qquad
-\sqrt{2}\cbrac{\ov{M}^{\dot{\alpha}}\eta\mu_{\dot{\alpha}}-\ov{\mu}^{\dot{\alpha}}\eta M_{\dot{\alpha}}}
-i\sqrt{2}\ov{\nu}_{\alpha}\phi\nu^\alpha
+i\sqrt{2}\ov{\mu}^{\dot{\alpha}}\ov{\phi}\mu_{\dot{\alpha}}\\
&\qquad\qquad\qquad\qquad\qquad
-{1\over\sqrt{2}}\cbrac{\ov{M}^{\dot{\alpha}}\chi_{\alpha\dot{\alpha}}\nu^\alpha
	-\ov{\nu}_\alpha\chi^{\dot{\alpha}\alpha}M_{\dot{\alpha}}}\left.\phantom{\over}\right]
+{i\theta\over 8\pi^2}\int_{M_4}\cbrac{F\wedge F}.
\end{aligned}
\end{equation}

Following the arguments used in \S\ref{sec:DWadiabatic}, we must have flat modified connections on the reduced directions on $C$, to ensure a finite action. Likewise, the monopole kinetic term, $-\ov{M}^{\dot{\alpha}}D_{\alpha\dot{\alpha}}D^{\dot{\beta}\alpha}M_{\dot{\beta}}$, must also be set to zero on $C$. So, we need to impose the conditions 
\begin{subequations}
	\label{con:adiabaticSW}
	\begin{eqnarray}
	\label{con:adiabaticSWa}
	&\CF_{ab}=0\\
	\label{con:adiabaticSWb}
	&\cbrac{\ov{M}^{\dot{\alpha}}D_{\alpha\dot{\alpha}}D^{\dot{\beta}\alpha}M_{\dot{\beta}}}_{ab}=0,
	\end{eqnarray}
\end{subequations}
where the $C$ components of (\ref{con:adiabaticSWb}) may be obtained by switching from the bispinor representation to the vector representation.% (See appendix \ref{appendix} for more details). 

After some simplifications, and writing in terms of complex coordinates defined in (\ref{eqn:complexcoords}), (\ref{con:adiabaticSW}) becomes
\begin{subequations}
	\label{eqn:almostvortex}
	\begin{eqnarray}
	&F_{34}={1\over 2}\cbrac{\vbrac{M_1}^2-\vbrac{M_2}^2}\\
	&D_{w}M_1=0\\
	&D_{\ov{w}}M_2=0.
	\end{eqnarray}
\end{subequations}

To obtain vortex equations, we may freeze out two monopole degrees of freedom.\footnote
{	
%	Recall that monopole equations were obtained as BPS equations in (\ref{eqn:SW}), which can be written as
%	\begin{subequations}
%		\label{eqn:SWfromaction}
%		\begin{eqnarray}
%		\label{eqn:SWfromactiona}
%		&F_{\mu\nu}-i\cbrac{\ov\sigma_{\mu\nu}}^{\dot{\alpha}\dot{\beta}}\ov{M}_{\left(\dot{\alpha}\phantom{\over}\right.}M_{\left.\dot{\beta}\phantom{\over}\right)}=0\\
%		\label{eqn:SWfromactionb}
%		&D^{\dot{\alpha}\alpha}M_{\dot{\alpha}}=0.
%		\end{eqnarray}
%	\end{subequations}

	We can perturb the SW equations(\ref{eqn:SW}) by adding a closed 2-form $\Omega={1\over 2}\cbrac{dx^1\wedge dx^2+dx^3\wedge dx^4}$ without affecting the corresponding correlation functions \cite{witten1994monopoles}, so that
	\begin{equation}
	F_{\mu\nu}=i\cbrac{\ov\sigma_{\mu\nu}}^{\dot{\alpha}\dot{\beta}}\ov{M}_{\left(\dot{\alpha}\phantom{\over}\right.}M_{\left.\dot{\beta}\phantom{\over}\right)}
	+\Omega_{\mu\nu}.
	\end{equation}
	Shrinking $C$, and ensuring that the action remains finite, the perturbed SW equations become
	\begin{subequations}
		\label{eqn:vortexfromSWalmost}
		\begin{eqnarray}
		&F_{34}={1\over 2}\cbrac{1+\vbrac{M_1}^2-\vbrac{M_2}^2}\\
		&D_w M_1=0\\
		&D_{\ov{w}}M_2=0.
		\end{eqnarray}
	\end{subequations}
	Since we have a TQFT, $C$ can be taken to very large, so that ``boundary conditions" can be imposed. If we take $\vbrac{M_2}\to 1$ when $\vbrac{w}\to \infty$, solutions to (\ref{eqn:vortexfromSWalmost}) then necessarily requires that $M_1=0$, so that $F_{34}\to 0$ on the boundary. This then gives us the vortex equations
	\begin{subequations}
		\label{eqn:vortexfromSW}
		\begin{eqnarray}
		&F_{34}={1\over 2}\cbrac{1-\vbrac{\varphi}^2}\\
		&D_{\ov{w}}\varphi=0,
		\end{eqnarray}
	\end{subequations}
	where $\varphi=M_2$ is now identified with the vortex field. Physically, this is tantamount to freezing out two monopole degrees of freedom.
	
	If we began with the unperturbed SW equations (\ref{eqn:SW}) instead, shrinking $C$ leads to (\ref{eqn:almostvortex}), which produces the same set of solutions. To obtain vortex equations (\ref{eqn:vortexfromSW}) from (\ref{eqn:almostvortex}), we may simply set $\vbrac{M_1}=1$ -- which, also suppresses two monopole degrees of freedom. This amounts to choosing the same boundary condition $\vbrac{M_2}\to 1$ in the limit $\vbrac{w}\to \infty$, so that $F_{34}$ also vanishes on the boundary. 
	We can then set $\vbrac{M_1}=1$ to obtain the vortex equations (\ref{eqn:vortex}).} 
To this end, let us conveniently choose $M_1=1$ and $M_2=\varphi$, so that (\ref{con:adiabaticSW}) becomes
\begin{subequations}
	\label{eqn:vortex}
	\begin{empheq}[box=\widefbox]{align}
\	&F_{w\ov{w}}={i\over 4}\cbrac{1-\vbrac{\varphi}^2}\\
	&D_{\ov{w}}\varphi=0
	\end{empheq}
\end{subequations}
which are indeed vortex equations on $C$. Similar derivations can be found in \S 3 of \cite{labastida2005topological}. 

%\bigskip\noindent\textit{A Slight Diversion -- Deriving Vortices from the Perturbed Seiberg-Witten Equations}

\bigskip\noindent\textit{The Moduli Space of Charge $q$ Vortices}

To look for nearby gauge-inequivalent solutions, vortex equations (\ref{eqn:vortex}) can be linearized as
\begin{subequations}
	\begin{eqnarray}
	&D_w\delta A_{\ov{w}}-D_{\ov{w}}\delta A_w={i\over 4} \cbrac{\delta\ov{\varphi}\cdot\varphi+\ov{\varphi}\cdot\delta\varphi}\\
	&D_{\ov{w}}\delta\varphi=i\delta A_{\ov{w}}\varphi.
	\end{eqnarray}
\end{subequations}
Further choosing the gauge-fixing condition \cite{tong2005tasi}
\begin{equation}
D_w\delta A_{\ov{w}}+D_{\ov{w}}\delta A_w={i\over 4} \cbrac{\delta\ov{\varphi}\cdot\varphi-\ov{\varphi}\cdot\delta\varphi},
\end{equation}
we may rewrite the linearized vortex equations as
\begin{subequations}
	\label{eqn:linearizedvortex}
	\begin{eqnarray}
	&D_w\delta A_{\ov{w}}={i\over 4} \delta\ov{\varphi}\cdot\varphi\\
	&D_{\ov{w}}\delta\varphi=i\delta A_{\ov{w}}\varphi.
	\end{eqnarray}
\end{subequations}
Solutions to (\ref{eqn:linearizedvortex}) then span a moduli space of vortices $\vortexmoduli{q}{C}$, where $q$ is the vortex charge. 

The vortex charge is defined very similarly to the monopole charge (\ref{charge:monopole}), with the difference being that it is defined in a 2d theory, say on a Riemann surface $C$. The vortex charge \cite{tong2005tasi} can be written as
\begin{equation}
\label{charge:vortex}
q={1\over 2\pi}\int_C \, F\in\integer.
\end{equation}
Referring to (\ref{charge:monopole}), if we restrict the surface ${\mathscr S}\subset M_4$ to $C$ -- i.e. set ${\mathscr S}=C$ -- the monopole charge (\ref{charge:monopole}) takes exactly the same form as (\ref{charge:vortex}). Hence, upon shrinking $C$, the monopole charge descends trivially to the vortex charge.

Since there are additional bosonic degrees of freedom on $C$ -- being those of the vortex fields $\varphi$ -- we can also define the variations $(\delta\varphi,\delta\ov{\varphi})$ in terms of additional basis cotangent vectors $(\beta_I,\ov{\beta}_I)$. Using the same technique used in \S\ref{sec:DWadiabatic}, we may then write the variations, in terms of collective coordinates $X:\Sigma\to\vortexmoduli{q}{C}$ and basis cotangent vectors $(\alpha_{I C},\beta_I)$, as
\begin{subequations}
	\label{eqn:variationsvortex}
	\begin{eqnarray}
	&{\partial A_C\over\partial X^I}=\alpha_{I C}+\partial_C E_I\\
	&{\partial \varphi\over\partial X^I}=\beta_I+iE_I\varphi\\
	&{\partial \ov{\varphi}\over\partial X^I}=\beta_I+iE_I\ov\varphi,
	\end{eqnarray}
\end{subequations}
where $E_I$ are gauge transformation parameters. The moduli space of vortices, $\vortexmoduli{q}{C}$,  is K\"ahler \cite{tong2005tasi}. Its metric can be written as
\begin{equation}
\label{met:vortex}
G^{\text{vort}}_{IJ}=\int_C d^2w\ \cbrac{\alpha_{I w}\alpha_{J\ov{w}}+\beta_I\ov{\beta}_J+\alpha_{I \ov{w}}\alpha_{Jw}+\ov{\beta}_I\beta_J},
\end{equation}
while the symplectic form \cite{dey2006geometric} is
\begin{equation}
\label{eqn:symplecticvortex}
\omega^{\text{vort}}_{IJ}=\int_C d^2w\ \cbrac{\alpha_{I w}\alpha_{J\ov{w}}+\beta_I\ov{\beta}_J-\alpha_{I \ov{w}}\alpha_{Jw}-\ov{\beta}_I\beta_J}.
\end{equation}

However, the reduced SW action appears to be unsuitable for deriving the sigma model action. This problem arises because we chose $\vbrac{M_1}=1$ in order to obtain the vortex equations in (\ref{eqn:vortex}), which means that the surviving monopole kinetic terms in (\ref{action:SW}) becomes \textit{only} linear in derivatives of $(\varphi,\ov\varphi)$. 

More explicitly, the surviving monopole kinetic terms are the ones with mixed indices -- i.e. 
\begin{equation}
\begin{aligned}
&\cbrac{\ov{M}^{\dot{\alpha}}D_{\dot{\alpha}\alpha}D^{\alpha\dot{\beta}}M_{\dot{\beta}}}_{Aa}+\cbrac{\ov{M}^{\dot{\alpha}}D_{\dot{\alpha}\alpha}D^{\alpha\dot{\beta}}M_{\dot{\beta}}}_{aA}\\
&=\cbrac{\sigma^A}_{\alpha\dot{\alpha}}\cbrac{\ov{\sigma}^a}^{\dot{\beta}\alpha}D_A\ov{M}^{\dot{\alpha}}D_aM_{\dot{\beta}}
+\cbrac{\sigma^a}_{\alpha\dot{\alpha}}\cbrac{\ov{\sigma}^A}^{\dot{\beta}\alpha}D_a\ov{M}^{\dot{\alpha}}D_AM_{\dot{\beta}}\\
%&=4D_{\ov{z}}\ov{M}_1D_{\ov{w}}M_2
%-4D_z\ov{M}_2D_w M_1
%-4D_{\ov{w}}\ov{M}_1D_{\ov{z}}M_2
%+4D_w\ov{M}_2 D_zM_1\\
&=-4iA_{\ov{z}}D_{\ov{w}}\varphi
+4iD_z\ov{\varphi} A_w
+4iA_{\ov{w}}D_{\ov{z}}\varphi
-4Di_w\ov{\varphi} A_z,
\end{aligned}
\end{equation}
which are indeed only linear in derivatives of $(\varphi,\ov\varphi)$.

To get around this problem -- i.e. to be able to write down the sigma model action -- we can rewrite spinor monopole fields as monopole vectors instead. Roughly speaking, monopole spinors can be combined with Pauli matrices (also viewed as gamma matrices) to define vector fields. In doing so, a modified gauge connection can be written, which is somewhat analogous to the case of GL-twisted \susy{4} theory \cite{kapustin2006electric}. Next, we will explain how this procedure can be carried out.

\bigskip\noindent\textit{The Bosonic Part of the Sigma Model Action}

In its current form, the $(A,M)$ part of the action (\ref{action:SWmodified}) cannot produce the target space metric (\ref{met:vortex}). Here, we shall show how the bosonic part of the sigma model can be obtained.

Instead of choosing $M_1=1$ and $M_2=\varphi$, we shall choose monopole vectors $\Upsilon$, so that (\ref{eqn:modifiedF}) can be rewritten as
\begin{equation}
\CF_{\mu\nu}=\partial_\mu \cbrac{A_\nu+\Upsilon_\nu}-\partial_\nu \cbrac{A_\mu+\Upsilon_\mu}.
\end{equation}
We may then define the corresponding \textit{modified} gauge connection $\CA$ with respect to the modified field strength $\CF$ -- i.e. take $\CA=A+\Upsilon$. This can be achieved by setting
%\begin{equation}
%\begin{aligned}
%&\partial_{\ov z}\Upsilon_{w}=\quar \ov M_2 M_1=-\partial_{\ov w}\Upsilon_{z}\\
%&\partial_{w}\Upsilon_{\ov z}=\quar \ov M_1 M_2=-\partial_{z}\Upsilon_{\ov w}\\
%&\partial_{z}\Upsilon_{w}=\partial_{\ov z}\Upsilon_{\ov w}=\partial_{w}\Upsilon_{z}=\partial_{\ov w}\Upsilon_{\ov z}=0.
%\end{aligned}
%\end{equation}
\begin{equation}
\p_\mu\Upsilon_\nu-\p_\nu\Upsilon_\mu=-i\cbrac{\ov\sigma_{\mu\nu}}^{\dot{\alpha}\dot{\beta}}\ov{M}_{\left(\dot{\alpha}\phantom{\over}\right.}M_{\left.\dot{\beta}\phantom{\over}\right)}.
\end{equation}
%where the monopole \textit{vectors} are real-valued.

Next, we may write the variations of $A_C$ and $\Upsilon_C$ in terms of basis cotangent vectors, which we shall now denote by $\alpha_{IC}$ and $\widetilde{\beta}_{IC}$, respectively. We may still define collective coordinates $X:\Sigma\to\vortexmoduli{q}{C}$ as in (\ref{eqn:variationsvortex}), since solutions to the SW equation descend to the same set of solutions upon shrinking $C$. In other words, variations of the pair $(A_C,\Upsilon_C)$ still obey the linearized vortex equations in (\ref{eqn:linearizedvortex}).

Instead of (\ref{eqn:variationsvortex}), the variations here can be written as
\begin{subequations}
	\begin{eqnarray}
	&{\partial A_C\over\partial X^I}=\alpha_{I C}+\partial_C E_I\\
	&{\partial \Upsilon_C\over \partial X^I}=\widetilde{\beta}_{IC}.
	\end{eqnarray}
\end{subequations}
The ensuing metric and symplectic forms can then be written as
\begin{align}
\label{met:vortexaction}
&G^{\text{vort}}_{IJ}=\int_C d^2w\ \cbrac{\alpha_{I w}\alpha_{J\ov{w}}+\widetilde{\beta}_{Iw}\widetilde{\beta}_{J\ov w}+\alpha_{I \ov{w}}\alpha_{Jw}+\widetilde{\beta}_{I\ov w}\widetilde{\beta}_{Jw}}\\
\label{eqn:symplecticvortexaction}
&\omega^{\text{vort}}_{IJ}=\int_C d^2w\ \cbrac{\alpha_{I w}\alpha_{J\ov{w}}+\widetilde{\beta}_{Iw}\widetilde{\beta}_{J\ov w}-\alpha_{I \ov{w}}\alpha_{Jw}-\widetilde{\beta}_{I\ov w}\widetilde{\beta}_{Jw}}.
\end{align}
Consequently, we may then write the mixed components of the modified field strength as
\begin{equation}
\label{eqn:mixedmodfieldstrength}
\CF_{\Sigma C}=\partial_\Sigma X^I\cbrac{\alpha_{IC}+\partial_{C}E_I+\widetilde{\beta}_{IC}}-\partial_C\CA_{\Sigma}.
\end{equation}
So the surviving kinetic Lagrangian in $\CA$ becomes
\begin{equation}
\label{action:almostsigmavortex}
\begin{aligned}
{L'}_{\text{SW}}^{(A,\Upsilon)}
&={1\over 2e^2}\CF_{Aa}\CF^{Aa}\\
&={1\over e^2}\left(\partial z X^I\partial_{\ov z}X^J
\cbrac{\alpha_{Iw}\alpha_{J\ov w}+\widetilde{\beta}_{Iw}\widetilde{\beta}_{J\ov w}+\alpha_{I\ov w}\alpha_{J w}+\widetilde{\beta}_{I\ov w}\widetilde{\beta}_{Jw}}\right.\\
&\left.\quad+\partial z X^I\partial_{\ov z}X^J
\cbrac{\widetilde{\beta}_{Iw}\alpha_{J\ov w}+\alpha_{Iw}\widetilde{\beta}_{J\ov w}+\widetilde{\beta}_{I\ov w}\alpha_{J w}+\alpha_{I\ov w}\widetilde{\beta}_{Jw}}
+\cdots\right),
\end{aligned}
\end{equation}
where the ellipses denote extra terms that contain terms with derivatives tangent to $C$.

The ensuing action can then be written as
\begin{equation}
\label{action:SWsigmabosonic}
\begin{aligned}
	{S'}_{\text{SW}}^{(A,\Upsilon)}&={1\over 2e^2}\int_\Sigma\cbrac{\int_C d^2w\ \CF_{Aa}\CF^{Aa}}\\
	&={1\over e^2}\int_\Sigma G^{\text{vort}}_{IJ}\partial z X^I\partial_{\ov z}X^J,
	\end{aligned}
\end{equation}
upon further choosing the gauge-fixing condition $\alpha_{IA}{\widetilde{\beta}_J}\phantom{ }^A=-\widetilde{\beta}_{IA}{\alpha_{J}}^A$. Note that terms in the ellipses have vanished since this integration amounts to suppressing the dependence on directions tangent to $C$ -- i.e. setting ${\p}_C=0$. This is indeed the bosonic part of a sigma model action, which is defined on a worldsheet $\Sigma$ and target $\vortexmoduli{q}{C}$.

\bigskip\noindent\textit{Pullback of the Symplectic Form}

Recall the topological term ${i\theta\over 8\pi^2}\int_{M_4}\cbrac{F\wedge F}$ in the action (\ref{action:SW}). Note that $F$ only provides the $\alpha$ part of the symplectic form in (\ref{eqn:symplecticvortexaction}), but not the $\widetilde{\beta}$ part. Furthermore, since the topological term contributes nothing to the dynamics of SW theory -- i.e. it merely produces some number -- it can brought outside of the path integral as a phase factor.

%%%%%Extracting toplogical term from Q-exact action%%%%%%

%Instead of working with ${F\wedge F}$, we shall exploit the self-dual property of ${\ov \sigma}_{\mu\nu}$ to write
%\begin{equation}
%\CF\wedge*\CF=\half\CF\wedge*\CF+\half\CF\wedge\CF.
%\end{equation}
%Then, we can write an appropriately rescaled kinetic term of $\CA$ as
%\begin{equation}
%\begin{aligned}
%{1\over 8\pi^2}\int_{M_4} d^4x \sqrt{G_{M_4}}\CF^{\mu\nu}\CF_{\mu\nu}
%&={1\over 16\pi^2}\int_{M_4} d^4x \sqrt{G_{M_4}}\CF^{\mu\nu}\CF_{\mu\nu}
%+{1\over 16\pi^2}\int_{M_4} d^4x \half\epsilon^{\mu\nu\rho\lambda}\CF_{\rho\lambda}\CF_{\mu\nu}\\
%&=S^{(\CA)}_{\text{KE}}+S_{\text{top}},
%\end{aligned}
%\end{equation}
%where the second term $S_{\text{top}}$ is now metric-independent -- i.e. $S_{\text{top}}$ is a topological term.

%%%%%%%%%%%%%%End of discussion%%%%%%%%%%%%%%%%%%%%

Consider adding to the action, a $\qcharge$-exact \textit{topological} term\footnote
{
	This $\CQ$-exact topological term can be added inconsequentially, since it does not contribute to the path integral. 
} 
of the form
\begin{equation}
\label{action:extratop}
\begin{aligned}
i\theta S_{\text{top}}&={i\theta\over 16\pi^2}\int_{M_4}d^4x\ \cbrac{-{i\over 2}\epsilon^{\mu\nu\rho\lambda}\ccbrac{\qcharge,\CF_{\rho\lambda}\lambda_{\mu\nu}}}\\
&={i\theta\over 8\pi^2}\int_{M_4} \CF\wedge\CF\\
	&\quad+{\theta\over 16\pi^2}\int_{M_4}d^4x\ \epsilon^{\mu\nu\rho\lambda}\cbrac{D_\rho\chi_\lambda\lambda_{\mu\nu}
		+{1\over\sqrt{2}}\cbrac{\ov\sigma_{\rho\lambda}}^{\dot{\alpha}\dot{\beta}}\cbrac{\ov{M}_{\left(\dot{\alpha}\phantom{\over}\right.}\mu_{\left.\dot{\beta}\phantom{\over}\right)}
		-\ov{\mu}_{\left(\dot{\alpha}\phantom{\over}\right.}M_{\left.\dot{\beta}\phantom{\over}\right)}}},
\end{aligned}
\end{equation}
to allow the appearance of $\widetilde{\beta}$. Note that in shrinking $C$, no kinetic terms involving the monopole fermions $\mu$ and $\nu$ can be obtained. Consequently, they become auxiliary and can be integrated out. Hence, we may set $\mu=0$ in (\ref{action:extratop}) in the ensuing sigma model on $\Sigma$ with target $\vortexmoduli{q}{C}$. Furthermore, since this term is topological, there are only fermionic zero modes, whence the fermionic term $D_\rho\chi_\lambda\lambda_{\mu\nu}$ can be ignored.

Next, (\ref{action:extratop}) can be expanded in terms of $\Sigma$ and $C$ directions. Then, shrinking $C$, and consequently setting $\CF_{ab}=0$, (\ref{action:extratop}) becomes
\begin{equation}
\label{eqn:SWtop1}
\begin{aligned}
S_{\text{top}}
&={1\over 8\pi^2}\int_{\Sigma\times \varepsilon C} d^4x \cbrac{\epsilon^{AaBb}\CF_{Aa}\CF_{Bb}}\\%+\cdots\\
&={1\over 4\pi^2}\int_{\Sigma\times \varepsilon C} d^4x \cbrac{\CF_{zw}\CF_{\ov{z}\ov{w}}-\CF_{z\ov{w}}\CF_{\ov{z}w}},%+\cdots.
\end{aligned}
\end{equation}
where the terms in $\cdots$ are ignored in the path integral. Finally, using (\ref{eqn:mixedmodfieldstrength}) and applying the same gauge-fixing condition $\alpha_{IA}{\widetilde{\beta}_J}\phantom{ }^A=-\widetilde{\beta}_{IA}{\alpha_{J}}^A$, (\ref{eqn:SWtop1}) becomes
\begin{equation}
\label{eqn:SWtop2}
\begin{aligned}
S_{\text{top}}&
={1\over 2\pi^2}\int_{\Sigma}d^2z(\partial_z X^I\partial_{\ov{z}}X^{J})\cbrac{\int_{C}d^2w\ \cbrac{\alpha_{I w}\alpha_{J\ov{w}}+\widetilde{\beta}_{Iw}\widetilde{\beta}_{J\ov w}-\alpha_{I \ov{w}}\alpha_{Jw}-\widetilde{\beta}_{I\ov w}\widetilde{\beta}_{Jw}}}\\
&={1\over 2\pi^2}\int_{\Sigma}d^2z\ \omega^{\text{vort}}_{IJ}(\partial_z X^I\partial_{\ov{z}}X^{J}),
\end{aligned}
\end{equation}
where the last equality makes use of the definition of the symplectic form in (\ref{eqn:symplecticvortexaction}).

This means that (\ref{eqn:SWtop2}) can also be written as a pullback of the symplectic form as
\begin{equation}
\label{charge:degreepullbackSW}
\boxed{S_{\text{top}}={1\over 8\pi^2}\int_{M_4}\ \CF\wedge \CF=\int_{\Sigma}\ X^*\omega_{\text{vort}}=k\in\integer}
\end{equation}

\bigskip\noindent\textit{The A-model}

The argument preceding (\ref{action:DWsigma}) also applies to the case of SW theory -- that is, the (twisted) SW theory on $M_4=\Sigma\times C$ must also become a twisted \susy{(2,2)} theory on $\Sigma$ when $C$ is small. This is due to the breaking of half the original supersymmetry when a 4d (non-twisted) supersymmetric theory on $\real^2\times C$ is reduced on a curved Riemann surface $C$. Since we began with 8 supercharges in the 4d \susy{2} theory, 4 supercharges remain in the 2d theory. 
Further carrying out a topological twist then allows us to replace $\real^2$ by an arbitrary (curved) Riemann surface $\Sigma$. Consequently, reducing SW theory on $C$ must lead to a twisted 2d \susy{(2,2)} theory on $\Sigma$. 

Furthermore, it was seen in (\ref{action:SWsigmabosonic}) that shrinking $C$ gives rise to a 2d sigma model on $\Sigma$ with target $\vortexmoduli{q}{C}$. This means that the 2d \susy{(2,2)} theory is also a sigma model, whose fermi fields can then be determined from supersymmetry.

As mentioned earlier, the hypermultiplet fermions $\mu,\nu$ become auxiliary upon shrinking $C$, since their derivatives on $C$ vanish. Hence, the only surviving fermions are the gauge fermions $\chi_C$ and $\lambda_{C\Sigma}$, which can now be written in terms of the basis cotangent vectors $\alpha$ and $\widetilde{\beta}$ as\footnote
{
	Note that in general, we may write a 4d fermion as an arbitrary linear combination of $\alpha$ and $\widetilde{\beta}$. For instance, we could have written $\lambda_{wz}={\rho_z}^{\ov I}\alpha_{\ov I w}+{\widetilde{\rho}_z}^{\ov I}{\widetilde{\beta}}_{\ov I w}$ instead, where generically, ${\rho_z}^{\ov I}\neq {\widetilde{\rho}_z}^{\ov I}$. However, since we expect the 2d theory to possess \susy{(2,2)} supersymmetry, ${\rho_z}^{\ov I}$ and ${\widetilde{\rho}_z}^{\ov I}$ cannot be independent of each other. Otherwise, extra fermionic degrees of freedom will be obtained. Hence, we may assume ${\rho_z}^{\ov I}= {\widetilde{\rho}_z}^{\ov I}$, so that the correct number of fermionic degrees of freedom is obtained.
	
	This argument also applies to the other worldsheet fermions.
}
\begin{equation}
\begin{aligned}
&\chi_w={\chi}^{\ov I}\cbrac{\alpha_{\ov{I}w}+{\widetilde{\beta}}_{\ov{I}w}}\\
&\chi_{\ov{w}}=\chi^{I}\cbrac{\alpha_{I\ov{w}}+{\widetilde{\beta}}_{I\ov{w}}}\\
&\lambda_{wz}={{\rho}_{z}}^{\ov I}\cbrac{\alpha_{\ov{I}w}+{\widetilde{\beta}}_{\ov{I}w}}\\
&\lambda_{\ov{w}\ov{z}}={\rho_{\ov{z}}}^{I}\cbrac{\alpha_{I\ov{w}}+{\widetilde{\beta}}_{I\ov{w}}}.
\end{aligned}
\end{equation}
%; these fermions descend to vectors on $\vortexmoduli{q}{C}$.

Next,  consider the 4d supersymmetry transformation $\lambda_{\mu\nu}=i\zeta\CF^+_{\mu\nu}$ (see \eqref{susy:twistedSW}). After shrinking $C$, the only surviving 2-form fermi terms are $\lambda_{wz}={{\rho}_{z}}^{\ov I}\cbrac{\alpha_{\ov{I}w}+{\widetilde{\beta}}_{\ov{I}w}}$ and $\lambda_{\ov{w}\ov{z}}={\rho_{\ov{z}}}^{I}\cbrac{\alpha_{I\ov{w}}+{\widetilde{\beta}}_{I\ov{w}}}$. Further recalling \eqref{eqn:mixedmodfieldstrength}, and using complex coordinates on $\vortexmoduli{q}{C}$, the ensuing 2d supersymmetry transformations can then be written as 
\begin{equation}
\begin{aligned}
&\delta{{\rho}_{z}}^{\ov I}=-{i\over 2}\p_{\ov z} X^I\\
&\delta{\rho_{\ov{z}}}^{I}=-{i\over 2}\p_z X^{\ov I}.
\end{aligned}
\end{equation}
Hence, the corresponding BPS equations of the 2d sigma model are
\begin{equation}
\begin{aligned}
&\p_{\ov z} X^I=0\\
&\p_z X^{\ov I}=0,
\end{aligned}
\end{equation}
which means the maps $X^I:\Sigma\to\vortexmoduli{q}{C}$ are \textit{holomorphic maps}.

%Finally, it was shown in (\ref{charge:degreepullbackSW}) that the topological term $S_{\text{top}}$ descends to the degree $k$ of maps $X$. This is characteristic of an A-twisted sigma model. Altogether, this means that the 2d \susy{(2,2)} theory on $\Sigma$ is an A-twisted sigma model with target $\vortexmoduli{q}{C}$. Since we have an A-model, this implies that the maps $X:\Sigma\to\vortexmoduli{q}{C}$ are also holomorphic maps.
Finally, it was shown in (\ref{charge:degreepullbackSW}) that the topological term $S_{\text{top}}$ descends to the pullback, which means $X^I$ are holomorphic maps of degree $k$. Altogether, this means that the 2d \susy{(2,2)} theory on $\Sigma$ is an A-twisted sigma model with target $\vortexmoduli{q}{C}$, which has the action
\begin{equation}
\boxed{\begin{aligned}
\label{action:SWsigma}
S_{\text{SW}}'&={1\over e^2}\int_{\Sigma} d^2z\ \left(G^{\text{vort}}_{I\ov J}\cbrac{
%	\quar\partial_{z}X^{I}\partial_{\ov{z}}{X}^{\ov J}
%	+\quar\partial_{\ov{z}}X^{I}\partial_z{X}^{\ov J}
	\half\partial_{z}X^{I}\partial_{\ov{z}}{X}^{\ov J}
	+\half\partial_{\ov{z}}X^{I}\partial_z{X}^{\ov J}
	+{{\rho}_{z}}^{\ov J}\nabla_{\ov{z}}\chi^{I}
	+{\rho_{\ov{z}}}^{I}\nabla_z{\chi}^{\ov J}}\right.\\
&\qquad\qquad\qquad\left.\phantom{\half}
-R_{I\ov JK\ov L}{{\rho}_{\ov z}}^{I}{\rho_{z}}^{\ov J}{\chi}^K\chi^{\ov L}\right)
+{i\theta}\int_{\Sigma}\ X^*\omega_{\text{vort}}
%-{2\pi^2\over e^2}\int_{\Sigma}\ X^*\omega_{\text{vort}}
\end{aligned}}
\end{equation}
where $R_{I\ov JK\ov L}$ is the Riemann curvature tensor on $\vortexmoduli{q}{C}$, and $\nabla_{\ov z}\chi^{I}=\partial_{\ov z}\chi^{I}+\chi^{J}\Gamma^{I}_{JK}\partial_{\ov z}X^{K}$. Here, $\Gamma^{I}_{JK}$ are the Christoffel symbols on $\vortexmoduli{q}{C}$. 

Therefore, for the case of SW theory on $M_4=\Sigma\times C$, upon shrinking $C$, an A-twisted sigma model with degree $k$ holomorphic maps from $\Sigma$ to $\vortexmoduli{q}{C}$ is obtained.

\section{Physical Proofs of Theorems and Conjectures on Floer Homologies, and their Higher Rank Generalizations \label{sec:4}}

\subsection{The Atiyah-Floer Conjecture: Relating Instanton to Lagrangian Floer Homology\label{sec:atiyah-floer}}

%Look at the specific case of SU(2) now

The Atiyah-Floer conjecture \cite{atiyah1988new} relates the critical points of the Chern-Simons functional to those of Lagrangian intersections $L_0\cap L_1$. It is given by 
\begin{equation}
\label{math:atiyahfloer}
	{\HFinst{Y_3}\cong\HFlagr{\flatmoduli{\Sigma},L_0,L_1}},
\end{equation}
where $L_0$ and $L_1$ are, in general, different Lagrangian submanifolds of $\flatmoduli{\Sigma}$. Here, $\HFinst{Y_3}$ is the usual instanton Floer homology of a closed, compact $Y_3$, while $\HFlagr{\flatmoduli{\Sigma},L_0,L_1}$ is known as the Lagrangian (intersection) Floer homology. $\Sigma$ is a Riemann surface of genus $g$ that Heegaard splits $Y_3$.

We would now like give a physical proof of (\ref{math:atiyahfloer}).

\bigskip\noindent\textit{A Heegaard Split, DW theory, and the A-model}

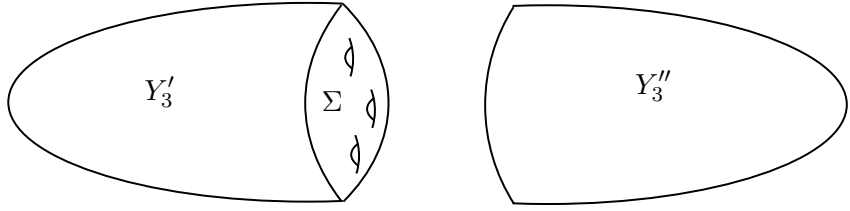
\begin{figure}[t]
	\centering	
	\tikzset{every picture/.style={line width=0.75pt}} %set default line width to 0.75pt        
	\begin{tikzpicture}[x=0.75pt,y=0.75pt,yscale=-1,xscale=1]
	%uncomment if require: \path (0,300); %set diagram left start at 0, and has height of 300
	%Shape: Arc [id:dp572457077374827] 
	\draw  [draw opacity=0] (262.55,207.2) .. controls (250.8,191.91) and (244.35,175.25) .. (244.35,157.8) .. controls (244.35,140.3) and (250.85,123.58) .. (262.67,108.25) -- (503.85,157.8) -- cycle ; \draw   (262.55,207.2) .. controls (250.8,191.91) and (244.35,175.25) .. (244.35,157.8) .. controls (244.35,140.3) and (250.85,123.58) .. (262.67,108.25) ;
	%Shape: Arc [id:dp8568192138250517] 
	\draw  [draw opacity=0] (263.59,207.52) .. controls (278.04,193.01) and (286.36,175.95) .. (286.36,157.7) .. controls (286.36,139.31) and (277.92,122.14) .. (263.28,107.56) -- (130.81,157.7) -- cycle ; \draw   (263.59,207.52) .. controls (278.04,193.01) and (286.36,175.95) .. (286.36,157.7) .. controls (286.36,139.31) and (277.92,122.14) .. (263.28,107.56) ;
	%Shape: Arc [id:dp23032787416413392] 
	\draw  [draw opacity=0] (263.28,207.16) .. controls (257.34,207.4) and (251.29,207.52) .. (245.16,207.52) .. controls (162.01,207.52) and (94.61,185.06) .. (94.61,157.36) .. controls (94.61,129.66) and (162.01,107.2) .. (245.16,107.2) .. controls (251.29,107.2) and (257.34,107.32) .. (263.28,107.56) -- (245.16,157.36) -- cycle ; \draw   (263.28,207.16) .. controls (257.34,207.4) and (251.29,207.52) .. (245.16,207.52) .. controls (162.01,207.52) and (94.61,185.06) .. (94.61,157.36) .. controls (94.61,129.66) and (162.01,107.2) .. (245.16,107.2) .. controls (251.29,107.2) and (257.34,107.32) .. (263.28,107.56) ;
	%Shape: Arc [id:dp6184571194109034] 
	\draw  [draw opacity=0] (349.65,208.6) .. controls (340.42,194.1) and (335.1,177.04) .. (335.1,158.8) .. controls (335.1,140.71) and (340.33,123.79) .. (349.42,109.36) -- (434.6,158.8) -- cycle ; \draw   (349.65,208.6) .. controls (340.42,194.1) and (335.1,177.04) .. (335.1,158.8) .. controls (335.1,140.71) and (340.33,123.79) .. (349.42,109.36) ;
	%Shape: Arc [id:dp11690074461743793] 
	\draw  [draw opacity=0] (348.9,208.36) .. controls (354.84,208.59) and (360.89,208.72) .. (367.03,208.72) .. controls (450.13,208.72) and (517.5,186.26) .. (517.5,158.56) .. controls (517.5,130.86) and (450.13,108.4) .. (367.03,108.4) .. controls (360.89,108.4) and (354.84,108.52) .. (348.9,108.76) -- (367.03,158.56) -- cycle ; \draw   (348.9,208.36) .. controls (354.84,208.59) and (360.89,208.72) .. (367.03,208.72) .. controls (450.13,208.72) and (517.5,186.26) .. (517.5,158.56) .. controls (517.5,130.86) and (450.13,108.4) .. (367.03,108.4) .. controls (360.89,108.4) and (354.84,108.52) .. (348.9,108.76) ;
	%Shape: Arc [id:dp7858917606780487] 
	\draw  [draw opacity=0] (266.59,124.96) .. controls (267.83,128.24) and (268.5,131.79) .. (268.5,135.5) .. controls (268.5,138.57) and (268.04,141.54) .. (267.18,144.33) -- (238.5,135.5) -- cycle ; \draw   (266.59,124.96) .. controls (267.83,128.24) and (268.5,131.79) .. (268.5,135.5) .. controls (268.5,138.57) and (268.04,141.54) .. (267.18,144.33) ;
	%Shape: Arc [id:dp7689612848210476] 
	\draw  [draw opacity=0] (268.18,140.43) .. controls (265.89,138.84) and (264.55,136.9) .. (264.55,134.8) .. controls (264.55,132.71) and (265.88,130.77) .. (268.15,129.18) -- (284.53,134.8) -- cycle ; \draw   (268.18,140.43) .. controls (265.89,138.84) and (264.55,136.9) .. (264.55,134.8) .. controls (264.55,132.71) and (265.88,130.77) .. (268.15,129.18) ;
	%Shape: Arc [id:dp48320022662203876] 
	\draw  [draw opacity=0] (277.59,150.96) .. controls (278.83,154.24) and (279.5,157.79) .. (279.5,161.5) .. controls (279.5,164.57) and (279.04,167.54) .. (278.18,170.33) -- (249.5,161.5) -- cycle ; \draw   (277.59,150.96) .. controls (278.83,154.24) and (279.5,157.79) .. (279.5,161.5) .. controls (279.5,164.57) and (279.04,167.54) .. (278.18,170.33) ;
	%Shape: Arc [id:dp7751036746093392] 
	\draw  [draw opacity=0] (279.18,166.43) .. controls (276.89,164.84) and (275.55,162.9) .. (275.55,160.8) .. controls (275.55,158.71) and (276.88,156.77) .. (279.15,155.18) -- (295.53,160.8) -- cycle ; \draw   (279.18,166.43) .. controls (276.89,164.84) and (275.55,162.9) .. (275.55,160.8) .. controls (275.55,158.71) and (276.88,156.77) .. (279.15,155.18) ;
	%Shape: Arc [id:dp5272090032432541] 
	\draw  [draw opacity=0] (269.59,173.96) .. controls (270.83,177.24) and (271.5,180.79) .. (271.5,184.5) .. controls (271.5,187.57) and (271.04,190.54) .. (270.18,193.33) -- (241.5,184.5) -- cycle ; \draw   (269.59,173.96) .. controls (270.83,177.24) and (271.5,180.79) .. (271.5,184.5) .. controls (271.5,187.57) and (271.04,190.54) .. (270.18,193.33) ;
	%Shape: Arc [id:dp41957159995504445] 
	\draw  [draw opacity=0] (271.18,189.43) .. controls (268.89,187.84) and (267.55,185.9) .. (267.55,183.8) .. controls (267.55,181.71) and (268.88,179.77) .. (271.15,178.18) -- (287.53,183.8) -- cycle ; \draw   (271.18,189.43) .. controls (268.89,187.84) and (267.55,185.9) .. (267.55,183.8) .. controls (267.55,181.71) and (268.88,179.77) .. (271.15,178.18) ;
	% Text Node
	\draw (171,152) node  [align=left] {$\displaystyle Y'_{3}$$ $};
	% Text Node
	\draw (420,149) node  [align=left] {$\displaystyle Y''_{3}$};
	% Text Node
	\draw (258,157) node  [align=left] {$\displaystyle \Sigma $};
	\end{tikzpicture}
	\caption{Heegaard splitting of $Y_3$ into $Y'_3$ and $Y''_3$, along $\Sigma$. Here, $\Sigma$ is a Riemann surface of genus $g$.}
	\label{fig:heegaardsplitting}
\end{figure}

To this end, let $M_4=\real^+\times Y_3$, and Heegaard split $Y_3$. This Heegaard split, $Y_3=Y'_3\cup_\Sigma Y''_3$, can be carried out along $\Sigma$, as shown in figure \ref{fig:heegaardsplitting}.

%\bigskip\noindent\textit{The Warped Metric in 3d and 4d}

Note that the compact three-manifolds $Y'_3$ and $Y''_3$ can be regarded as a fibration of a two-manifold $\Sigma$ with genus $g$, over an interval $I$ -- i.e. we may write ${{Y_3'}^{,}}\phantom{}'' = {{I'}^{,}}\phantom{}''\times_f \Sigma$. In other words, we can express the metric on ${{Y_3'}^{,}}\phantom{}''$ as a warped metric, which takes the form
\begin{equation}
\label{met:warped3d}
{ds_{{{Y_3 ' }^{,}}''}}^2=\cbrac{dx^2}^2+f(x^2)\cbrac{G_\Sigma}_{ab}dx^adx^b,
\end{equation}
where $x^2\in {{I ' }^{,}}\phantom{}''$, and $f(x^2)$ is an arbitrary $x^2$-dependent function.
Correspondingly, the metric on ${{M_4 ' }^{,}}\phantom{}'' =\real^+\times {{Y_3 ' }^{,}}\phantom{}''$ takes the form
\begin{equation}
\label{met:warped4d}
{ds_{{{M_4 ' }^{,}}''}}^2=\cbrac{dx^1}^2+\cbrac{dx^2}^2+f(x^2)\cbrac{G_\Sigma}_{ab}dx^adx^b,
\end{equation}
where $x^1$ is identified with the time-direction $\real^+$.

Let us consider DW theory on $M_4$. Then, we will have a TQFT on ${{M_4 '} ^{,}}\phantom{} ''$, whence one can carry out a Weyl rescaling on (\ref{met:warped4d}), so that the metric becomes
\begin{equation}
\label{met:warped4dweyl}
{ds_{{{M_4 ' }^{,}}''}}^2={1\over f(x^2)}\sbrac{\cbrac{dx^1}^2+\cbrac{dx^2}^2}+\cbrac{G_\Sigma}_{ab}dx^adx^b.
\end{equation}
This describes ${{M_4 ' }^{,}}\phantom{}''=(\real^+\times {{I ' }^{,}}\phantom{}'')_f\times \Sigma$. As the factor $1/f(x^2)$ leaves the topology of $\real^+\times {{I '} ^{,}}\phantom{}''$ unchanged, we may simply write $(\real^+\times {{I '} ^{,}}\phantom{}'')_f$ as $\real^+\times {{I '} ^{,}}\phantom{}''$. This means that upon shrinking $\Sigma$, we will obtain, from the DW theory on ${{M_4 '} ^{,}}\phantom{}''$, an A-model on $\real^+\times {{I '} ^{,}}\phantom{}''$ with target $\flatmoduli{\Sigma}$.

%In doing so, we may define the DW theory on each half of the four-manifold $M_4$ -- i.e. $M'_4=\real^+\times Y'_3$ and $M''_4=\real^+\times Y''_3$. The metrics on $M'_4$ and $M''_4$ both take the same form as that on $M_4$, which is the warped metric (\ref{met:warped4d}). Hence, instanton Floer homology can also be defined on each half $Y'_3$ and $Y''_3$.

%Furthermore, it was discussed in (\ref{met:warped4dweyl}) that we may carry out a Weyl rescaling to convert the fibration into a product manifold $M_4=\real^+\times I\times \Sigma$. Then, if a Heegaard split was carried out prior to the Weyl rescaling, shrinking $\Sigma$ leads to two different A-models on $\real^+\times I'$ and $\real^+\times I''$. The primes on the interval $I$ reminds the reader that a Heegaard split was carried out. Both sigma models have the same target space $\flatmoduli{\Sigma}$. So, a Heegaard split of $Y_3$ amounts to splitting the original A-model on $\real^+\times I$, into two other A-models on $\real^+\times I'$ and $\real^+\times I''$ respectively.

\bigskip\noindent\textit{Physical Proof of the Atiyah-Floer Conjecture}

For DW theory on $M_4 =\real^+\times Y_3$, it was explained in \S\ref{sec:DW} that the partition function sums classes of $\HFinst{Y_3}$. This gives us the left hand side of (\ref{math:atiyahfloer}).

Now, shrinking $\Sigma$ will leave the topological DW theory on ${{M_4 ' }^{,}}\phantom{}''$ and thus $M_4$, invariant. Therefore, one can also compute the left hand side of (\ref{math:atiyahfloer}) in terms of the A-model on $\real^+\times {{I ' }^{,}}\phantom{}''$ with target $\flatmoduli{\Sigma}$.

\begin{figure}[t]
\centering
\tikzset{every picture/.style={line width=0.75pt}} %set default line width to 0.75pt        
\begin{tikzpicture}[x=0.75pt,y=0.75pt,yscale=-1,xscale=1]
%uncomment if require: \path (0,300); %set diagram left start at 0, and has height of 300
%Shape: Rectangle [id:dp5439009320165964] 
\draw  [fill={rgb, 255:red, 155; green, 155; blue, 155 }  ,fill opacity=1 ] (213,47.6) -- (259.1,47.6) -- (259.1,200.6) -- (213,200.6) -- cycle ;
%Shape: Rectangle [id:dp7374650168978218] 
\draw  [fill={rgb, 255:red, 155; green, 155; blue, 155 }  ,fill opacity=1 ] (402,48.6) -- (448.1,48.6) -- (448.1,201.6) -- (402,201.6) -- cycle ;
%Curve Lines [id:da3888884632572316] 
\draw    (260.1,110.6) .. controls (300.1,80.6) and (362.1,145.6) .. (402.1,115.6) ;
%Shape: Circle [id:dp25561323272670866] 
\draw  [fill={rgb, 255:red, 0; green, 0; blue, 0 }  ,fill opacity=1 ] (258.11,110.46) .. controls (258.18,109.36) and (259.14,108.53) .. (260.24,108.61) .. controls (261.34,108.68) and (262.17,109.64) .. (262.09,110.74) .. controls (262.02,111.84) and (261.06,112.67) .. (259.96,112.59) .. controls (258.86,112.52) and (258.03,111.56) .. (258.11,110.46) -- cycle ;
%Shape: Circle [id:dp12780860285769569] 
\draw  [fill={rgb, 255:red, 0; green, 0; blue, 0 }  ,fill opacity=1 ] (400.11,115.46) .. controls (400.18,114.36) and (401.14,113.53) .. (402.24,113.61) .. controls (403.34,113.68) and (404.17,114.64) .. (404.09,115.74) .. controls (404.02,116.84) and (403.06,117.67) .. (401.96,117.59) .. controls (400.86,117.52) and (400.03,116.56) .. (400.11,115.46) -- cycle ;
% Text Node
\draw (236.05,124.1) node  [align=left] {$\displaystyle \widetilde{L}'$};
% Text Node
\draw (424,119) node  [align=left] {$\displaystyle L'$};
% Text Node
\draw (330,137) node  [align=left] {$\displaystyle \mathbb{R}^{+} \times I'$};
\end{tikzpicture}
\caption{An A-model open string ends on two identical Lagrangian branes of opposite orientations $L'$ and $\widetilde{L}'$. It should be noted that while the diagram depicts two non-intersecting Lagrangian branes, in general $L'$ and $\widetilde{L}'$ are allowed to intersect.}
\label{fig:twolagangianbranes}
\end{figure}
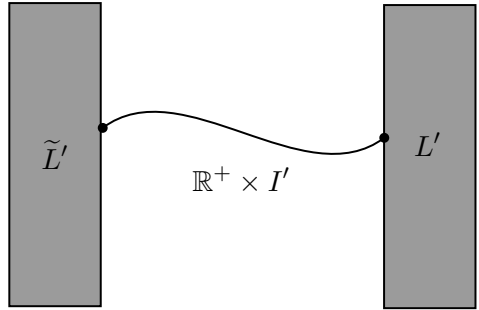

Since the A-model really describes an \textit{open} string propagating in the target space $\flatmoduli{\Sigma}$, we are required to specify its boundary conditions. The open string starts and ends on two Lagrangian branes, which are half-space filling objects -- i.e. branes with spacetime dimensions $\half\dim(\flatmoduli{\Sigma})$ -- that, as shown in figure \ref{fig:twolagangianbranes}, we shall denote by $\widetilde{L}'$ and $L'$, for the A-model on $\real^+\times I'$.  Here, tilde denotes an opposite orientation -- so, these Lagrangian branes are of opposite orientations, but are otherwise identical. 

\begin{figure}[t]
	\centering
\tikzset{every picture/.style={line width=0.75pt}} %set default line width to 0.75pt        
\begin{tikzpicture}[x=0.75pt,y=0.75pt,yscale=-1,xscale=1]
%uncomment if require: \path (0,300); %set diagram left start at 0, and has height of 300
%Shape: Rectangle [id:dp5439009320165964] 
\draw  [fill={rgb, 255:red, 155; green, 155; blue, 155 }  ,fill opacity=1 ] (152,51.6) -- (198.1,51.6) -- (198.1,204.6) -- (152,204.6) -- cycle ;
%Shape: Rectangle [id:dp7374650168978218] 
\draw  [fill={rgb, 255:red, 155; green, 155; blue, 155 }  ,fill opacity=1 ] (276,48.6) -- (322.1,48.6) -- (322.1,201.6) -- (276,201.6) -- cycle ;
%Curve Lines [id:da3888884632572316] 
\draw    (199.1,114.6) .. controls (239.1,84.6) and (235.1,144.6) .. (275.1,114.6) ;
%Shape: Circle [id:dp25561323272670866] 
\draw  [fill={rgb, 255:red, 0; green, 0; blue, 0 }  ,fill opacity=1 ] (197.11,114.46) .. controls (197.18,113.36) and (198.14,112.53) .. (199.24,112.61) .. controls (200.34,112.68) and (201.17,113.64) .. (201.09,114.74) .. controls (201.02,115.84) and (200.06,116.67) .. (198.96,116.59) .. controls (197.86,116.52) and (197.03,115.56) .. (197.11,114.46) -- cycle ;
%Shape: Circle [id:dp12780860285769569] 
\draw  [fill={rgb, 255:red, 0; green, 0; blue, 0 }  ,fill opacity=1 ] (273.11,114.46) .. controls (273.18,113.36) and (274.14,112.53) .. (275.24,112.61) .. controls (276.34,112.68) and (277.17,113.64) .. (277.09,114.74) .. controls (277.02,115.84) and (276.06,116.67) .. (274.96,116.59) .. controls (273.86,116.52) and (273.03,115.56) .. (273.11,114.46) -- cycle ;
%Shape: Rectangle [id:dp10940760184866227] 
\draw  [fill={rgb, 255:red, 155; green, 155; blue, 155 }  ,fill opacity=1 ] (323,48.6) -- (369.1,48.6) -- (369.1,201.6) -- (323,201.6) -- cycle ;
%Shape: Rectangle [id:dp17490628869555547] 
\draw  [fill={rgb, 255:red, 155; green, 155; blue, 155 }  ,fill opacity=1 ] (439.33,49.6) -- (485.43,49.6) -- (485.43,202.6) -- (439.33,202.6) -- cycle ;
%Curve Lines [id:da2535108673638209] 
\draw    (370.1,111.6) .. controls (410.1,81.6) and (399.07,143.73) .. (439.07,113.73) ;
%Shape: Circle [id:dp5682459473787369] 
\draw  [fill={rgb, 255:red, 0; green, 0; blue, 0 }  ,fill opacity=1 ] (368.11,111.46) .. controls (368.18,110.36) and (369.14,109.53) .. (370.24,109.61) .. controls (371.34,109.68) and (372.17,110.64) .. (372.09,111.74) .. controls (372.02,112.84) and (371.06,113.67) .. (369.96,113.59) .. controls (368.86,113.52) and (368.03,112.56) .. (368.11,111.46) -- cycle ;
%Shape: Circle [id:dp9618978834594385] 
\draw  [fill={rgb, 255:red, 0; green, 0; blue, 0 }  ,fill opacity=1 ] (437.07,113.59) .. controls (437.15,112.49) and (438.11,111.66) .. (439.21,111.74) .. controls (440.31,111.82) and (441.14,112.77) .. (441.06,113.88) .. controls (440.98,114.98) and (440.03,115.81) .. (438.92,115.73) .. controls (437.82,115.65) and (436.99,114.69) .. (437.07,113.59) -- cycle ;
% Text Node
\draw (175.05,128.1) node  [align=left] {$\displaystyle \widetilde{L}'$};
% Text Node
\draw (299.05,125.1) node  [align=left] {$\displaystyle L'$};
% Text Node
\draw (237,150) node  [align=left] {$\displaystyle \mathbb{R}^{+} \times I'$};
% Text Node
\draw (405,150) node  [align=left] {$\displaystyle \mathbb{R}^{+} \times I''$};
% Text Node
\draw (346.05,125.1) node  [align=left] {$\displaystyle \widetilde{L}''$};
% Text Node
\draw (464,123.6) node  [align=left] {$\displaystyle L''$};
\end{tikzpicture}
	\caption{We may glue $Y'_3$ and $Y''_3$ along their common boundary $\Sigma$ so that the original three-manifold $Y_3$ can be obtained. This is tantamount to identifying $L'=\widetilde{L}''$, so that we get a single open string starting and ending on $\widetilde{L}'$ and $L''$.}
	\label{fig:gluedlagangianbranes}
\end{figure}
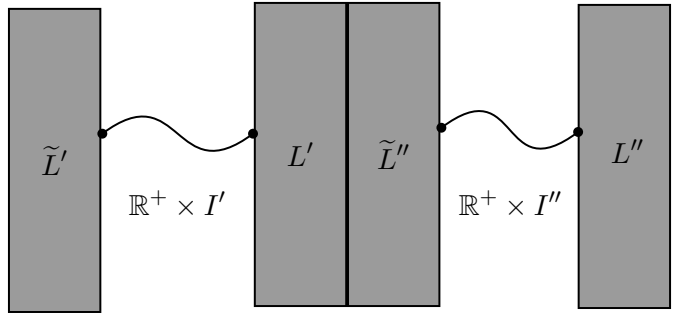

That we have two sigma models on $\Sigma$, obtained from DW theory on $M'_4$ and $M''_4$, means that we now have two different strings ending on different pairs of Lagrangian branes, $(\widetilde{L}',L')$ and $(\widetilde{L}'',L'')$. To relate to the DW theory on $M_4$ which underlies the left hand side of (\ref{math:atiyahfloer}), it is clear that we just need to `glue' them so that we get a \textit{single} sigma model with a pair of \textit{different} Lagrangian branes $\widetilde{L}'$ and ${L}''$. This is shown in figure \ref{fig:gluedlagangianbranes}, where we merge the adjacent Lagrangian branes $L'$ and $\widetilde{L}''$, so that the two strings merge into a single open string, which now extends between $\widetilde{L}'$ and $L''$ instead.

Further relabeling $L_0=\widetilde{L}'$ and $L_1=L''$ to be consistent with mathematical notation, states in the 2d sigma model can then be identified with states of the Lagrangian Floer homology $\HFlagr{\flatmoduli{\Sigma},L_0,L_1}$. This is just the right hand side of (\ref{math:atiyahfloer}).

%Here, the supersymmetric ground states of the DW theory are identified with those of the 2d A-model with target $\flatmoduli{\Sigma}$.

The physical equivalence between the partition function of the DW theory on $M_4$ which sums classes in $\HFinst{Y_3}$, and the partition function of the A-twisted sigma model on $\real^+\times I$ which sums  classes $\HFlagr{\flatmoduli{\Sigma},L_0,L_1}$, means that we have
\begin{equation}
\label{physics:atiyahfloer}
	\boxed{\HFinst{Y_3}\cong\HFlagr{\flatmoduli{\Sigma},L_0,L_1}}
\end{equation}
where $\HFinst{Y_3}$ and $\HFlagr{\flatmoduli{\Sigma},L_0,L_1}$ are graded by the instanton number and degree of maps, respectively. This completes our physical proof of the Atiyah-Floer conjecture. 

\subsection{The Monopole Analog of the Atiyah-Floer Conjecture}

There is a monopole analog of the Atiyah-Floer conjecture \cite{kutluhan2010hf} that relates the critical points of the Chern-Simons-Dirac functional with those of Lagrangian intersections $L_0\cap L_1$. It is given by
\begin{equation}
\label{math:atiyahfloermonopole}
\HFmono{q,Y_3}\cong\HFheeg{\vortexmoduli{q}{\Sigma},L_0,L_1},
\end{equation}
where $\HFmono{q,Y_3}$ is the usual monopole Floer homology, and $\HFheeg{\vortexmoduli{q}{\Sigma},L_0,L_1}$ is known as the Heegaard Floer homology -- which is the monopole analog of the Lagrangian Floer homology described earlier. 

We would now like to provide a physical proof of (\ref{math:atiyahfloermonopole}).

\bigskip\noindent\textit{Physical Proof of the Monopole Atiyah-Floer Conjecture}

Let us now consider SW theory on $M_4=\real^+\times Y_3$. Since, as explained in \S\ref{sec:SW}, partition function sums classes of $\HFmono{q,Y_3}$, the left hand side of (\ref{math:atiyahfloermonopole}) is obtained.

Next, we will Heegaard split $Y_3=Y'_3\cup_\Sigma Y''_3$, and further take the Weyl rescaled warped metric in (\ref{met:warped4dweyl}), so that $\Sigma$ can be trivially shrunken away. In doing so, an A-model on $\real^+\times {{I'}^{,}}\phantom{}''$ with target $\vortexmoduli{q}{\Sigma}$ is obtained. 

This A-model describes an \textit{open} string propagating through $\vortexmoduli{q}{\Sigma}$, which is taken to start and end on two Lagrangian branes of opposite orientations, that are denoted by ${\widetilde{L}'}\phantom{}^,\phantom{}''$ and ${{L'}^{,}}\phantom{}''$ (see figure \ref{fig:twolagangianbranes}). To relate to the SW theory on $M_4$ which underlies the left hand side of (\ref{math:atiyahfloermonopole}), it is clear that we just need to `glue' them so that we get a \textit{single} sigma model with a pair of \textit{different} Lagrangian branes $\widetilde{L}'=L_0$ and ${L}''=L_1$ (see figure \ref{fig:gluedlagangianbranes}).

States of the A-model are then identified with classes of the Heegaard Floer homology\\ $\HFheeg{\vortexmoduli{q}{\Sigma},L_0,L_1}$. This gives the right hand side of (\ref{math:atiyahfloermonopole}).

Since the partition function of SW theory on $M_4=\real^+\times Y_3$, which sums classes in $\HFmono{q,Y_3}$, can be identified with the partition function of the A-model on $\real^+\times I$, which sums classes in \\$\HFheeg{\vortexmoduli{q}{\Sigma},L_0,L_1}$, we have
\begin{equation}
\boxed{\HFmono{q,Y_3}\cong\HFheeg{\vortexmoduli{q}{\Sigma},L_0,L_1}}
\end{equation}
where $\HFmono{Y_3}$ and $\HFheeg{\vortexmoduli{q}{\Sigma},L_0,L_1}$ are graded by the number of solutions of the SW equations and degree of maps, respectively. This completes our physical proof of the monopole Atiyah-Floer conjecture.

\subsection{Mu\~noz's Theorem: Relating Instanton Floer to Quantum Cohomology\label{sec:munoz}}

Mu\~noz's theorem \cite{munoz1999ring} is the statement that 
\begin{equation}
\label{math:munoz}
	\Qch{\flatmoduli{\Sigma}}\cong\cHFsymp{\flatmoduli{\Sigma}}\cong\cHFinst{\Sigma\times S^1},
\end{equation}
where $\Qch{\flatmoduli{\Sigma}}$ is the quantum cohomology (ring) of $\flatmoduli{\Sigma}$, while $\text{HF}^*$'s are the corresponding Floer \textit{cohomologies}. Here, $\cHFsymp{\flatmoduli{\Sigma}}$ is the symplectic Floer cohomology, which we distinguish from the Lagrangian and Heegaard Floer cohomologies.

We would now like to give a physical proof of (\ref{math:munoz}).
%The second equality is merely the Poincar\'e dual of the Atiyah-Floer conjecture, which we discussed in the previous subsection. 

\bigskip\noindent\textit{Physical Proof of Mu\~noz's Theorem}

Let us now consider DW theory on $M_4=\Sigma\times S^1\times \real^+$. Shrinking $\Sigma$ away, an A-model on $S^1\times \real^+$ with target $\flatmoduli{\Sigma}$ is obtained. This describes a \textit{closed} string propagating through $\flatmoduli{\Sigma}$, starting from time $t=0$.
It is known from \cite{sadov1995equivalence} that for \textit{any} \textit{closed} topological A-model with target $T$, there is an isomorphism between the \textit{quantum cohomology} $\Qch{T}$, and symplectic Floer cohomology $\cHFsymp{T}$. This tells us that our A-model possesses the isomorphism $\Qch{\flatmoduli{\Sigma}} \cong \cHFsymp{\flatmoduli{\Sigma}}$, which is just the first equality of (\ref{math:munoz}).

Since $\p M_4=\Sigma\times S^1$, the partition function of DW theory on $M_4$ will sum classes in the instanton Floer cohomology $\cHFinst{\Sigma\times S^1}$. Because the partition function of DW theory is equivalent to the partition function of the A-model on $S^1\times\real^+$ which sums classes in $\cHFsymp{\flatmoduli{\Sigma}}$, we can write
\begin{equation}
\label{math:munozboxed}
\boxed{\Qch{\flatmoduli{\Sigma}}\cong\cHFsymp{\flatmoduli{\Sigma}}\cong\cHFinst{\Sigma\times S^1}}
\end{equation}
where $\cHFsymp{\flatmoduli{\Sigma}}$ and $\cHFinst{\Sigma\times S^1}$ are graded by the degree of maps and instanton number, respectively. This furnishes a physical proof of Mu\~noz's theorem.

\subsection{Monopole Analog of Mu\~noz's Theorem\label{sec:munozmonopole}}

It was suggested in \cite{munoz2005seiberg} that there is an isomorphism between $\HFmono{q,Y_3}$ and $\vortexmoduli{q}{\Sigma}$. We will now show how this isomorphism can be obtained physically.

To that end, we will now consider SW theory on $M_4=\Sigma\times S^1\times \real^+$. Shrinking $\Sigma$, we obtain an A-model on $S^1\times \real^+$ with target $\vortexmoduli{q}{\Sigma}$. This describes a \textit{closed} string propagating through $\vortexmoduli{q}{\Sigma}$, starting from time $t=0$. %The $\CQ$-cohomology of this A-model then corresponds to the quantum cohomology $\Qch{\vortexmoduli{q}{\Sigma}}$.

Since the partition function of SW theory on $M_4$ which sums classes in $\HFmono{q,\Sigma\times S^1}$, equals the partition function of the  A-model on $S^1\times \real^+$ which sums classes in $\cHFsymp{\vortexmoduli{q}{\Sigma}}$, we can identify $\HFmono{q,\Sigma\times S^1}$ with $\cHFsymp{\vortexmoduli{q}{\Sigma}}$. 

Furthermore, note that from \cite{sadov1995equivalence},  $\Qch{\vortexmoduli{q}{\Sigma}}\cong\cHFsymp{\vortexmoduli{q}{\Sigma}}$. Altogether, these relations can be written as
\begin{equation}
\label{math:munozmonopole}
	\boxed{\Qch{\vortexmoduli{q}{\Sigma}}\cong\cHFsymp{\vortexmoduli{q}{\Sigma}}\cong\cHFmono{q,\Sigma\times S^1}}
\end{equation}
where $\cHFsymp{\vortexmoduli{q}{\Sigma}}$ and $\cHFmono{q,\Sigma\times S^1}$ are graded by the degree of maps and number of solutions of the SW equations, respectively. This furnishes a {\it mathematically novel}, monopole version of Mu\~noz's theorem (\ref{math:munoz}).

\subsection{Higher Rank Generalizations}

We can also generalize our approach hitherto to DW and SW theory defined with higher rank gauge groups $G$.% We can also generalize our approach hitherto to SW theory with $U(1)^R$ gauge groups, where $R > 1$. 

%Let us briefly outline how this change will be effected on the relevant observables -- i.e. the $\CQ$-cohomologies of DW and SW theories, and the corresponding $\CQ$-cohomologies on the ensuing A-models.

\bigskip\noindent\textit{Higher Rank Atiyah-Floer Conjecture and Mu\~noz's Theorem}

Let us consider DW theory with gauge group $G$. All arguments about the relevant Floer homologies hold, since we may simply define the gauge group to be $G$ instead of $SU(2)$, and the rest of the analysis remains the same.  

This suggests that the Atiyah-Floer conjecture can be generalized to $G$, whereby we obtain an isomorphism between instanton Floer homology and Lagrangian Floer homology for $G$. 
Likewise, we should also be able to generalize Mu\~noz's theorem to $G$, by starting with higher rank DW theory, whilst noting that Sadov's results in \cite{sadov1995equivalence} are valid for any $G$. %$\cHFsymp{\flatmoduli{\Sigma}}\cong\Qch{\flatmoduli{\Sigma}}$, which is derived in our case by shrinking $\Sigma$ to obtain an A-model on $S^1\times\real^+$ with target $\flatmoduli{\Sigma}$. The last isomorphism can be deduced by taking $\real^+$ to be the time direction, so that $SU(N)$ instanton Floer homology can be defined on $Y_3=\Sigma\times S^1$.

\bigskip\noindent\textit{Higher Rank Monopole Atiyah-Floer Conjecture and Mu\~noz's Theorem}

%\bigskip\noindent\textit{Nonabelian Seiberg-Witten Theory}

We can also consider a nonabelian version of the SW theory, for which the BPS solutions lead to nonabelian monopoles \cite{labastida2005topological}. Then, if the four-manifold is taken to be $M_4=\Sigma\times C$, upon shrinking $C$, we should also obtain \textit{nonabelian} vortices on $C$. 

We will be able to show this explicitly if we can write a modified version of the nonabelian field strength $\CF$, in a manner analogous to (\ref{eqn:modifiedF}). Then, shrinking $C$ will force the modified gauge connections to be flat on $C$ -- i.e. we need to set $\CF_{ab}=0$ to ensure that the action remains finite. This condition should correspond to the nonabelian version of the vortex equations on $C$. The solutions to these equations will span the moduli space of nonabelian vortices.
Then, the kinetic term $\CF\wedge *\CF$ of nonabelian SW theory will descend to the action of a sigma model on $\Sigma$ with target moduli space of nonabelian vortices. We can then identify states of the A-model with states of nonabelian SW theory, to obtain the relevant mathematical identities.
%Then, it follows that the topological term $\CF\wedge \CF$ descends to the symplectic form as described in \S\ref{sec:SWadiabatic}.

%Attempts to obtain the nonabelian version of the monopole Floer homology by mathematicians such as Kronheimer and Mwroka \cite{doob} have been unsuccessful so far. 

For nonabelian SW theory on $M_4=\real^+\times Y_3$, there should also exist a nonabelian monopole Floer homology on $Y_3$ -- i.e. its classes will be identified with critical points of the nonabelian version of the Chern-Simons-Dirac functional  on the space of nonabelian connections on $Y_3$. Thus, it is also possible to obtain a nonabelian version of the monopole Atiyah-Floer conjecture and the monopole analog of Mu\~noz's theorem.

%The upshot is that any \susy{2} theory, with or without matter, should descend to an A-model on a Riemann surface with target moduli space of \textit{nonabelian} vortices. Perhaps this analysis can also be extended to quiver theory, which contains multiple matter hypermultiplets, each transforming in different representations in general.

\section{Relating Instanton Floer Homology to Affine Algebras via 4d \susy{2} TQFT\label{sec:5}}

\subsection{Relating Instanton Floer Homology of $\Sigma \times S^1$ to Affine Algebras\label{sec:DWloop}}

Let us now consider the case in which DW theory is defined on $M_4=\Sigma\times D$, where $D$ is a disk. Let us also take the gauge group to be $G=SU(2)$. Here, as in \cite{popov2015loop}, we assume the flat connections on $D$ to be irreducible.%, despite it being a genus $0$ Riemann surface. %There are some properties of this particular choice of $M_4$ which are interesting to us. 

%Firstly, the boundary \cite{munoz1999ring} of such a product manifold is simply $Y_3=\Sigma\times S^1$, for which the remaining direction can be taken to be the time direction. An equivalent picture involves the fact that $D$ is also topologically equivalent to a semi-infinite cylinder $S^1\times \real^+$ -- i.e. $D\cong S^1\times \real^+$. In this case, $\real^+$ can be taken to be the time direction, so that we can define instanton Floer homology on the spatial three-manifold $\Sigma\times S^1$.
	
%Also interesting is the appearance of the based loop group $\Omega G$ when $D$ is shrunken away. More precisely, we have the isomorphism $\flatmoduli{D}\cong\Omega G$ upon choosing appropriate boundary conditions, so the target space of the corresponding sigma model on $\Sigma$ is now the based loop group.
	
\bigskip\noindent{\textit{Instanton Floer Homology of $\Sigma  \times S^1$}}

%The boundary \cite{munoz1999ring} of such a product manifold is simply $Y_3=\Sigma\times S^1$, for which the remaining direction can be taken to be the time direction. Then, the ensuing boundary conditions on $Y_3$ can be identified with classes of the instanton Floer homology $\HFinst{\Sigma\times S^1}$.

Note that $D$ is also topologically equivalent to a semi-infinite cylinder, i.e. $D\cong S^1\times \real^+$, so with regard to DW theory, the four-manifold can also be taken as $M_4=\Sigma\times S^1\times \real^+$. In this case, $\real^+$ can be regarded as the time direction, and we can define instanton Floer homology on the spatial three-manifold $\Sigma\times S^1$. Hence, the partition function of DW theory sums classes in $\HFinst{\Sigma\times S^1}$.
	
%Since $D\cong S^1\times \real^+$, the four-manifold can also be written in the form $M_4=\Sigma\times S^1\times \real^+$, the problem can be analyzed in the same manner as it was discussed in \S\ref{sec:instantonfloer}. Taking $\real^+$ to be the time direction, the instanton Floer homology can then be defined on $Y_3=\Sigma\times S^1$, on which a Chern-Simons functional can be obtained upon integrating over $Y_3$. 
%
%Like with before, we can identify its critical points, which turn out to be flat connections on $Y_3$. Hence, classes of $\HFinst{\Sigma\times S^1}$ can be identified with classes of $\qcharge$-cohomology on DW theory.

%There is nothing exceptional about the instanton Floer homology of DW theory on such a $M_4$. However, it becomes interesting when we try to shrink either Riemann surfaces.
	
\bigskip\noindent{\textit{The Adiabatic Limit and a 2d A-Model on $\Sigma$ with Target Based Loop Group}}
	
Like (\ref{met:blockdiagonalgen}), the metric of the four-manifold may be written in a block diagonal form
\begin{equation}
	ds^2=\cbrac{G_{\Sigma}}_{ab}dx^adx^b+\cbrac{G_{D}}_{AB}dx^Adx^B.
\end{equation}
Hence, the rest of the discussion in \S\ref{sec:DWadiabatic} follows, and we obtain an A-twisted sigma model with target $\flatmoduli{\Sigma}$ or $\flatmoduli{D}$, depending on which Riemann surface we take to be small.
	
\bigskip\noindent\textit{$\Omega G$ Sigma Models}
	
Let us consider the case in which we shrink the disk $D$, whence we will have an A-twisted sigma model with target $\flatmoduli{D}$. 

Note that we need to impose boundary conditions on $\partial D=S^1$. In particular, imposing the generic trivialization condition on the boundary of $D$ allows us to identify $\flatmoduli{D}$ with the \textit{based loop group} $\Omega G$ \cite{bershadsky1995topological,popov2015loop}, which is the group of contractible based loops into $G$ \cite{salamon1998notes}, i.e.
\begin{equation}
	\Omega G=\ccbrac{\gamma:S^1\to G|\gamma(1)=\id},
\end{equation}
where elements of $\Omega G$ are $G$-valued functions with $S^1$ parameters. 

Thus, we have an A-model on $\Sigma$ with target $\Omega G$. 
	
%We now have an isomorphism between the moduli space of $k$ instantons and the moduli space of degree $k$ holomorphic maps $X:\Sigma\to\Omega G$. In a more compact form, we have $\instantonmoduli{k}{\Sigma\times D}\cong \mapsmoduli{k}{\Sigma\to\Omega G}$.
	
\bigskip\noindent\textit{Instanton Floer Homology of $\Sigma \times S^1$ and Affine Algebra}
	
%Next, we expand the fields on the sigma model in terms of generators of the gauge group $G$ -- i.e. write $X^I=X^{\tilde{a}I}T^{\tilde{a}}$ and ${X}^{\ov I}={X}^{{\tilde{a}}\ov I}T^{\tilde{a}}$, where $T^{\tilde{a}}$ are generators of the gauge group $G$. In doing so, the action (\ref{action:DWsigma}) can be written as

It was shown in \cite{ashwinkumar2018little} that the sigma model exhibits affine symmetry on the worldsheet $\Sigma$, which is generated by an affine Lie algebra $\mathfrak{g}_{\text{aff}}$. In addition, A-model states form modules of $\mathfrak{g}_{\text{aff}}$. Thus, since the partition function of the A-model which sums these states, can be identified with the partition function of DW theory which sums classes in $\HFinst{\Sigma\times S^1}$, we have the {\it mathematically novel} isomorphism 
\begin{equation}
\label{math:affine-iso-HFi}
\boxed{ \HFinst{\Sigma\times S^1}\cong \mathfrak{G}_{\text{mod}}(\Sigma)}
\end{equation}
where $\mathfrak{G}_{\text{mod}}(\Sigma)$ is the space of $\frak g_{\textrm {aff}}$-modules on $\Sigma$. The $\integer_8$ grading on the LHS by instanton solutions \cite{floer1988instanton} corresponds to the $\integer_8$ grading on the RHS by energy level. Let us now elaborate on why this is the case.

\bigskip\noindent\textit{Relevant Gradings}

On the LHS of \eqref{math:affine-iso-HFi}, it is known that gauge equivalence leads to a $\integer_8$ grading between the sets of critical points of the Chern-Simons functional which plays the role of the Morse function \cite{floer1988instanton}. In other words, the relative Morse index can only take 8 distinct (gauge-\textit{inequivalent}) values, which implies that there are only 8 distinct sets of critical points. In turn, this means that the ground states of DW theory on $M_4=\Sigma\times S^1\times \real^+$ are $\integer_8$-graded, and that there are 8 distinct sets of instanton solutions. 

Furthermore, since DW ground states are $\integer_8$-graded, A-model states must likewise be $\integer_8$-graded. Thus, $\mathfrak{G}_{\text{mod}}(\Sigma)$ is also $\integer_8$-graded. Let us now explain this.

As established in \S\ref{sec:DWadiabatic}, upon shrinking $D\cong S^1\times \real^+$, instanton solutions descend to pullback solutions (of the symplectic form on $\flatmoduli{D}\cong\Omega G$). Also, $L_0$ can be interpreted as a translation operator on $\Sigma$. When an appropriate normalization is chosen, $L_0$ determines the number of times $\Sigma$ can be wrapped around $\Omega G$ which is just the degree $k$ of maps $X:\Sigma\to\Omega G$. As $\mathfrak{G}_{\text{mod}}(\Sigma)$ is graded by $L_0$, and the degree of the aforementioned map which is associated with the  pullback solutions, like the instanton solutions, are only distinct mod $\integer_8$, it will mean that $\mathfrak{G}_{\text{mod}}(\Sigma)$ is $\integer_8$-graded. 
%Since eigenstates of $L_0$ can be identified with A-model states, these states are in turn identified with classes of $\HFinst{\Sigma\times S^1}$. This means that 

Hence, the grading on the LHS of \eqref{math:affine-iso-HFi} is identified with the grading of the RHS of \eqref{math:affine-iso-HFi} in the manner described above, as it should.  

\subsection{Relating Instanton Floer Homology of Seifert Manifolds to Affine Algebras}

\bigskip\noindent\textit{Nontrivial Seifert Manifolds}

Up to now, we have considered a four-manifold of the form $M_4=\Sigma\times D \cong \Sigma \times S^1 \times \mathbb R_+$, which has  boundary $Y_3=\Sigma\times S^1$. This three-manifold is really the trivially-fibered Seifert manifold $M_{g,0}$, where the subscript `$g,0$' refers to a  $\Sigma$ of genus $g$ with an $S^1$-bundle that has first Chern class equal to 0. %We will describe nontrivial Seifert manifolds in what follows.
Nonetheless, the discussion in \S\ref{sec:DWloop} can be generalized to a four-manifold with a nontrivial Seifert manifold boundary. To this end, let us consider a nontrivial disk fibration over $\Sigma$, i.e. $M_4=\Sigma\times_f D \cong \Sigma \times_f S^1 \times \mathbb R_+$, for which boundary $\Sigma \times_f S^1$ is therefore a nontrivial Seifert manifold $M_{g,p}$. 

\bigskip\noindent\textit{Instanton Floer Homology of $M_{g,p}$ and Affine Algebra}

We would now like to ascertain $\HFinst{M_{g,p}}$. We can start from the trivially-fibered case of $M_4=\Sigma\times D \cong M_{g,0} \times \mathbb R_+$, and make use of the \textit{fibering operator} \cite{blau2006chern}, $\CP$, which shifts the Chern number  $p_0\to p_0+1$. 

Specifically, the DW partition function on $M_{g,p} \times \mathbb R_+$ may be written as~\footnote
{To justify this, first, note that the metric of $M_4= M_{g,1}\times \real^+$ can be written as a sum  $G_{M_{g,1}\times \real^+}=G_{M_{g,0}\times \real^+}+ G_{\textrm{deformed}}$, where $G_{\textrm{deformed}}$ contains the remaining terms of the metric that characterize the nontriviality of the fibration. This means that the action (\ref{S}) over $M_4=M_{g,1}\times \real^+$ can be written as a sum of an action over $M_{g,0}\times \real^+$ and a $\CQ$-exact deformation term defined over $G_{\textrm{deformed}}$, where we can ignore the latter since it will not contribute to the path integral (as it is $\CQ$-exact). 

Second, note that the topological term $F\wedge F$ of the underlying DW theory takes the form of a Chern-Simons functional on the boundary $M_{g,1}$. This can also be written as a sum of the Chern-Simons functional on $M_{g,0}$ and a deformation term. It is this extra term in the action that can be interpreted as (the phase of the exponential that defines) a fibering operator $\CP$ in the path integral \cite{blau2006chern}.

Altogether, this means we can write $\abrac{1}_{M_{g,1}} = \abrac{\CP}_{M_{g,0}}$. It is now clear that we can also insert $p$ copies of $\CP$ and write $\abrac{1}_{M_{g,p}} = \abrac{\CP \dots \CP}_{M_{g,0}}$, as claimed.    

} 
\begin{equation}
\label{eqn:fiberingop}
\vev{1}_{M_{g,p}} =\abrac{\CP\cdots\CP}_{M_{g,0}},
\end{equation}
where we have inserted $p$ copies of $\CP$. In other words, starting from $\abrac{1}_{M_{g,0}}$, which sums classes in the instanton Floer homology $\HFinst{\Sigma\times S^1}$, we can insert $p$ copies of the fibering operator $\CP$ as shown in (\ref{eqn:fiberingop}) to obtain $\abrac{1}_{M_{g,p}}$, which sums classes in the instanton Floer homology $\HFinst{M_{g,p}}$.

Via \eqref{ob:DWfloer}, we see that \eqref{eqn:fiberingop} means that a sum over classes in $\HFinst{M_{g,p}}$ must be given by a sum over classes in $\HFinst{M_{g,0}}$ which {\it {each}} has been acted upon by $\CP \dots \CP$. In turn, from (\ref{math:affine-iso-HFi}), we have the {\it mathematically novel} isomorphism 
\begin{equation}
\label{math:affine-iso-HFi-p}
\boxed{ \HFinst{M_{g,p}}\cong \mathfrak{G}_{\text{mod},p}(\Sigma) }
\end{equation}
where $p$ on the right hand side denotes that each basis component of the original  space $\mathfrak{G}_{\text{mod}}(\Sigma)$ of $\frak g_{\textrm {aff}}$-modules on $\Sigma$ has been acted upon $p$ times by a suitable representation of $\cal P$. The grading on the LHS by instanton solutions corresponds to the grading on the RHS by energy level, in a manner analogous to that in \eqref{math:affine-iso-HFi}.

%Hence, we have obtained a relationship between instanton Floer homology of a general Seifert manifold with affine algebra using physical arguments. This was achieved via the trivial case with the insertion of $p$ copies of the fibering operator along the disk boundary $S^1$.%, which is equivalent to inserting a single $p$-fibering operator $\CP_p$.

%\bigskip\noindent\textit{The Solid Torus}
%
%Let us briefly look at a special case of the trivial Seifert manifold, for which $\Sigma=D$. The corresponding four-manifold is then $M_4=D\times S^1\times\real^+\cong D\times D$, and on its boundary we have the solid torus $Y_3=D\times S^1$. We can then shrink either disk $D$ -- take one of them to be much larger than the other -- to obtain identical 2d sigma models on the large disk with target $\Omega G$.
%
%This sigma model possesses affine symmetry as discussed previously. We may also define instanton Floer homology on the solid torus $D\times S^1$. Furthermore, symplectic Floer homology can also be defined on $\Omega G$, since we may interpret the worldsheet $D\cong\real^+$ as that of a freely propagating closed string. Altogether, this means that there exists an isomorphism between modules of $\su{N}_{\text{aff}}$, and classes of instanton and symplectic Floer homologies.

\subsection{Relating Quantum Cohomology to Affine Algebras}%: A Sanity Check}

Let us revisit DW theory on $M_4=\Sigma\times D\cong\Sigma\times S^1\times\real^+$, in which the DW partition function sums classes of $\cHFinst{\Sigma\times S^1}$ (or $\HFinst{\Sigma\times S^1}$).

It is straightforward to see, from \eqref{math:munoz} and \eqref{math:affine-iso-HFi}, that there is yet another \textit{mathematically novel} identity of the form
\begin{equation}
\label{math:symp-aff}
\boxed{\Qch{\flatmoduli{\Sigma}}\cong \mathfrak{G}_{\text{mod}}(\Sigma)}
\end{equation}
where the $\integer_8$ grading of cohomology classes \cite{munoz1999ring} on the LHS corresponds to the $\integer_8$ grading on the RHS by energy level.

\bigskip\noindent\textit{Mathematical Consistency}

In hindsight, this result is not surprising. As mentioned briefly in \S\ref{sec:DWadiabatic}, for every \textit{4d instanton} of charge $k$ on the underlying four-manifold $M_4=\Sigma\times D\cong\Sigma\times S^1\times\real^+$ in DW theory, there is a corresponding \textit{2d holomorphic map} $X:S^1\times \real^+\to\flatmoduli{\Sigma}$ of degree $k$ \cite{popov2015loop,dostoglou1994self} in the A-model on $D\cong S^1\times\real^+$. This means that there is an isomorphism $\instantonmoduli{k}{M_4}\cong\mapsmoduli{k}{S^1\times \real^+\to\flatmoduli{\Sigma}}$.

It is known that for each instanton solution corresponding to $k$, there is an action of $\mathfrak{g}_{\text{aff}}$ on homology cycles of $\instantonmoduli{k}{M_4}$ (cf. \cite{luo2017four}). Since $\instantonmoduli{k}{M_4}\cong\mapsmoduli{k}{S^1\times \real^+\to\flatmoduli{\Sigma}}$, it must also be true that there is an action of $\mathfrak{g}_{\text{aff}}$ on homology cycles of $\mapsmoduli{k}{S^1\times \real^+\to\flatmoduli{\Sigma}}$.

Moreover, by Poincar\'e duality, homology cycles of $\mapsmoduli{k}{S^1\times \real^+\to\flatmoduli{\Sigma}}$, must correspond to differential forms on $\mapsmoduli{k}{S^1\times \real^+\to\flatmoduli{\Sigma}}$ which generate the quantum cohomology of $\flatmoduli{\Sigma}$. Hence, there must be an action of $\mathfrak{g}_{\text{aff}}$ on $\Qch{\flatmoduli{\Sigma}}$ -- i.e. classes of $\Qch{\flatmoduli{\Sigma}}$ can be viewed as modules of $\mathfrak{g}_{\text{aff}}$.

This gives an independent verification of the result in \eqref{math:symp-aff}, which was deduced using purely physical arguments.

\section{A Derivation of the Verlinde Formula via 4d \susy{2} TQFT\label{sec:6}}

\subsection{SQM on the Moduli Space of Flat Connections\label{sec:SQM}}% and Affine Algebras on $\Sigma$\label{sec:SQM}}

Looking at the sigma model on $D\cong \real^+\times S^1$ with target $\flatmoduli{\Sigma}$, we may further shrink $S^1$ so that we get a 1d sigma model -- this really becomes supersymmetric quantum mechanics (SQM) with $\real^+$ being identified with the temporal direction. To see this, we shall carry out the usual dimensional reduction on $S^1$, so that the action in (\ref{action:DWsigma}) becomes
\begin{equation}
\label{action:SQM}
\begin{aligned}
S_{1d}&=S_{\text{SQM}}\\
&={1\over e^2}\int d\tau\ 
\cbrac{
	\dot{X}^I\dot{ X}_I
	+G^{\text{flat}}_{I\ov J}\cbrac{{{\rho}_{z}}^{\ov J}\nabla_\tau\chi^I
	+{\rho_{\ov z}}^I\nabla_\tau{\chi}^{\ov J}}
}
	-R_{I\ov JK\ov L}{{\rho}_{\ov z}}^{I}{\rho_{z}}^{\ov J}{\chi}^K\chi^{\ov L} + \cdots,
\end{aligned}
\end{equation}
where we have relabeled $\tau=x^1$, having chosen it as the temporal direction, and frozen out the $x^2$ dependence. A dotted field denotes a derivative over $\tau$, and the ellipses here correspond to topological terms which we can ignore.

Notice that the geometry of the target space $\flatmoduli{\Sigma}$ is dependent on the geometry of $\Sigma$, since its metric was defined with respect to $\Sigma$ (recall the definition of the moduli space metric in (\ref{eqn:kahlermetric})). Since the geometry of $\Sigma$ is inconsequential in a TQFT, we may scale it such that $\flatmoduli{\Sigma}$ becomes flat -- i.e. we may set $G^{\text{flat}}_{I\ov J}$ to be the flat metric, and hence also set $R_{I\ov J K\ov L}=0$. Crucially, since the Christoffel symbols $\Gamma^I_{JK}$ are defined by derivatives of the metric, they can also be taken to be zero.

In doing so, (\ref{action:SQM}) can be written as a free action. We shall take $N_{\text{I}}=0$ so that there are no fermionic zero modes. The only fermionic contributions come from the free fermionic action, which we can integrate out in the path integral to just give a constant. 

Let us analyze the bosonic part, which is written as
%Next, we may integrate out the fermionic fluctuations (non-zero modes), in the same manner as (\ref{eqn:nonzeromodes}) so that they become a mere factor $\prod\Varepsilon$ that can be brought outside of the integral in (\ref{action:SQM}). In doing so, the action may be rewritten as
\begin{equation}
\label{action:QM}
\boxed{\begin{aligned}
S_{\text{QM}}&={2\over e^2}\int d\tau\ 
	\half\dot{X}^I\dot{X}_I\\
	&={1\over \hbar}\int d\tau\ L_{\text{QM}}
\end{aligned}}
\end{equation}
where the Planck's constant of the QM model is identified as ${1\over\hbar}={2\over e^2}$. 
Then we may define the conjugate momenta by $P^I={\partial L_{\text{QM}}/ \partial \dot{X}_I}={\dot{{ X}}}^I$ which, by definition, satisfy the Poisson brackets, which are written as
\begin{equation}
\begin{aligned}
&\{X^I,P^J\}=\hbar\delta^{IJ},
\end{aligned}
\end{equation}
and zero otherwise. Canonical quantization amounts to replacing the canonical coordinates $(X,P)$ with operators $(\hat{X},\hat{P})$ so that the Poisson brackets become commutator brackets instead. These relations take the form
\begin{equation}
\label{eqn:commutatorspi}
\boxed{[\hat{X}^I,\hat{P}^J]=\hbar\delta^{IJ}}
\end{equation}

Since $X^I$ are defined to be coordinates on the target $\flatmoduli{\Sigma}$, the operators $\hat{X}^I$ describe \textit{quantized} coordinates on $\flatmoduli{\Sigma}$. In doing so, we have quantized $\flatmoduli{\Sigma}$, which can be taken to be of finite volume \cite{witten1989quantum}.

\subsection{Deriving the Verlinde Formula}

Generally, the Verlinde formula computes the dimension of the space of conformal blocks, which can be defined in any 2d conformal field theory (CFT) with affine symmetry. There are several definitions of the Verlinde formula, because there are various ways of computing the dimensions of the space of conformal blocks. One definition is by Faltings \cite{faltings1994proof,beauville1994conformal}, in which the dimension of the space of conformal blocks on $\Sigma$ is the same as the number of holomorphic sections of (an integer power of) the determinant line bundle over $\flatmoduli{\Sigma}$, where $\Sigma$ is a compact Riemann surface. We would now like to outline a physical proof of this mathematical statement, and hence derive the Verlinde formula.

We shall consider $M_4=\Sigma \times D$, so that we can shrink either $\Sigma$ or $D$ to obtain two different A-models. The upshot is that we can describe each side of the Verlinde formula by separately shrinking each Riemann surface. Since both sigma models descend from the same TQFT in 4d, their respective partition functions can be identified with each other. It is this identification that will allow us to derive the Verlinde formula.

\bigskip\noindent\textit{Physical Proof of the Verlinde Formula -- LHS}

To begin the physical proof of this result, let us consider the simple case in which the virtual dimension of $\instantonmoduli{k}{M_4}$ is zero -- i.e. $N_{\text{I}}=0$ -- so that the only surviving observables are the partition functions.

Let us first shrink $D$, so that an A-model on $\Sigma$ with target $\Omega G$ can be obtained. Such an A-model, as discussed in \S\ref{sec:DWloop}, exhibits affine symmetry on $\Sigma$, which is generated by $\mathfrak{g}_{\text{aff}}$ at level $\ell$. Because this 2d sigma model possesses affine symmetry, its states, $\kappa$, will be  modules of $\mathfrak{g}_{\text{aff}}$ on $\Sigma$. The partition function can then be written as
\begin{equation}
\label{ob:Amodel-pfnnooperators}
\vev{1}=\sum_{v}\kappa_v\ov{\kappa}_v,
%\abrac{0|0}
\end{equation}
where $v$ labels the energy eigenstates.

Next, note that since we are dealing with modules of $\mathfrak{g}_{\text{aff}}$ on $\Sigma$, we can also write (\ref{ob:Amodel-pfnnooperators}) in terms of conformal blocks $\mathscr{F}$ \cite{francesco2012conformal} as
\begin{equation}
\label{ob:pfnnooperators}
%\vev{1}
\sum_{v}\kappa_v\ov{\kappa}_v=\sum_{v}\mathscr{F}_v\ov{\mathscr{F}}_v.
%\abrac{0|0}
\end{equation}
%where $v$ labels the energy eigenfunctions of the theory, and $\ket{0}$ are the vacuum states. 
%(\ref{ob:Amodel-pfnnooperators}) and (\ref{ob:pfnnooperators}) amount to saying 
Thus, $\kappa_v$ can be identified with the holomorphic conformal blocks $\mathscr{F}_v$. 
% (or antiholomorphic conformal blocks $\ov{\mathscr{F}}_v$). %, which are clearly themselves $\qcharge$-cohomology states of the A-model on $\Sigma$. 
Let us denote the space of $\kappa_v$ states -- i.e. the \textit{space of zero-point conformal blocks on $\Sigma$} -- by $V_\ell (\Sigma)$.

\bigskip\noindent\textit{A Slight Excursion}

It will be useful for later, to describe the affine algebra of the A-model in greater detail. 
%The standard formula for the conserved current may be written as
%\begin{equation}
%\label{eqn:affinecurrent}
%J^{A}={\p\CL_{\sigma}\over\p (\p_A\Phi)}\delta\Phi,
%\end{equation}
%where $\Phi$ describes all fields in the A-model, the index $A$ describes worldsheet coordinates on $\Sigma$ and $\CL_\sigma$ here is the Lagrangian density of the A-model in (\ref{action:Amodelaffine}) -- not to be confused with the determinant line bundle discussed earlier. The ensuing affine algebra can be written as
This affine algebra can be written as
\begin{equation}
\label{com:affine}
[J^{\widetilde{a}m},J^{\widetilde{b}n}]=if^{\widetilde{a}\widetilde{b}}_{\widetilde{c}}J^{\widetilde{c},(m+n)}+ L m\delta^{\widetilde{a}\widetilde{b}}\delta^{(m+n),0},
\end{equation}
where indices $m,n\in\integer$ describe the order of the Laurent expansion of the conserved currents $J$, $L$ is the level, indices $\widetilde{a},\widetilde{b},\widetilde{c}$ runs over generators of the gauge group $G$, and $f^{\widetilde{a}\widetilde{b}}_{\widetilde{c}}$ are the structure constants.

The affine algebra in (\ref{com:affine}) is obtained by taking a contour integral over the operator expansion product (OPE), which can be written as
\begin{equation}
\label{eqn:OPE}
J^{\widetilde{a}}(z)J^{\widetilde{b}}(\zeta)-{if^{\wta\wtb}_{\wtc} J^{\wtc}(\zeta)\over (z-\zeta)}
={L\delta^{\wta\wtb}\over(z-\zeta)^2},
\end{equation}
where $\zeta,z\in\Sigma$. %Next, we note that operator insertions are position-independent in a TQFT such as our A-model on $\Sigma$. 
Let us carry out a topological deformation on the underlying two-manifold $\Sigma$, such that the coordinates become $\zeta\to\widetilde{\zeta}=\sqrt{\ell} \zeta$ and $z\to\widetilde{z}=\sqrt{\ell} z$. In doing so, (\ref{eqn:OPE}) becomes 
\begin{equation}
\label{eqn:OPEdeformedfully}
J^{\widetilde{a}}(\widetilde{z})J^{\widetilde{b}}(\widetilde{\zeta})-{if^{\wta\wtb}_{\wtc} J^{\wtc}(\widetilde{\zeta})\over (\widetilde{z}-\widetilde{\zeta})}
={L\delta^{\wta\wtb}\over(\widetilde{z}-\widetilde{\zeta})^2}.
\end{equation}

Since operator insertions are position-independent in a TQFT, the expressions on the left hand side of equations (\ref{eqn:OPE}) and (\ref{eqn:OPEdeformedfully}) are equivalent and can be equated. Hence, the left hand side of (\ref{eqn:OPE}) can be replaced with the left hand side of (\ref{eqn:OPEdeformedfully}), so that we obtain the relation

\begin{equation}
\label{eqn:OPEdeformed}
J^{\widetilde{a}}(\widetilde{z})J^{\widetilde{b}}(\widetilde{\zeta})-{if^{\wta\wtb}_{\wtc} J^{\wtc}(\widetilde{\zeta})\over (\widetilde{z}-\widetilde{\zeta})}={\cbrac{\ell L}\delta^{\wta\wtb}\over(\widetilde{z}-\widetilde{\zeta})^2}.
\end{equation}
Further carrying out a Laurent expansion on $J$, and taking a contour integral over $\Sigma$, (\ref{eqn:OPEdeformed}) becomes
\begin{equation}
\label{com:affinedeformed}
[J^{\widetilde{a}m},J^{\widetilde{b}n}]=if^{\widetilde{a}\widetilde{b}}_{\widetilde{c}}J^{\widetilde{c},(m+n)}+ \cbrac{\ell L} m\delta^{\widetilde{a}\widetilde{b}}\delta^{(m+n),0},
\end{equation}
which can be viewed as a ``topologically deformed" affine algebra at level $\cbrac{\ell L}$ instead. Hence, deforming the geometry of $\Sigma$ by $\cbrac{ds_\Sigma}^2\to \ell\cbrac{ds_\Sigma}^2$ amounts to tuning the level of the affine algebra by $L\to \ell L$. 

The topological deformation of $\Sigma$ can be understood to originate from a topological deformation of the \textit{four-manifold}. Meanwhile, the topological deformation of the four-manifold can alternatively be interpreted as the modification of the 4d coupling. Altogether, this means that modifying the 4d coupling by $\cbrac{1\over e^2}'\to \ell \cbrac{1\over e^2}'$ is tantamount to effecting a topological deformation $z\to\widetilde{z}$ on $\Sigma$. 

Let us set $L=1$. Henceforth, we shall take $\cbrac{1\over e^2}'$ to be the 4d coupling constant corresponding to an affine algebra at level $L=1$. Then, the 4d coupling $\ell\cbrac{1\over e^2}'$ corresponds to the affine algebra at level $\ell$.

\bigskip\noindent\textit{Physical Proof of the Verlinde Formula -- RHS}

If we were to shrink $\Sigma$ instead, we obtain another 2d sigma model, this time on $D\cong \real^+\times S^1$, and with target $\flatmoduli{\Sigma}$. By further dimensionally reducing on $S^1$ as we did in \S\ref{sec:SQM}, we may quantize $\flatmoduli{\Sigma}$, to obtain a Hilbert space over $\flatmoduli{\Sigma}$. The QM space of states can then be identified as the \textit{space of holomorphic sections} of $\CL$, raised to a power $k$ \cite{witten1989quantum,segal1988definition}, where $\CL$ is the determinant line bundle over $\flatmoduli{\Sigma}$. We shall denote this space by $H^0(\flatmoduli{\Sigma},\CL^{k})$.

The integer $k$ can be interpreted as the coupling constant of the QM model. To see that, let us define $k={1\over \hbar}={2\over e^2}$, so that the QM action (\ref{action:QM}) can be rewritten as
\begin{equation}
S_{\text{QM}}=k\int d\tau\ \dot{X}^I\dot{X}_I.
\end{equation}
Note that $k$ is now taken to be a large, positive integer, since $\hbar$ is small.

\bigskip\noindent\textit{The Verlinde Formula}

As discussed earlier, 
%there is an association between the 4d coupling constant $1\over e^2$, and the affine algebra of the A-model on $\Sigma$ with target $\Omega G$. T
tuning the value of the 4d coupling also tunes the level $\ell$ of $\mathfrak{g}_{\text{aff}}$. Meanwhile, the Planck's constant $\hbar$, which was obtained by quantization in \S\ref{sec:SQM}, also descended from the 4d coupling. This then implies that there is a connection between the level $\ell$ of the affine algebra, and the power $k$ of the determinant line bundle.

%Likewise, changing the value of $e$ also tunes the value of $\hbar$, hence this establishes an indirect relation to the level $\ell$ of the affine algebra associated with the \textit{other} sigma model. It should be noted that there is no concept of ``levels" in the QM model -- instead, the level $\ell$ can be interpreted as the coupling constant for the QM model.
%This is analogous to the canonical quantization of the Chern-Simons theory in \cite{witten1989quantum}, in which the coupling constant is identified with the Chern-Simons level $k$.

To further investigate the connection between $\ell$ and $k$, we first write the 4d coupling constant as ${1\over e^2}=\ell\cbrac{1\over {e}^2}'$. Consequently, $\ell$ then corresponds to the level of $\mathfrak{g}_{\text{aff}}$ on the A-model on $\Sigma$ with target $\Omega G$. 
On the other side, we have an A-model on $D\cong\real^+\times S^1$, where the usual QM is obtained upon further shrinking $S^1$ (See (\ref{action:QM})). Recall that the Planck's constant, which descended from the 4d coupling ${1\over e^2}$, can be expressed as $k={1\over \hbar}={2\over e^2}$. Hence, we may now write
\begin{equation}
\label{eqn:QMk}
k=2\ell\cbrac{1\over e^2}'=\ell k',
\end{equation}
where $k'=2\cbrac{1\over e^2}'$. This then shows the explicit relation between the level $\ell$ and QM coupling $k$. Further setting $k'=1$ then allows us to make the identification $k=\ell$, and the Hilbert space may then be written as $H^0(\flatmoduli{\Sigma},\CL^\ell)$.

Finally, since states of each physical theory can be identified with each other, we see that the two spaces of states -- the space of zero-point conformal blocks obtained from the A-model on $\Sigma$, and the Hilbert space obtained from the QM model -- are equivalent. Hence, we may write this relation as
\begin{equation}
\label{math:verlindeunderlying}
\boxed{V_\ell (\Sigma)\cong H^0(\flatmoduli{\Sigma},\CL^\ell)}
\end{equation}
which is Faltings's result \cite{faltings1994proof,beauville1994conformal} -- the statement that underlies the Verlinde formula. Thus, the dimension of $V_\ell$ is obtained by counting the number of holomorphic sections of $\CL^\ell$. This completes our derivation of the Verlinde formula.

\bigskip\noindent\textit{With Extra Operator Insertions}

There is also a more general result found by Pauly \cite{pauly1996espaces}, in which $n$ operators are inserted in the 2d CFT. The positions of insertions are denoted by $\vec{p}=(p_1,\cdots,p_n)$, where $p_1,\cdots,p_n\in\Sigma$. The statement is that there is an isomorphism between the space ${V_\ell}(\Sigma, \vec{p})$ of $n$-point conformal blocks on $\Sigma$, and holomorphic sections of determinant line bundles over the moduli space of parabolic vector bundles on $\Sigma$, which we denote by $\CM_{\text{para}}\cbrac{\Sigma,\vec{p}}$. The space of holomorphic sections in this case is $H^0(\CM_{\text{para}}\cbrac{\Sigma,\vec{p}},\CL^\ell)$, and so here, Pauly's result is the isomorphism
\begin{equation}
\label{math:verlindepauly}
{V_\ell}(\Sigma, \vec{p}) \cong H^0(\CM_{\text{para}}\cbrac{\Sigma,\vec{p}},\CL^\ell).
\end{equation}
We will now show the relation (\ref{math:verlindepauly}) physically.

Let us now insert $n$ \textit{scalar} operators in DW theory, and consider $N_{\text{I}}=0$ which is still valid, since the scalar operators have no zero modes -- i.e. there is no $U(1)_R$ anomaly. 
In doing so, observables of the 4d theory are no longer just partition functions, but now take the form in (\ref{CF}) -- i.e. they are $n$-point correlation functions. Starting from the 4d DW theory, we assume that the $n$ operators $\CO_1,\cdots,\CO_n$ are inserted at points \textit{orthogonal} to $D$, so that they are defined only on $\Sigma$. We further note that operator insertions at points $\vec{p}$ in a CFT on $\Sigma$ is equivalent to defining the \textit{same} CFT \textit{without insertions}, albeit on $\Sigma$ with $n$ punctures -- i.e. $\Sigma$ can be replaced with $\Sigma-\vec{p}$. We may then carry out the same shrinking procedure to obtain two 2d sigma models, on worldsheets $D\cong \real^+\times S^1$ and $\Sigma-\vec{p}$, respectively.

Let us first shrink $D$ to obtain a sigma model that possesses an affine algebra at level $\ell$, on the worldsheet $\Sigma$ with target $\Omega G$. In this part of the analysis, we shall take $\Sigma$ to be the worldsheet with $n$ insertions, rather than a Riemann surface $\Sigma$ with $n$ punctures. Consequently, the observables are correlation functions of the form
\begin{equation}
\label{ob:npointsigma}
\abrac{\prod_{r=1}^n \CO_r(z_r,{\ov z}_r)},%=\sum_v \kappa_{v;n}(z_1,\cdots,z_n)\ov{\kappa}_{v;n}({\ov z}_1,\cdots,{\ov z}_n),
\end{equation}
where $(z,\ov z)$ are the holomorphic and antiholomorphic coordinates on $\Sigma$.

Since the sigma model is topological, the positions of operator insertions $\vec{p}$ are irrelevant. This means that we can bring a pair of operators close together and make use of fusion rules to merge them into a single operator. This can be repeated until all $n$ operators coalesce into a single operator $\CO'$.

Furthermore, the position-independence of operator insertions also implies that the scalar operator $\CO'$ is a constant. (\ref{ob:npointsigma}) then becomes 
\begin{equation}
\label{ob:Amodel-1pointsigma}
\abrac{\prod_{r=1}^n \CO_r(z_r,{\ov z}_r)}=\abrac{\CO'} = \abrac{1}.
%=\abrac{\widetilde{\CO}^\dagger\widetilde{\CO}},
\end{equation}

Then, (\ref{ob:Amodel-1pointsigma}) can be written as
\begin{equation}
\label{ob:Amodel-1pointsigma-states}
%\vev{\widetilde{1}}=
%\vev{\widetilde{0}|\widetilde{0}}
%\abrac{\widetilde{\CO}'^\dagger(\ov z)\widetilde{\CO}'(z)}=
\abrac{1}=\sum_{v}\kappa_{v;n}\ov{\kappa}_{v;n},
\end{equation}
where $\kappa_{v;n}$ are A-model eigenstates. Here, $n$ has been included in the subscript of $\kappa$ to distinguish these A-model states from the case in (\ref{ob:Amodel-pfnnooperators}) with no insertions -- i.e. $\kappa_{v;0}=\kappa_v$. Since we dealing with modules of $\mathfrak{g}_{\text{aff}}$ on $\Sigma$, (\ref{ob:Amodel-1pointsigma-states}) may also be written in terms of conformal blocks \cite{francesco2012conformal} as 
\begin{equation}
%\vev{\widetilde{1}}=
%\vev{\widetilde{0}|\widetilde{0}}
\sum_{v}\kappa_{v;n}\ov{\kappa}_{v;n}=\sum_{v}\mathscr{F}_{v;n}\ov{\mathscr{F}}_{v;n}.
\end{equation}
In doing so, the $n$-point conformal blocks $\mathscr{F}_{v;n}(z_1,\cdots,z_n)$ can now be identified with eigenstates of the A-model on $\Sigma$ with target $\Omega G$. Hence, $\kappa_{v;n}$ states span the \textit{space of $n$-point conformal blocks} on $\Sigma$, $V_\ell(\Sigma,\vec{p})$. In the same way, the level $\ell$ of $\mathfrak{g}_{\text{aff}}$ can also be interpreted in terms of the 4d coupling ${1\over e^2}=\ell\cbrac{1\over e^2}'$.

Let us now take the view that $\Sigma$ with $n$ insertions is just $\Sigma-\vec{p}$, so that we can shrink $\Sigma-\vec{p}$ instead. In doing so, we obtain a sigma model on $D\cong \real^+\times S^1$ with target $\flatmoduli{\Sigma-\vec{p}}$ which can be quantized upon further dimensionally reducing on $S^1$. We further note that $\flatmoduli{\Sigma-\vec{p}}$, is the same as the moduli space of parabolic bundles on $\Sigma$, $\CM_{\text{para}}\cbrac{\Sigma,\vec{p}}$, where the parabolic structures are identified with punctures on $\Sigma$. In other words, $\flatmoduli{\Sigma-\vec{p}}\cong\CM_{\text{para}}\cbrac{\Sigma,\vec{p}}$.

Like before, $\CL$ is raised to a power $k$, whereby $k$ is identified with the coupling constant of the QM model on $\CM_{\text{para}}\cbrac{\Sigma,\vec{p}}$. The space of QM states on $\CM_{\text{para}}\cbrac{\Sigma,\vec{p}}$ can be identified with the space of holomorphic sections of $\CL^k$.

Since both spaces of states are derived from the same 4d TQFT, they must be equivalent. Furthermore, the same arguments made for the case \textit{without} insertions dictates that the level $\ell$ of $\mathfrak{g}_{\text{aff}}$ is the same as the QM coupling constant $k$ -- i.e. they both descended from the same 4d coupling constant -- and we may write $k=\ell$. We may now identify both spaces of states, to write down the relation
\begin{equation}
\label{math:verlindepauly-physics}
\boxed{{V_\ell}(\Sigma, \vec{p})\cong H^0(\CM_{\text{para}}\cbrac{\Sigma,\vec{p}},\CL^\ell)}
\end{equation}
which is just (\ref{math:verlindepauly}). Then, the dimension of ${V_\ell}$ is obtained by counting the number of holomorphic sections of $\CL^\ell$. This then completes our physical derivation of the Verlinde formula with extra operator insertions.

%\newpage
\appendix
\section{Representations of 4d Euclidean spinors\label{appendix}}
In this paper, we have used the conventions seen in \cite{labastida2005topological}. Since we are working with Euclidean spaces, the Lorentz group in 4d is just the rotation group $SO(4)\cong SU(2)_+\times SU(2)_-$, where $\pm$ denotes the independent spins of each $SU(2)$ rotation group. We shall take spinors with (un)dotted indices to transform under ($SU(2)_-$) $SU(2)_+$. In this way, fields in the vector representation of $SO(4)$ -- fields with spacetime indices $\mu,\nu=1,2,3,4$ -- can be rewritten in the bispinor representation using Clebsch-Gordan coefficients. The Clebsch-Gordan coefficients can be written as
\begin{equation}
\begin{aligned}
&\cbrac{\sigma_\mu}_{\alpha\dot{\alpha}}=\ccbrac{\sigma_1,\sigma_2,\sigma_3,i}_{\alpha\dot{\alpha}}\\
&\cbrac{\ov{\sigma}_\mu}^{\dot{\alpha}\alpha}=\ccbrac{-\sigma_1,-\sigma_2,-\sigma_3,i}^{\dot{\alpha}\alpha},
\end{aligned}
\end{equation}
where $\sigma_1,\sigma_2,\sigma_3$ are the usual Pauli matrices
\begin{equation}
\sigma_1=
\begin{pmatrix}
0&1\\
1&0
\end{pmatrix},\qquad
\sigma_2=
\begin{pmatrix}
0&-i\\
i&0
\end{pmatrix},\qquad
\sigma_3=
\begin{pmatrix}
1&0\\
0&-1
\end{pmatrix}.
\end{equation}

Explicitly, vectors $V$ can be written as
\begin{equation}
\begin{aligned}
&V^{\dot{\alpha}\alpha}=\cbrac{\sigma_\mu}^{\dot{\alpha}\alpha}V^\mu\\
&V_{\alpha\dot{\alpha}}=\cbrac{\ov\sigma_\mu}_{\alpha\dot{\alpha}}V^\mu.
\end{aligned}
\end{equation}

To write antisymmetric matrices $K$ in the bispinor representation, we first need to define the matrix
\begin{equation}
\cbrac{\ov\sigma_{\mu\nu}}_{\dot{\alpha}\dot{\beta}}=\quar\cbrac{\ov\sigma_\mu\sigma_\nu-\ov\sigma_\nu\sigma_\mu}_{\dot{\alpha}\dot{\beta}},
\end{equation}
which has components
\begin{equation}
\sbrac{\ov\sigma_{\mu\nu}}={i\over 2}
\begin{pmatrix}
	0 & -\sigma_3 & \sigma_2 & -\sigma_1\\
	\sigma_3 & 0 & -\sigma_1 & -\sigma_2\\
	-\sigma_2 & \sigma_1 & 0 &-\sigma_3\\
	\sigma_1 & \sigma_2 & \sigma_3 & 0
\end{pmatrix}.
\end{equation}
Further note that $\ov\sigma_{\mu\nu}$ is self-dual -- i.e.
\begin{equation}
\label{con:selfdualsigma}
\ov\sigma_{\mu\nu}=\half\epsilon_{\mu\nu\rho\lambda}\ov\sigma^{\rho\lambda}.
\end{equation}
We can write antisymmetric matrices as
\begin{equation}
\label{eqn:CGmatrix}
K_{\dot{\alpha}\dot{\beta}}
=\cbrac{\ov\sigma^{\mu\nu}}_{\dot{\alpha}\dot{\beta}}K_{\mu\nu}
=\cbrac{\ov\sigma_{\mu\nu}}_{\dot{\alpha}\dot{\beta}}K^+_{\mu\nu},
\end{equation}
where the second equality makes use of the self-dual condition (\ref{con:selfdualsigma}). 

We can show the second equality of (\ref{eqn:CGmatrix}), by multiplying a factor of $\cbrac{\ov\sigma^{\mu\nu}}_{\dot{\alpha}\dot{\beta}}$ to the self-dual matrix $K_{\mu\nu}^+$ so that we get
\begin{equation*}
\begin{aligned}
\cbrac{\ov\sigma^{\mu\nu}}_{\dot{\alpha}\dot{\beta}}K^+_{\mu\nu}
&={\half}\cbrac{\cbrac{\ov\sigma^{\mu\nu}}_{\dot{\alpha}\dot{\beta}}K_{\mu\nu}
+{\half}\epsilon_{\mu\nu\rho\lambda}\cbrac{\ov\sigma^{\mu\nu}}_{\dot{\alpha}\dot{\beta}}K^{\rho\lambda}}\\
&={\half}\cbrac{\cbrac{\ov\sigma^{\mu\nu}}_{\dot{\alpha}\dot{\beta}}K_{\mu\nu}
+{\half}\epsilon_{\rho\lambda\mu\nu}\cbrac{\ov\sigma^{\rho\lambda}}_{\dot{\alpha}\dot{\beta}}K^{\mu\nu}}\\
&=\cbrac{\ov\sigma^{\mu\nu}}_{\dot{\alpha}\dot{\beta}}K_{\mu\nu}.
\end{aligned}
\end{equation*}
\qed

\bibliographystyle{ieeetr}
\bibliography{bib}

\begin{thebibliography}{10}

\bibitem{witten1988topological}
E.~Witten, ``{Topological quantum field theory},'' {\em Communications in
  Mathematical Physics}, vol.~117, no.~3, pp.~353--386, 1988.

\bibitem{marcolli1999seiberg}
M.~Marcolli, {\em {Seiberg Witten Gauge Theory}}, vol.~17.
\newblock Springer, 1999.

\bibitem{bershadsky1995topological}
M.~Bershadsky, A.~Johansen, V.~Sadov, and C.~Vafa, ``{Topological reduction of
  4D {SYM} to 2D $\sigma$-models},'' {\em Nuclear Physics B}, vol.~448,
  no.~1-2, pp.~166--186, 1995.

\bibitem{witten1988topologicalsig}
E.~Witten, ``{Topological sigma models},'' {\em Communications in Mathematical
  Physics}, vol.~118, no.~3, pp.~411--449, 1988.

\bibitem{gukov2009surface}
S.~Gukov, ``Surface operators and knot homologies,'' in {\em New Trends in
  Mathematical Physics}, pp.~313--343, Springer, 2009.

\bibitem{sadov1995equivalence}
V.~Sadov, ``{On equivalence of Floer's and quantum cohomology},'' {\em
  Communications in mathematical physics}, vol.~173, no.~1, pp.~77--99, 1995.

\bibitem{ashwinkumar2018little}
M.~Ashwinkumar, J.~Cao, Y.~Luo, M.-C. Tan, and Q.~Zhao, ``{Little strings,
  quasi-topological sigma model on loop group, and toroidal Lie algebras},''
  {\em Nuclear Physics B}, vol.~928, pp.~469--498, 2018.

\bibitem{blau2006chern}
M.~Blau and G.~Thompson, ``{Chern-Simons theory on $S^1$-bundles:
  Abelianisation and q-deformed Yang-Mills theory},'' {\em Journal of High
  Energy Physics}, vol.~2006, no.~05, p.~003, 2006.

\bibitem{faltings1994proof}
G.~Faltings, ``{A proof for the Verlinde formula},'' {\em Journal of Algebraic
  Geometry}, vol.~3, no.~2, p.~347, 1994.

\bibitem{beauville1994conformal}
A.~Beauville, ``{Conformal blocks, fusion rules and the Verlinde formula},''
  {\em arXiv preprint alg-geom/9405001}, 1994.

\bibitem{pauly1996espaces}
C.~Pauly, ``{Espaces de modules de fibr{\'e}s paraboliques et blocs
  conformes},'' {\em Duke Mathematical Journal}, vol.~84, no.~1, pp.~217--235,
  1996.

\bibitem{lozano2001donaldson}
C.~Lozano and M.~Mari{\~n}o, ``{Donaldson Invariants of Product Ruled Surfaces
  and Two-Dimensional Gauge Theories},'' {\em Communications in Mathematical
  Physics}, vol.~220, no.~2, pp.~231--261, 2001.

\bibitem{labastida2005topological}
J.~Labastida and M.~Mari{\~n}o, {\em {Topological quantum field theory and four
  manifolds}}, vol.~25.
\newblock Springer, 2005.

\bibitem{moore1997integration}
G.~Moore and E.~Witten, ``{Integration over the u-plane in Donaldson theory},''
  {\em arXiv preprint hep-th/9709193}, 1997.

\bibitem{atiyah1978self}
M.~Atiyah, N.~Hitchin, and I.~Singer, ``{Self-duality in four-dimensional
  Riemannian geometry},'' {\em Proceedings of the Royal Society of London. A.
  Mathematical and Physical Sciences}, vol.~362, no.~1711, pp.~425--461, 1978.

\bibitem{atiyah1988new}
M.~Atiyah, ``{New invariants of 3-and 4-dimensional manifolds},'' {\em The
  mathematical heritage of Hermann Weyl (Durham, NC, 1987)}, vol.~48,
  pp.~285--299, 1988.

\bibitem{donaldson2002floer}
S.~Donaldson, {\em {Floer homology groups in Yang-Mills theory}}, vol.~147.
\newblock Cambridge University Press, 2002.

\bibitem{popov2015loop}
A.~Popov, ``{Loop groups in Yang--Mills theory},'' {\em Physics Letters B},
  vol.~748, pp.~439--442, 2015.

\bibitem{hori2003mirror}
K.~Hori, S.~Katz, C.~Vafa, and R.~Pandharipande, {\em {Mirror symmetry}},
  vol.~1.
\newblock American Mathematical Soc., 2003.

\bibitem{salamon1998notes}
D.~Salamon, ``{Notes on flat connections and the loop group},'' {\em Preprint,
  University of Warwick}, 1998.

\bibitem{elbistan2017weyl}
M.~Elbistan, ``{Weyl semimetal and topological numbers},'' {\em International
  Journal of Modern Physics B}, vol.~31, no.~29, p.~1750221, 2017.

\bibitem{shnir2006magnetic}
Y.~Shnir, {\em {Magnetic monopoles}}.
\newblock Springer Science \& Business Media, 2006.

\bibitem{witten1994monopoles}
E.~Witten, ``{Monopoles and four-manifolds},'' {\em arXiv preprint
  hep-th/9411102}, 1994.

\bibitem{tong2005tasi}
D.~Tong, ``{TASI lectures on solitons},'' {\em arXiv preprint hep-th/0509216},
  2005.

\bibitem{dey2006geometric}
R.~Dey, ``{Geometric prequantization of the moduli space of the vortex
  equations on a Riemann surface},'' {\em Journal of mathematical physics},
  vol.~47, no.~10, p.~103501, 2006.

\bibitem{kapustin2006electric}
A.~Kapustin and E.~Witten, ``{Electric-magnetic duality and the geometric
  Langlands program},'' {\em arXiv preprint hep-th/0604151}, 2006.

\bibitem{kutluhan2010hf}
C.~Kutluhan, Y.-J. Lee, and C.~Taubes, ``{HF= HM I: Heegaard Floer homology and
  Seiberg--Witten Floer homology},'' {\em arXiv preprint arXiv:1007.1979},
  2010.

\bibitem{munoz1999ring}
V.~Mu{\~n}oz, ``{Ring structure of the Floer cohomology of {$\Sigma\times
  S^1$}},'' {\em Topology}, vol.~38, no.~3, pp.~517--528, 1999.

\bibitem{munoz2005seiberg}
V.~Mu{\~n}oz and B.-L. Wang, ``{Seiberg--Witten--Floer homology of a surface
  times a circle for non-torsion spinC structures},'' {\em Mathematische
  Nachrichten}, vol.~278, no.~7-8, pp.~844--863, 2005.

\bibitem{floer1988instanton}
A.~Floer, ``{An instanton-invariant for 3-manifolds},'' {\em Communications in
  mathematical physics}, vol.~118, no.~2, pp.~215--240, 1988.

\bibitem{dostoglou1994self}
S.~Dostoglou and D.~Salamon, ``{Self-dual instantons and holomorphic curves},''
  {\em Annals of Mathematics}, vol.~139, no.~3, pp.~581--640, 1994.

\bibitem{luo2017four}
Y.~Luo, M.-C. Tan, P.~Vasko, and Q.~Zhao, ``{Four-dimensional N=2
  supersymmetric theory with boundary as a two-dimensional complex Toda
  theory},'' {\em Journal of High Energy Physics}, vol.~2017, no.~5, p.~121,
  2017.

\bibitem{witten1989quantum}
E.~Witten, ``{Quantum field theory and the Jones polynomial},'' {\em
  Communications in Mathematical Physics}, vol.~121, no.~3, pp.~351--399, 1989.

\bibitem{francesco2012conformal}
P.~Francesco, P.~Mathieu, and D.~S{\'e}n{\'e}chal, {\em {Conformal field
  theory}}.
\newblock Springer Science \& Business Media, 2012.

\bibitem{segal1988definition}
G.~Segal, ``{The definition of conformal field theory},'' in {\em Differential
  geometrical methods in theoretical physics}, pp.~165--171, Springer, 1988.

\end{thebibliography}
\end{document}